\documentclass[11pt]{article}
\usepackage{comment,soul}
\usepackage{hyperref}
\hypersetup{
  colorlinks=true,
  linkcolor=blue!60!black,
  citecolor=blue!60!black,
  urlcolor=blue!60!black,
}
\usepackage{enumitem}
\usepackage{amsmath,amssymb,epsf,cite,graphicx,subfigure}
\usepackage{braket,tikzsymbols}
\usepackage[bbgreekl]{mathbbol}
\usepackage[export]{adjustbox}
\usepackage{dsfont}
\usepackage{faktor}
\usepackage{mathrsfs}
\usepackage{float}
\usepackage{empheq}
\usepackage{fancyhdr}

\usepackage{comment}

\usepackage{amsmath,amssymb,amsthm}
\usepackage[margin=1in]{geometry}
\usepackage[numbers,sort&compress]{natbib}
\usepackage{hyperref}
\usepackage{booktabs}
\usepackage{graphicx}
\usepackage[section]{placeins}
\usepackage{slashed}
\usepackage{tabularx}
\usepackage[dvipsnames]{xcolor}
\usepackage[normalem]{ulem}
\usepackage{tikz}
\usepackage[compat=1.1.0]{tikz-feynman}
\tikzfeynmanset{warn luatex=false}

\usetikzlibrary{arrows.meta,decorations.markings}

\tikzset{
  fermion line/.style={very thick,
    postaction={decorate,
      decoration={markings, mark=at position 0.55 with {\arrow{Stealth[length=6pt,width=5pt]}}}}},
  scalar line/.style={thick, dashed, red},
  dressed scalar line/.style={thick, double, double distance=2pt, dashed, red},
  fermion bubble/.style={very thick,
    postaction={decorate,
      decoration={markings,
        mark=at position 0.25 with {\arrow{Stealth[length=6pt,width=5pt]}},
        mark=at position 0.75 with {\arrow{Stealth[length=6pt,width=5pt]}}}}}
}

\newcommand{\bubblechain}{%
\begin{tikzpicture}[>=Stealth]
  \def\R{2.2}        
  \def\rb{0.28}      
  \def\xM{2.5}       
  \def\arcStart{150}  
  \def\arcEnd{30}     
  \pgfmathsetmacro{\yC}{-\R*sin(\arcEnd)}
  \def\tA{120}  
  \def\tB{90}   
  \def\tC{60}   
  \pgfmathsetmacro{\delta}{asin(\rb/\R)}
  \pgfmathsetmacro{\xL}{\xM+\R*cos(\arcStart)}
  \pgfmathsetmacro{\xR}{\xM+\R*cos(\arcEnd)}
  \draw[fermion line] (0,0) -- (\xL,0);
  \draw[fermion line] (\xL,0) -- (\xR,0);
  \draw[fermion line] (\xR,0) -- (5,0);
  \fill (\xL,0) circle (1.5pt);
  \fill (\xR,0) circle (1.5pt);
  \draw[scalar line] (\xM,\yC) ++(\arcStart:\R) arc[start angle=\arcStart, end angle={\tA+\delta}, radius=\R];
  \draw[scalar line] (\xM,\yC) ++({\tA-\delta}:\R) arc[start angle={\tA-\delta}, end angle={\tB+\delta}, radius=\R];
  \draw[scalar line] (\xM,\yC) ++({\tB-\delta}:\R) arc[start angle={\tB-\delta}, end angle={\tC+\delta}, radius=\R];
  \draw[scalar line] (\xM,\yC) ++({\tC-\delta}:\R) arc[start angle={\tC-\delta}, end angle=\arcEnd, radius=\R];
  \draw[fermion bubble] ({\xM+\R*cos(\tA)},{\yC+\R*sin(\tA)}) circle (\rb);
  \draw[fermion bubble] ({\xM+\R*cos(\tB)},{\yC+\R*sin(\tB)}) circle (\rb);
  \draw[fermion bubble] ({\xM+\R*cos(\tC)},{\yC+\R*sin(\tC)}) circle (\rb);
  \node[below] at (0.3,0) {$p$};
  \node[below] at (4.7,0) {$p$};
\end{tikzpicture}%
}

\newcommand{\selfenergy}{%
\begin{tikzpicture}[>=Stealth]
  \def\R{1.4}
  \def\xM{2.0}
  \def\arcStart{150}
  \def\arcEnd{30}
  \pgfmathsetmacro{\yC}{-\R*sin(\arcEnd)}
  \pgfmathsetmacro{\xL}{\xM+\R*cos(\arcStart)}
  \pgfmathsetmacro{\xR}{\xM+\R*cos(\arcEnd)}
  \draw[fermion line] (0,0) -- (\xL,0);
  \draw[fermion line] (\xL,0) -- (\xR,0);
  \draw[fermion line] (\xR,0) -- (4,0);
  \fill (\xL,0) circle (1.5pt);
  \fill (\xR,0) circle (1.5pt);
  \draw[dressed scalar line] (\xM,\yC) ++(\arcStart:\R) arc[start angle=\arcStart, end angle=\arcEnd, radius=\R];
  \node[below] at (0.4,0) {$p$};
  \node[below] at (2.0,0) {$p{+}k$};
  \node[below] at (3.6,0) {$p$};
  \node[above] at (\xM,{\yC+\R*sin(90)}) {$k$};
\end{tikzpicture}%
}

\tikzset{
  soft fermion/.style={very thick, blue!70!black,
    postaction={decorate,
      decoration={markings, mark=at position 0.55 with
        {\arrow{Stealth[length=5pt,width=4pt]}}}}},
  soft fermion label/.style={blue!70!black, font=\small},
  operator x/.style={blue!75!black, font=\Large, inner sep=0pt},
  operator label/.style={blue!75!black, font=\scriptsize},
  operator leg/.style={very thick, blue!75!black}
}

\newcommand{\CintLocalFull}{%
\begin{tikzpicture}[>=Stealth, baseline=-0.5ex]
  \draw[fermion line] (0,0) -- (1.5,0);
  \draw[fermion line] (1.5,0) -- (3.0,0);
  \fill (1.5,0) circle (1.5pt);
  \draw[scalar line] (1.5,0) -- (1.5,0.85);
  \fill (1.5,0.85) circle (1.5pt);
  \draw[soft fermion] (0.65,1.55) -- (1.5,0.85);
  \draw[soft fermion] (1.5,0.85) -- (2.35,1.55);
  \node[above, font=\scriptsize] at (0,0.10) {$0$};
  \node[below, font=\scriptsize] at (1.5,-0.05) {$y$};
  \node[above, font=\scriptsize] at (3.0,0.10) {$x$};
  \node[below, orange!85!black, font=\scriptsize] at (0.55,1.52) {$p_a$};
  \node[below, orange!85!black, font=\scriptsize] at (2.45,1.52) {$p_b$};
\end{tikzpicture}%
}

\newcommand{\CintLocalFullB}{%
\begin{tikzpicture}[>=Stealth, baseline=-0.5ex]
  \draw[fermion line] (0,0) -- (1.0,0);
  \draw[scalar line] (1.0,0) -- (2.0,0);
  \draw[fermion line] (2.0,0) -- (3.0,0);
  \fill (1.0,0) circle (1.5pt);
  \fill (2.0,0) circle (1.5pt);
  \draw[soft fermion] (1.0,0) -- (1.0,1.25);
  \draw[soft fermion] (2.0,1.25) -- (2.0,0);
  \node[above, font=\scriptsize] at (0,0.10) {$0$};
  \node[below, font=\scriptsize] at (1.0,-0.05) {$z$};
    \node[below, font=\scriptsize] at (2.0,-0.05) {$y$};
  \node[above, font=\scriptsize] at (3.0,0.10) {$x$};
  \node[below, orange!85!black, font=\scriptsize] at (0.7,0.7) {$p_b$};
  \node[below, orange!85!black, font=\scriptsize] at (2.3,0.7) {$p_a$};
\end{tikzpicture}%
}

\newcommand{\CintLocalOPE}{%
\begin{tikzpicture}[>=Stealth, baseline=-0.5ex]
  \def\Rop{0.18}   
  \def\Rloop{0.46} 
  \def\topY{0.88}  
  \draw[very thick, blue!75!black] (1.5,0) circle (\Rop);
  \node[operator x] at (1.5,0) {$\times$};
  \draw[fermion line] ({1.5+\Rop},0) arc[start angle=-67, end angle=90, radius=\Rloop];
  \draw[fermion line] (1.5,\topY) arc[start angle=90, end angle=247, radius=\Rloop];
  \fill (1.5,\topY) circle (1.5pt);
  \draw[scalar line] (1.5,\topY) -- (1.5,{\topY+0.65});
  \fill (1.5,{\topY+0.65}) circle (1.5pt);
  \draw[soft fermion] (0.65,{\topY+1.35}) -- (1.5,{\topY+0.65});
  \draw[soft fermion] (1.5,{\topY+0.65}) -- (2.35,{\topY+1.35});
  \node[below, orange!85!black, font=\scriptsize] at (0.55,{\topY+1.32}) {$p_a$};
  \node[below, orange!85!black, font=\scriptsize] at (2.45,{\topY+1.32}) {$p_b$};
\end{tikzpicture}%
}

\newcommand{\CintLocalOPEB}{%
\begin{tikzpicture}[>=Stealth, baseline=-0.5ex]
  \def\Rop{0.18}
  \def\Rloop{0.55}

  \tikzset{
    soft open ext/.style={very thick, blue!70!black,
      postaction={decorate,
        decoration={markings, mark=at position 0.55 with
          {\arrow{Stealth[length=5pt,width=4pt]}}}}},
    soft open arc/.style={very thick, blue!70!black,
      postaction={decorate,
        decoration={markings, mark=at position 0.50 with
          {\arrow{Stealth[length=5pt,width=4pt]}}}}}
  }

  \coordinate (L) at (0.95,0);
  \coordinate (R) at (2.05,0);
  \coordinate (O) at (1.50,-\Rloop);

  \draw[soft open ext] (0.20,0) -- (L);
  \draw[soft open ext] (R) -- (2.80,0);

  \pgfmathsetmacro{\opgap}{2*asin(\Rop/(2*\Rloop))}
  \pgfmathsetmacro{\angL}{270-\opgap}
  \pgfmathsetmacro{\angR}{270+\opgap}

  \draw[soft open arc] (L) arc[start angle=180,end angle=\angL,radius=\Rloop];
  \draw[soft open arc] ({1.50+\Rloop*cos(\angR)},{\Rloop*sin(\angR)})
    arc[start angle=\angR,end angle=360,radius=\Rloop];

  \draw[very thick, blue!75!black] (O) circle (\Rop);
  \node[operator x] at (O) {$\times$};

  \draw[scalar line] (L) arc[start angle=180,end angle=0,radius=\Rloop];

  \fill (L) circle (1.5pt);
  \fill (R) circle (1.5pt);

  \node[above, orange!85!black, font=\scriptsize] at (0.28,0.06) {$p_a$};
  \node[above, orange!85!black, font=\scriptsize] at (2.72,0.06) {$p_b$};
\end{tikzpicture}%
}

\newcommand{\CtwoLOfull}{%
\begin{tikzpicture}[>=Stealth, baseline=-0.5ex]
  \draw[fermion line] (0,0) -- (0.9,0);
  \draw[fermion line] (0.9,0) -- (2.1,0);
  \draw[fermion line] (2.1,0) -- (3.0,0);
  \fill (0.9,0) circle (1.5pt);
  \fill (2.1,0) circle (1.5pt);
  \draw[scalar line] (0.9,0) -- (0.9,0.75);
  \draw[scalar line] (2.1,0) -- (2.1,0.75);
  \fill (0.9,0.75) circle (1.5pt);
  \fill (2.1,0.75) circle (1.5pt);
  \draw[soft fermion] (0.35,1.45) -- (0.9,0.75);
  \draw[soft fermion] (0.9,0.75) -- (1.35,1.45);
  \draw[soft fermion] (1.65,1.45) -- (2.1,0.75);
  \draw[soft fermion] (2.1,0.75) -- (2.65,1.45);
  \node[below] at (0.3,0) {$p$};
  \node[below] at (2.7,0) {$p$};
\end{tikzpicture}%
}

\tikzset{
  mom arrow/.style={-{Stealth[length=6pt,width=5pt]}, orange!85!black, very thick}
}

\newcommand{\freeleftA}{%
\begin{tikzpicture}[>=Stealth, baseline=0.8ex, scale=0.55]
  \fill[blue!70!black] (0,1.8) circle (3pt);
  \fill[blue!70!black] (1.8,1.8) circle (3pt);
  \node[above, blue!70!black, font=\scriptsize] at (0,1.8) {$x$};
  \node[above, blue!70!black, font=\scriptsize] at (1.8,1.8) {$0$};
  \fill[blue!70!black] (0,0.4) circle (3pt);
  \fill[blue!70!black] (1.8,0.4) circle (3pt);
  \draw[soft fermion] (0,0.4) -- (0,1.8);
  \draw[soft fermion] (1.8,1.8) -- (1.8,0.4);
  \draw[mom arrow] (-0.65,-0.15) -- (-0.05,0.35);
  \draw[mom arrow] (2.45,-0.15) -- (1.85,0.35);
  \node[below left, orange!85!black, font=\scriptsize] at (-0.55,-0.1) {$p_a$};
  \node[below right, orange!85!black, font=\scriptsize] at (2.35,-0.1) {$p_b$};
\end{tikzpicture}%
}

\newcommand{\freeleftB}{%
\begin{tikzpicture}[>=Stealth, baseline=0.8ex, scale=0.55]
  \fill[blue!70!black] (0,1.8) circle (3pt);
  \fill[blue!70!black] (1.8,1.8) circle (3pt);
  \node[above, blue!70!black, font=\scriptsize] at (0,1.8) {$x$};
  \node[above, blue!70!black, font=\scriptsize] at (1.8,1.8) {$0$};
  \fill[blue!70!black] (0,0.4) circle (3pt);
  \fill[blue!70!black] (1.8,0.4) circle (3pt);
  \draw[soft fermion] (0,1.8) -- (1.8,1.8);
  \draw[soft fermion] (0,0.4) -- (1.8,0.4);
  \draw[mom arrow] (-0.65,-0.15) -- (-0.05,0.35);
  \draw[mom arrow] (2.45,-0.15) -- (1.85,0.35);
  \node[below left, orange!85!black, font=\scriptsize] at (-0.55,-0.1) {$p_a$};
  \node[below right, orange!85!black, font=\scriptsize] at (2.35,-0.1) {$p_b$};
\end{tikzpicture}%
}

\newcommand{\scalarselfenergy}{%
\begin{tikzpicture}[>=Stealth, baseline=-0.5ex,scale=1.3]
  \def\rb{0.3}
  \draw[scalar line] (0,0) -- (0.4,0);
  \draw[scalar line] ({0.4+2*\rb},0) -- ({0.8+2*\rb},0);
  \draw[fermion bubble] ({0.4+\rb},0) circle (\rb);
  \fill (0.4,0) circle (1.5pt);
  \fill ({0.4+2*\rb},0) circle (1.5pt);
  \node[below] at (0.15,0) {$p$};
  \node[below] at ({0.65+2*\rb},0) {$p$};
\end{tikzpicture}%
}

\newcommand{\sigmasquaredvev}{%
\begin{tikzpicture}[>=Stealth, baseline=-0.5ex]
  \def\Rop{0.18}
  \def\Rloop{0.5}
  \draw[dressed scalar line] (1.0,\Rloop) circle (\Rloop);
  \filldraw[very thick, fill=white, draw=blue!75!black] (1.0,0) circle (\Rop);
  \node[operator x] at (1.0,0) {$\times$};
\end{tikzpicture}%
}

\newcommand{\selfenergySigma}{%
\begin{tikzpicture}[>=Stealth, baseline=-0.5ex]
  \draw[fermion line] (0,0) -- (0.8,0);
  \draw[fermion line] (0.8,0) -- (2.2,0);
  \draw[fermion line] (2.2,0) -- (3.0,0);
  \fill (0.8,0) circle (1.5pt);
  \fill (2.2,0) circle (1.5pt);
  \draw[scalar line] (0.8,0) .. controls (0.8,0.9) and (2.2,0.9) .. (2.2,0);
  \node[above,red] at (1.5,0.7) {$\sigma$};
\end{tikzpicture}%
}

\newcommand{\selfenergyTadpole}{%
\begin{tikzpicture}[>=Stealth, baseline=-0.5ex]
  \def\rb{0.30}
  \draw[fermion line] (0,0) -- (1.5,0);
  \draw[fermion line] (1.5,0) -- (3.0,0);
  \fill (1.5,0) circle (1.5pt);
  \draw[fermion bubble] (1.5,{0.0+\rb}) circle (\rb);
\end{tikzpicture}%
}

\newcommand{\bilinearVC}{%
\begin{tikzpicture}[>=Stealth, baseline=0.8ex]
  \def\Rop{0.18}
  \node[operator x] at (1.5,1.4) {$\times$};
  \draw[very thick, blue!75!black] (1.5,1.4) circle (\Rop);
  \draw[fermion line] (0.5,0.5) -- ({1.5-0.13},{1.4-0.13});
  \draw[fermion line] ({1.5+0.13},{1.4-0.13}) -- (2.5,0.5);
  \fill (0.5,0.5) circle (1.5pt);
  \fill (2.5,0.5) circle (1.5pt);
  \draw[scalar line] (0.5,0.5) -- (2.5,0.5);
  \draw[fermion line] (0,0) -- (0.5,0.5);
  \draw[fermion line] (2.5,0.5) -- (3.0,0);
  \node[above, red] at (1.5,0.55) {$\sigma$};
\end{tikzpicture}%
}

\newcommand{\rhotriangle}{%
\begin{tikzpicture}[scale=1.0, >=stealth]
  \coordinate (O) at (0,0);
  \coordinate (L) at (3.5,0);        
  \coordinate (Le) at (-1.75,3.03);  
  \fill[blue!5] (O) -- (L) arc[start angle=0, end angle=120, radius=3.5] -- cycle;
  \draw[gray, dashed, thin] (-2.5,0) -- (5.2,0);
  \draw[gray, dashed, thin] (0,-0.5) -- (0,3.8);
  \draw[thick, blue, ->] (O) -- (1.75,0);
  \draw[thick, blue] (1.75,0) -- (L);
  \draw[thick, red, ->] (L) arc[start angle=0, end angle=60, radius=3.5];
  \draw[thick, red] (L) arc[start angle=0, end angle=120, radius=3.5];
  \draw[thick, green!60!black, ->] (Le) -- (-0.875,1.515);
  \draw[thick, green!60!black] (-0.875,1.515) -- (O);
  \draw[green!60!black, dashed, ->] (Le) -- (-2.25,3.9)
    node[above left, font=\small] {$\infty e^{i\alpha}$};
  \fill (O) circle (2pt) node[below left] {$0$};
  \fill (L) circle (2pt) node[below] {$\ln\Lambda$};
  \fill (Le) circle (2pt) node[above left] {$\ln\Lambda\cdot e^{i\alpha}$};
  \node[blue, below, font=\small] at (1.75,-0.45) {real axis};
  \node[red, right, font=\small] at (3.2,1.8)
    {arc $=\dfrac{\Lambda^{1-t}}{1-t}$};
  \node[green!60!black, left, font=\small] at (-1.4,1.0)
    {ray $\to\dfrac{1}{t-1}$};
  \draw[->] (0.8,0) arc[start angle=0, end angle=120, radius=0.8];
  \node[font=\small] at (0.15,0.7) {$\alpha$};
  \node[gray, font=\small] at (5.0,-0.3) {$\re\rho$};
  \node[gray, font=\small] at (0.35,3.8) {$\im\rho$};
\end{tikzpicture}%
}

\makeatletter
\def\fps@figure{t}
\makeatother

\newcommand{\Ein}{\mathrm{Ein}}

\newcommand{\im}{\mathrm{Im}\,}
\newcommand{\re}{\mathrm{Re}\,}
\newcommand{\psibar}{\overline{\psi}}
\newcommand{\psibartilde}[2]{\widetilde{\psibar}_{#2}^{\raisebox{-1.0ex}{$\scriptstyle #1$}}}
\newcommand{\Tr}{\mathrm{Tr}}
\newcommand{\tr}{\mathrm{tr}}

\newcommand{\cO}{\mathcal{O}}
\newcommand{\cL}{\mathcal{L}}
\newcommand{\cB}{\mathcal{B}}

\newcommand{\resy}{\underset{y = 1}{\text{Res}}}
\newcommand{\reszz}{\underset{z = 1}{\text{Res}}}
\newcommand{\opone}{ \mathbf{1}}
\newcommand{\bare}{\text{bare}}
\newcommand{\meas}[2]{\frac{d^{#2}#1}{(2\pi)^{#2}}}

\definecolor{darkred}{rgb}{0.5,0.0,0.0}
\definecolor{darkblue}{rgb}{0.0,0.0,0.9}
\definecolor{darkerblue}{rgb}{0.0,0.0,0.5}
\definecolor{darkgreen}{rgb}{0.0,0.5,0.0}
\definecolor{darkpurple}{rgb}{0.5, 0.2, 0.8}

\begin{document}

\begin{flushright}
    DESY-26-120 
\end{flushright}

\begin{center}
{\LARGE \bf Renormalon Saddles in the OPE}
\vskip 1cm

\textbf{Arindam Bhattacharya$^{1,a}$, Jordan Cotler$^{2,b}$, Aur\'{e}lien Dersy$^{2,c}$, \\ and Matthew D.~Schwartz$^{2,d}$}

\vspace{0.5cm}

{\it ${}^1$ Deutsches Elektronen-Synchrotron DESY, Notkestr. 85, 22607 Hamburg, Germany \\}
{\it ${}^2$ Department of Physics, Harvard University, Cambridge, MA 02138, USA\\}

\vspace{0.3cm}

{\tt  
${}^a$arindam.bhattacharya@desy.de
${}^b$jcotler@fas.harvard.edu, \\ ${}^c$adersy@g.harvard.edu
${}^d$schwartz@g.harvard.edu\\}

\medskip
\end{center}

\vskip1cm

\begin{center}
{\bf Abstract}
\end{center}
\hspace{.3cm} 

The cancellation of renormalon ambiguities in the operator product expansion is usually formulated as a matching between perturbative Borel ambiguities and prescription dependence in non-perturbative condensates. We give this cancellation a contour realization in the two-dimensional Gross--Neveu model at large $N$. The exact order-$1/N$ fermion self-energy in the massive vacuum is a finite real integral; ambiguities arise only after its large-momentum behavior is decomposed into products of perturbative Wilson coefficients and the matrix elements of the associated operators. A reduced integral isolates the leading renormalon and reveals a two-dimensional contour geometry in which the perturbative series is generated at the boundary corner where the exchanged momentum reaches the external hard scale, while its Stokes discontinuity is controlled by a distinct boundary critical point at zero exchanged momentum. After renormalizing the ultraviolet divergences of the condensate scale integral, we derive its compensating lateral ambiguity without fixing it by matching to the perturbative sector, distinguish the renormalon saddle from the apparent IR Landau pole, and obtain the reduced contour variables from the large-$N$ auxiliary-field path integral. The renormalon saddle is a boundary critical point of this reduced mode-space integral. The perturbative Wilson coefficient and the renormalized condensate thus define complementary relative cycles of the same reduced integrand whose complex tails cancel, reconstructing the original real integration cycle.

\newpage

\newpage
\tableofcontents

\section{Introduction}
\label{sec:intro}

The operator product expansion (OPE) separates the short-distance dependence of a correlation function from matrix elements of renormalized local operators that encode its long-distance physics~\cite{Wilson:1969zs,SVZ1979,SVZ1979b}. In an asymptotically free theory, the perturbative expansions of the Wilson coefficients appearing in the OPE are generally factorially divergent so that the series is asymptotic. Singularities on the positive axis in their Borel transforms make the resummation prescription dependent, with the two lateral prescriptions differing by imaginary terms whose scaling matches the corresponding power corrections in the OPE~\cite{SVZ1979,SVZ1979b,David:1982qv,David1984,Mueller:1984vh,David1986,Beneke:1998ui}. The corresponding operator matrix elements must carry compensating prescription dependence, since only the complete OPE contribution defines a physical observable. Power counting and dimensional analysis identify the operator sector capable of canceling a given renormalon, but do not by themselves explain how the renormalized matrix element acquires the required imaginary part with the correct normalization and sign.

For large-order behavior associated with instantons, such cancellations admit a geometric description. The complexified path integral decomposes into Lefschetz thimbles, and Stokes jumps in perturbation theory around one saddle are canceled by corresponding jumps of other saddle contributions~\cite{BogomolnyZinnJustin1980,ZinnJustin2004,Aniceto2019,Dunne2016,Dersy:2026ncf,Dersy:2026jat}. Renormalons ~\cite{tHooft1977,Lautrup1977,Parisi1978,Parisi1979}, by contrast, are typically associated with factorial growth generated by loop-momentum integrations, as in bubble-chain diagrams, with the corresponding ambiguities canceled by operator matrix elements in the OPE~\cite{SVZ1979,SVZ1979b,David1984,Mueller:1984vh,David1986,Maiani:1991az,Luke:1994xd,Beneke:1998ui}. Although the perturbative Wilson coefficient and the compensating condensate are known to carry equal and opposite ambiguities, they are not usually represented as associated steepest descent contour contributions controlled by a common critical point.

Recently, there has been a body of work analyzing the  renormalons in lower dimensional field theories with known exact results from the trans-series perspective~\cite{Marino:2025ido,Marino:2022ykm,Marino:2020dgc,Marino:2019fvu,Marino:2019eym,Marino:2021six} and in compactified dimensions~\cite{Fujimori:2018kqp,Ishikawa:2019tnw}. These studies either provide valuable tests of renormalon cancellation for observables that admit an OPE, or are restricted to quantities such as the free energy, which to our knowledge does not admit an OPE. For the latter the fate of the cancellation of the renormalons in theories with a compactified dimension is still to be resolved~\cite{Anber:2014sda,Morikawa:2020agf,Ashie:2020bvw,Ishikawa:2019oga}. Thus, a semiclassical geometric interpretation of renormalons, and in particular a contour description of their cancellation within the OPE, remains to be developed.

In this work, we develop such a contour description for the leading renormalon in the fermion self-energy of the two-dimensional Gross--Neveu model at large $N$. The model is asymptotically free, generates a fermion mass $m = \mu e^{-1/\lambda(\mu)}$ by dimensional transmutation, and permits an exact treatment of the fermion self-energy at leading nontrivial order in $1/N$~\cite{GrossNeveu1974,CampostriniRossi1992,Marino2024}. We focus on the form factor $A(p^2)$ multiplying the $\slashed p$ structure of the fermion self-energy. In the massive vacuum, this form factor is given by a finite and real integral. Renormalon ambiguities enter only after that integral is decomposed into asymptotic perturbative Wilson coefficients and power-suppressed matrix elements in the OPE.

Perturbation theory about the chirally symmetric configuration (when the fermion mass is set to its classical value of zero) produces the usual IR renormalons in the form factor, either via tracking bubble-chain diagrams, or via integration over the renormalized coupling with its one-loop running. Its Borel transform $\cB(t)$, where $t$ is conjugate to a scaled 't~Hooft coupling, has a leading pole at $t = 1$, giving an ambiguity of order $m^2/p^2$. The large-momentum trans-series of the exact massive-vacuum integral contains the same perturbative pole together with a dimension-two power correction of equal and opposite residue, so the cancellation is already manifest in the Mellin--Barnes representation of the exact answer~\cite{Marino2024}. We identify the contour geometry responsible for this relation and derive the compensating ambiguity directly from the renormalized condensate integral, without fixing it by matching to the perturbative ambiguity.

The full self-energy contains several renormalons together with the associated tower of power corrections, whose detailed structure obscures the contour geometry governing the leading renormalon. We isolate the first cancellation by replacing the exact angular kernel with its leading small-$m/p$ form. The resulting reduced observable retains the perturbative pole and the compensating dimension-two term with their exact relative normalization, while discarding higher power sectors and details of the full integral that do not affect the leading Stokes discontinuity. The reduced integral therefore isolates the leading renormalon cancellation without reproducing the complete large-momentum expansion of the self-energy.

As in Ref.~\cite{BenekeBraunKivel1998}, we find that by expressing the asymptotic series in terms of a non-perturbative coupling without a Landau pole, the reduced integral can be recast as a two-dimensional contour problem. Introducing a logarithmic scale coordinate $\rho$, with $0 \leq \rho \leq \ln Z$ and $Z$ the image of the external hard scale $p$, the integration domain consists of two adjacent rectangles in the $(t,\rho)$ plane, with $0 \leq t \leq 1$ and $1 \leq t \leq 2$, entering with a relative minus sign. Both are governed by the renormalon-sensitive reduced action $S(t,\rho) = (t - 1)\rho$. Within this two-dimensional contour problem, the perturbative expansion and its Stokes discontinuity are controlled by distinct points. The factorial coefficients are generated at the hard-scale corner $(t,\rho) = (0,\ln Z)$, whereas the Stokes discontinuity is controlled by the boundary critical point $(t,\rho) = (1,0)$, which we call the renormalon saddle. Thus the hard corner generates the asymptotic series, whereas the renormalon saddle controls the ambiguity of its lateral resummation.

We refer to the relative steepest-descent class associated with the hard-scale corner as the `perturbative boundary thimble'. Convenient convergent representatives of this class give the two lateral Borel sums. Their lateral ambiguity is controlled locally by a half-thimble through the renormalon saddle at $(t,\rho) = (1,0)$, while additional real contributions depend on the global completion of the contour.

The reduced $(t,\rho)$ contour representation also distinguishes the renormalon saddle from the Landau pole of the running coupling.  For the $\theta$-reduced observable, expressing the answer in terms of a non-perturbative coupling without an IR Landau pole reorganizes the power-suppressed contributions: the infinite sequence of sectors generated when the answer is re-expanded in the original coupling is recast into the finite set displayed below.  This simplification is a property of the reduced leading-renormalon problem; the full self-energy contains additional renormalons and their associated power sectors.  By contrast, the leading positive-axis Borel singularity at $t=1$ and the corresponding critical point of the reduced action persist under this reorganization, since the two couplings agree to all orders in ordinary perturbation theory.

The OPE identifies the dimension-two operator responsible for canceling the ambiguity associated with the renormalon saddle. For the form factor $A(p^2)$, the relevant scalar operator is $V = g(\psibar\psi)^2 = \sigma^2/g$ with $\sigma$ denoting the auxiliary field that becomes dynamical. Its vacuum expectation value carries the corresponding renormalon ambiguity, which cancels that of the Wilson coefficient. At the leading order in $1/N$ relevant here, $V$ is the only dimension-two operator in the scalar-singlet channel that gives a noncontact contribution to $A(p^2)$ with the normalization required for this cancellation.

The fluctuation contribution to $\langle V\rangle$ is governed by the same scale integration kernel as the reduced self-energy, but its integration range extends to arbitrarily high momentum. At finite ultraviolet cutoff, the condensate integral is real because the apparent singularities in its Borel representation cancel within the regulated expression. After renormalization, the operator's Borel representation retains a pole at $t = 1$. Its two lateral continuations carry opposite imaginary parts whose magnitude and sign are precisely those required to cancel the corresponding ambiguities of the perturbative Wilson coefficient; their average gives the principal-value prescription.

The cancellation between the perturbative Wilson-coefficient ambiguity and the condensate ambiguity can be stated directly as an identity between relative cycles. The perturbative cycle is anchored at the upper boundary $\rho = \ln{Z}$ of the finite scale interval, whereas the condensate cycle is anchored at the lower boundary $\rho = 0$. The two cycles extend toward complex infinity in the same asymptotic direction and encounter the renormalon saddle with opposite orientations. When their contributions are added, the complex tails cancel and the finite real strip of the reduced $(t,\rho)$ integral is reconstructed. The same cancellation is explicit after performing the $\rho$ (scale) integration: the two boundary contributions combine to replace the apparent pole $1/(1-t)$ by $(Z^{1-t}-1)/(1-t)$, which is regular at $t = 1$. The perturbative Wilson coefficient and the condensate can therefore be viewed as the two prescription-dependent relative cycles obtained by decomposing the original real integration cycle.

The reduced contour variables also have a direct interpretation in the large-$N$ auxiliary-field path integral. The coordinate $\rho$ parametrizes the momentum scale of the exchanged auxiliary-field fluctuation, and $t$ parametrizes its amplitude after normalization by the quadratic fluctuations about the massive vacuum. In these variables, the Gaussian mode action takes the simple form $t\rho$, and the momentum-space Jacobian modifies the effective exponent to the reduced action $S(t,\rho) = (t - 1)\rho$. At the renormalon saddle $(t,\rho) = (1,0)$, both the exchanged momentum and the amplitude of the original auxiliary-field mode vanish; $t$ remains finite only because the normalization becomes singular in this limit. The renormalon saddle is therefore a boundary critical point of the reduced mode integral rather than a new saddle of the full auxiliary-field path integral.

The large-$N$ $O(N)$ nonlinear sigma model provides a useful comparison with our Gross-Neveu analysis, which suggests how our results generalize to other models. For the sigma model, its auxiliary-field propagator contains similar logarithmic scale dependence that appears in the Gross--Neveu analysis, and the leading IR renormalon in the propagator of the auxiliary mode is canceled by the condensate of the dimension-two Lagrange-multiplier field~\cite{David:1982qv,David1984,David1986,Novikov1985,BenekeBraunKivel1998}. Near the leading singularity, the perturbative and condensate contributions are again described locally by the reduced action $(t - 1)\rho$, with opposite lateral orientations. The global structures are different: the momentum integral of the two-point function does not admit the two-rectangle contour representation found in the Gross--Neveu model, and the perturbative identity coefficient contains an infinite sequence of positive-axis IR renormalons. The comparison therefore concerns the local saddle geometry governing the leading renormalon cancellation, for which we find the same reduced structure in the Borel variable $t$ and an appropriate logarithmic scale coordinate $\rho$.

Our construction uses the continuum-renormalized OPE. After the local operator and its matrix element have been renormalized, the ultraviolet regulator is removed, leaving the Wilson coefficient and condensate separately dependent on correlated lateral prescriptions. A Wilsonian OPE instead keeps a finite factorization scale that partitions momentum regions between coefficients and matrix elements, with the factorization-scale dependence canceling in the physical observable. In the continuum organization adopted here, the two prescription-dependent contributions remain explicit and can be represented directly as complementary relative cycles.  Thus, while the underlying cancellation is scheme independent, the continuum-renormalized OPE is the organization in which it becomes manifest as a recombination of complementary relative cycles.

The paper is organized as follows. Section~\ref{sec:GN} reviews the large-$N$ Gross--Neveu model, its massive saddle, and the exact order-$1/N$ fermion self-energy. Section~\ref{sec:theta} isolates the leading renormalon in the perturbative sector and develops its representations in reduced variables, after which Section~\ref{sec:thimble} analyzes the perturbative boundary thimble, the renormalon half-thimble, and the additional finite contribution required to complete the contour. Section~\ref{sec:OPE} identifies the relevant dimension-two operator, derives its Wilson coefficient, renormalizes its condensate, and establishes the contour identity that reconstructs the finite real integration domain of the reduced integral. Section~\ref{sec:PI} shows how the Borel and scale variables $t$ and $\rho$ come from the auxiliary-field path integral and determines the boundary character of the renormalon saddle, while Section~\ref{sec:ON} compares the local saddle geometry with the corresponding leading renormalon cancellation in the large-$N$ $O(N)$ nonlinear sigma model. We conclude in Section~\ref{sec:discussion}.  Appendix~\ref{app:OPE_details} contains details of the OPE matching and operator renormalization, and Appendix~\ref{app:ON_details} provides additional details for the $O(N)$ sigma-model analysis.

\section{Gross--Neveu model and its self-energy}
\label{sec:GN}

In this section we set up the large-$N$ Gross--Neveu model and review the self-energy calculation. The important point for the later analysis is that the physical self-energy in the massive vacuum is a finite real integral. Renormalon ambiguities appear only after this real quantity is decomposed into perturbative Wilson coefficients and power-suppressed matrix elements in the OPE.

In four dimensions, Coleman and Gross showed that asymptotic freedom requires non-abelian gauge bosons~\cite{ColemanGross1973}. In two dimensions, however, four-fermion couplings are marginal rather than irrelevant, and can be asymptotically free on their own. The Gross--Neveu (GN) model is the simplest example~\cite{GrossNeveu1974}. We work in Euclidean spacetime with Hermitian Euclidean gamma matrices, and write $\psi=(\psi_1,\ldots,\psi_N)$ for $N$ flavors of two-dimensional Dirac fermions. Flavor indices are contracted implicitly, so that the Euclidean Lagrangian is
\begin{equation}
\label{eq:GN_Lagrangian}
\cL_E = \psibar \slashed{\partial}\psi - \frac{g}{2}(\psibar\psi)^2\,.
\end{equation}
The model has a $U(N)$ flavor symmetry and a discrete chiral symmetry under which $\psi\to\gamma_5\psi$ and $\psibar\to-\psibar\gamma_5$. Although it has no gauge fields, the GN model retains enough of the short-distance and non-perturbative structure of QCD to provide a useful solvable setting for our analysis.

\subsection{Model and running coupling}

The four-fermion coupling is renormalized at one loop by the scalar-singlet fermion bubble. In the large-$N$ limit, with $gN$ held fixed, this is the only one-loop correction to the four-fermion vertex that contributes at leading order. The closed fermion loop gives a flavor trace $N$, so the bubble graph is of order $Ng^2$, the same large-$N$ order as the tree-level vertex $g$. By contrast, crossed channels and non-singlet flavor structures do not carry this flavor enhancement and are suppressed by $1/N$. Working in $d=2-2\varepsilon$ dimensions with massless fermions, the leading logarithmically divergent graph is
\begin{equation}
\label{eq:bubble_direct}
\vcenter{\hbox{\begin{tikzpicture}
\begin{feynman}
  \vertex (i1) at (-2, 0.5);
  \vertex (i2) at (-2, -0.5);
  \vertex (v1) at (-0.6, 0);
  \vertex (v2) at (0.6, 0);
  \vertex (o1) at (2, 0.5);
  \vertex (o2) at (2, -0.5);
  \diagram*{
    (i1) -- [fermion] (v1),
    (v1) -- [fermion] (i2),
    (v1) -- [fermion, half left, edge label=$\psi_j$] (v2),
    (v2) -- [fermion, half left] (v1),
    (v2) -- [fermion] (o1),
    (o2) -- [fermion] (v2),
  };
  \filldraw (v1) circle (2pt);
  \filldraw (v2) circle (2pt);
\end{feynman}
\end{tikzpicture}}}
\;=\; g^2 N\!\int\!\frac{d^dk}{(2\pi)^d}\,
\frac{\mathrm{tr}[\slashed{k}(\slashed{k}+\slashed{p})]}{k^2(k+p)^2}
\;=\; \frac{g^2 N}{2\pi\varepsilon}+\text{finite}\,.
\end{equation}
Equivalently, this is the one-loop correction to the inverse propagator of the Hubbard-Stratonovich field introduced below. Writing the bare coupling as $g_0=\mu^{2\varepsilon}Z_g g$, the counterterm required to subtract Eq.~\eqref{eq:bubble_direct} is
\begin{equation}
\label{eq:Zg_gn}
Z_g = 1 - \frac{N}{2\pi\varepsilon}g + \cO(g^2).
\end{equation}
The beta function is therefore, at leading order in large $N$,
\begin{equation}
\label{eq:beta_g}
\beta(g)=\mu\frac{dg}{d\mu} = -\frac{Ng^2}{\pi} + \cO(g^3).
\end{equation}
The negative sign gives asymptotic freedom. Integrating the one-loop RG equation gives
\begin{equation}
\frac{\pi}{Ng(\mu)} = \ln\frac{\mu}{\Lambda_{\rm GN}},
\end{equation}
or
\begin{equation}
\label{eq:Lambda_def}
\Lambda_{\rm GN} = \mu\, e^{-\pi/(N g(\mu))}.
\end{equation}
Thus $\Lambda_{\rm GN}$ is the scale at which the one-loop running coupling would diverge, and it is the analogue of $\Lambda_{\rm QCD}$ in this model.

It is convenient to define the Gross--Neveu 't~Hooft coupling
\begin{equation}
  \lambda \equiv \frac{N}{\pi} g,
  \label{thooft_lambda}
\end{equation}
which is held fixed as $N\to\infty$. In terms of $\lambda$, the beta function becomes
\begin{equation}
\label{eq:beta_lambda}
\mu\frac{d\lambda}{d\mu} = -\lambda^2 + \cO(\lambda^3),
\end{equation}
and
\begin{equation}
\Lambda_{\rm GN} = \mu\, e^{-1/\lambda(\mu)}.
\end{equation}
Bubble-chain diagrams are naturally organized in powers of $Ng\sim\lambda$ where each fermion bubble contributes a flavor trace $N$, while each four-fermion vertex contributes a factor of $g$. This large-$N$ counting is especially transparent in the auxiliary-field formulation, to which we now turn.

\subsection{Mass gap and the auxiliary-field propagator}
\label{sec:effective_action}

To solve the model at large $N$, we introduce a Hubbard--Stratonovich field $\sigma$ and rewrite the Lagrangian as
\begin{equation}
\label{eq:GN_sigma}
\cL_\sigma = \psibar \slashed{\partial}\psi + \frac{1}{2g}\sigma^2 + \sigma\psibar\psi\,.
\end{equation}
The equation of motion $\sigma=-g\psibar\psi$ reproduces Eq.~\eqref{eq:GN_Lagrangian}. The advantage of Eq.~\eqref{eq:GN_sigma} is that the fermions appear only quadratically. Integrating out the $N$ fermion flavors gives
\begin{equation}
\label{eq:S_eff}
S_{\rm eff}[\sigma] = \int d^2x\,\frac{\sigma^2}{2g} - N\Tr\ln\bigl(\slashed{\partial}+\sigma\bigr),
\end{equation}
where the trace is over spacetime and Dirac indices. At fixed $\lambda=Ng/\pi$, the action is proportional to $N$, so the $\sigma$ path integral is dominated by saddle points as $N\to\infty$.

When $\sigma$ is taken to be constant, the effective action can be reduced to a simpler form. Indeed, the Yukawa term in Eq.~\eqref{eq:GN_sigma} gives a fermion mass $m = \sigma$, so the $\Tr\ln$ reduces to a standard Coleman--Weinberg integral $\mathrm{tr}\,\ln(i \slashed{k} + \sigma) = \ln(k^2 + \sigma^2)$ giving the effective potential
\begin{equation}
\label{eq:Veff}
V_{\rm eff}(\sigma) = \frac{\sigma^2}{2g} - N\int \frac{d^2k}{(2\pi)^2}\,\ln(k^2 + \sigma^2)\,.
\end{equation}
Evaluating the integral in dimensional regularization and using the same counterterm that defined $Z_g$ above to absorb the $1/\varepsilon$ pole, the renormalized effective potential in $\overline{\rm{MS}}$ is
\begin{equation}
\label{eq:Veff_ren}
V_{\rm eff}(\sigma) = \frac{\sigma^2}{2g(\mu)} + \frac{N \sigma^2}{4\pi}\left(\ln\frac{\sigma^2}{\mu^2} - 1\right)\,.
\end{equation}
Setting $\partial V_{\rm eff}/\partial \sigma^2 = 0$ gives the gap equation
\begin{equation}
\label{eq:gap}
\frac{1}{g(\mu)} = -\frac{N}{2\pi}\ln\frac{\sigma^2}{\mu^2}\,.
\end{equation}
Choosing the positive saddle, the vacuum expectation value of $\sigma$ is therefore
\begin{equation}
\label{eq:sigma_vev}
\langle\sigma\rangle = \mu\, e^{-\tfrac{\pi}{Ng(\mu)}} = \mu\, e^{-\tfrac{1}{\lambda(\mu)}} = \Lambda_\text{GN} = m \,,
\end{equation}
where $m$ is the fermion mass in the constant $\sigma$ background and $\Lambda_\text{GN}$ is the RG-invariant scale defined in Eq.~\eqref{eq:Lambda_def}. This is a non-perturbative result since $m \sim e^{-1/\lambda}$ is invisible to any finite order of perturbation theory.

The fluctuation propagator around the massive saddle is the only ingredient from the true vacuum that we will need for the self-energy analysis below. We write
\begin{equation}
\sigma(x) = m+\frac{\alpha(x)}{\sqrt{N}}\,,
\end{equation}
where the normalization by $1/\sqrt{N}$ makes the quadratic action for $\alpha$ order $N^0$. Expanding the effective action about the saddle gives
\begin{equation}
S_{\rm eff}\!\left[m+\frac{\alpha}{\sqrt N}\right] = S_{\rm eff}[m] +\frac{1}{2}\int\frac{d^2p}{(2\pi)^2}\,\alpha(p)\,\Gamma_m^{(\alpha)}(p^2)\,\alpha(-p) + \cO(N^{-1/2}\alpha^3)\,,
\end{equation}
where the term linear in $\alpha$ vanishes by the gap equation. In the quadratic kernel it is useful to restore the bare coupling $g_0$:
\begin{equation}
\Gamma_m^{(\alpha)}(p^2) = \frac{1}{N g_0}
+ \mu^{2\varepsilon}\int\!\frac{d^dq}{(2\pi)^d}\,\frac{\tr\!\left[(-i\slashed q+m)(-i(\slashed q+\slashed p)+m)\right]}
{(q^2+m^2)((q+p)^2+m^2)}\,.
\end{equation}
Equivalently, we have
\begin{align}
\Gamma_m^{(\alpha)}(p^2)
&= \frac{1}{N g_0} + 2\mu^{2\varepsilon}\!\int\!\frac{d^dq}{(2\pi)^d}\, \frac{m^2-q\cdot(q+p)}{(q^2+m^2)((q+p)^2+m^2)} \nonumber\\
&= \frac{1}{N g_0} - \frac{1}{2\pi} \left[\frac{1}{\varepsilon} - \ln\frac{m^2}{\bar\mu^2} \right] + \frac{1}{2\pi}\sqrt{1+\frac{4m^2}{p^2}}\, \ln\frac{\sqrt{p^2+4m^2}+\sqrt{p^2}}{\sqrt{p^2+4m^2}-\sqrt{p^2}} .
\label{Gamma2_sigma}
\end{align}
The bare gap equation can be written as
\begin{equation}
\frac{1}{N g_0} = \frac{1}{2\pi}\left[\frac{1}{\varepsilon} - \ln\frac{m^2}{\bar\mu^2}
\right],
\label{Ng0}
\end{equation}
so the divergent and scale-dependent terms cancel in Eq.~\eqref{Gamma2_sigma}. Thus the inverse propagator for the large-$N$ fluctuation field is finite, and the $\alpha$ propagator around the true vacuum is
\begin{equation}
\label{eq:alpha_propagator}
\Delta_\alpha(p^2) = \left[ \frac{1}{2\pi}\, \xi\ln\frac{\xi+1}{\xi-1} \right]^{-1},
\quad \xi = \sqrt{1+\frac{4m^2}{p^2}}\,.
\end{equation}

We will use Eq.~\eqref{eq:alpha_propagator} below as the dressed auxiliary-field propagator in the order-$1/N$ fermion self-energy around the massive vacuum. Fermion bubbles and tadpoles attached only to $\alpha$ have already been resummed into this propagator and should not be inserted again. Diagrams with external fermion lines, however, are computed using $\Delta_\alpha$.

\subsection{Fermion self-energy}
\label{sec:fermion_self_energy}

The observable we use throughout the paper is the order-$1/N$ fermion self-energy, which we decompose as
\begin{equation}
\label{eq:self_energy_decomp}
\Sigma(p) = -\frac{i\slashed p}{N}\,A(p^2) + m\,B(p^2) .
\end{equation}
Both form factors have large-momentum trans-series expansions, but we focus on $A(p^2)$ for simplicity.

\subsubsection{Perturbation theory around $\sigma = 0$}
\label{sec:pert_series}

We first recall the perturbative expansion around the chirally symmetric saddle $\sigma = 0$. This is not the true vacuum of the theory, but it is the expansion that defines the perturbative Wilson coefficient in the OPE. In the auxiliary-field formulation of Eq.~\eqref{eq:GN_sigma}, the bare $\sigma$ propagator is just $g$. The one-loop correction to the $\sigma$ propagator is the massless fermion bubble
\begin{equation}
\label{eq:sigma_1loop}
\Pi_2(p^2) = \vcenter{\hbox{\scalarselfenergy}} = N\!\int \frac{d^d k}{(2\pi)^d} \frac{\tr[\slashed{k}(\slashed{k}+\slashed{p})]}{k^2(k+p)^2} = \frac{N}{2\pi} \left( \frac{1}{\varepsilon} + \ln\frac{\mu^2}{p^2} \right) .
\end{equation}
Sewing these bubbles into the fermion self-energy gives the usual bubble-chain expansion, and the contribution with $n$ bubbles is
\begin{equation}
\label{eq:Sigma_bubblechain}
\Sigma_n^\text{pert}(p) = \vcenter{\hbox{\bubblechain}} = \int\frac{d^d k}{(2\pi)^d} \frac{-ig}{\slashed p+\slashed k} \left[g\Pi_2(k^2)\right]^n .
\end{equation}
After adding the counterterm and projecting onto the $A$ form factor by tracing with $i\slashed p$, one obtains
\begin{align}
A^\text{pert}(p^2)
&= \sum_{n = 0}^\infty \frac{g^{n+1}N^{n+1}}{(2\pi)^n} \int\frac{d^d k}{(2\pi)^d} \frac{p^2+p\cdot k}{p^2(p+k)^2} \ln^n\frac{\mu^2}{k^2} \nonumber\\
&= \frac{1}{2p^2} \sum_{n = 0}^\infty \frac{\lambda(\mu)^{n+1}}{2^{n+1}} \int_0^{p^2} dk^2\, \ln^n\frac{\mu^2}{k^2}\,,
\label{eq:A_pert}
\end{align}
where we used the two-dimensional angular integral
\begin{equation}
\frac{1}{2\pi}\int_0^{2\pi}d\theta\, \frac{p\cdot(p+k)}{p^2(p+k)^2} = \frac{1}{p^2}\Theta(p^2-k^2) .
\end{equation}
The remaining radial integral gives
\begin{equation}
A^\text{pert}(p^2) = \frac{\mu^2}{2p^2} \sum_{n = 0}^\infty \left(\frac{\lambda(\mu)}{2}\right)^{n+1} \Gamma\!\left(n+1,\ln\frac{\mu^2}{p^2}\right),
\end{equation}
where the incomplete gamma function makes the factorial growth manifest. Choosing the renormalization scale $\mu^2 = p^2$ and defining $\lambda_p \equiv \lambda(\mu^2 = p^2)$, the above can be written as
\begin{equation}
\label{eq:A_pert_final}
A^\text{pert}(p^2) = \frac{1}{2} \sum_{n = 0}^\infty \left(\frac{\lambda_p}{2}\right)^{n+1} n!\, .
\end{equation}
Thus the Borel transform, with $t$ conjugate to $\lambda_p/2$, is
\begin{equation}
\mathcal B[A](t) = \frac{1}{2}\sum_{n = 0}^\infty t^n = \frac{1}{2}\frac{1}{1-t} .
\end{equation}
The pole at $t = 1$ lies on the positive real axis, so the inverse Borel transform requires a lateral prescription,
\begin{equation}
\int_0^{\infty\pm i0} dt\, e^{-2t/\lambda_p}\,\mathcal B[A](t) .
\end{equation}
The two lateral sums differ by an imaginary part in the perturbative contribution,
\begin{equation}
\label{E:todisplay1}
\operatorname{Im} A^\text{pert}_\pm(p^2) = \pm \frac{\pi}{2}e^{-2/\lambda_p} = \pm\frac{\pi}{2}\frac{m^2}{p^2} \,,
\end{equation}
which is the leading IR-renormalon ambiguity. As we will show, it is not an ambiguity of the exact self-energy around the true massive vacuum, but rather an ambiguity of the perturbative Wilson coefficient that will ultimately be canceled by the OPE.

The choice $\mu^2 = p^2$ only makes the result transparent. For a general subtraction point, the Borel transform of Eq.~\eqref{eq:A_pert} is
\begin{equation}
\mathcal B_\mu[A](t) = \frac{1}{2}\, \frac{\exp\!\left[t\ln(\mu^2/p^2)\right]}{1-t}\,,
\end{equation}
and the residue at $t = 1$ gives
\begin{equation}
\operatorname{Im} A^\text{pert}_\pm(p^2) = \pm\frac{\pi}{2}\frac{\mu^2}{p^2} e^{-2/\lambda(\mu)} = \pm\frac{\pi}{2}\frac{m^2}{p^2}\,,
\end{equation}
showing that the ambiguity is RG invariant.

\subsubsection{Exact massive-vacuum self-energy}
\label{sec:massive_self_energy}

The same self-energy can also be computed in the true massive vacuum. Using the fluctuation field $\sigma = m+\alpha/\sqrt N$ and the dressed propagator $\Delta_\alpha$ in Eq.~\eqref{eq:alpha_propagator}, the order-$1/N$ fermion self-energy receives a single contribution,
\begin{equation}
\label{eq:Sigma_diagram}
\Sigma^{(1)}(p) = \vcenter{\hbox{\selfenergy}} = \frac{1}{N} \int\!\frac{d^2k}{(2\pi)^2}\, \frac{1}{i(\slashed p+\slashed k)+m}\, \Delta_\alpha(k^2)\,,
\end{equation}
where the two Yukawa vertices supply the overall factor $1/N$. Projecting onto the form factor $A$ in Eq.~\eqref{eq:self_energy_decomp} gives
\begin{equation}
\label{eq:A_Mink}
A(p^2/m^2) = \frac{1}{p^2} \int\!\frac{d^2k}{(2\pi)^2}\, \frac{p^2+k\cdot p}{(p+k)^2+m^2}\, \Delta_\alpha(k^2) .
\end{equation}
Since $\Delta_\alpha(k^2)$ depends only on $k^2$, the angular integral can be done explicitly:
\begin{equation}
\label{Ikpm}
I(k,p,m) = \int_0^{2\pi} d\theta\, \frac{p^2+pk\cos\theta}{p^2+k^2+m^2+2pk\cos\theta} = \pi\left[ 1 + \frac{p^2-k^2-m^2}{\sqrt{(p^2+k^2+m^2)^2 - 4p^2k^2}} \right] .
\end{equation}
Introducing dimensionless variables
\begin{equation}
y = \frac{k^2}{m^2}, \quad x = \frac{p^2}{m^2}, \quad \xi_y = \sqrt{1+\frac{4}{y}},
\end{equation}
the exact large-$N$ expression for the form factor is
\begin{equation}
\label{eq:A_integral}
A(x) = \frac{1}{4x} \int_0^\infty dy\, \frac{1}{\xi_y\ln\frac{\xi_y+1}{\xi_y-1}} \left[ 1 + \frac{x-y-1}{\sqrt{(x+y+1)^2 - 4xy}} \right] .
\end{equation}
We emphasize that~\eqref{eq:A_integral} is finite and real, and it is the physical self-energy before any asymptotic expansion or OPE decomposition has been introduced. Thus the imaginary parts encountered below are not ambiguities of the observable itself, and they arise only after this real quantity is split into a perturbative Wilson coefficient and power-suppressed matrix elements.

\subsubsection{Mellin--Barnes trans-series}
\label{sec:MB_transseries}

The large-$x$ expansion of Eq.~\eqref{eq:A_integral} can be extracted systematically using Mellin--Barnes techniques~\cite{BenekeBraunKivel1998,Marino2024}. We only need the structure of the result and the location of the leading cancellation, so we summarize the derivation briefly. The first step is to represent the $\alpha$ propagator as

\begin{equation}
\label{mellinstart}
\frac{\Delta_\alpha(m^2y)}{2\pi} = \frac{1}{\xi_y\ln\frac{\xi_y+1}{\xi_y-1}} = \int_0^\infty dt\, \frac{1}{\xi_y} \left( \frac{\xi_y-1}{\xi_y+1} \right)^t = \int_0^\infty dt\, \frac{1}{2\pi i} \int_\gamma ds\, K(s,t)y^{-s}\,,
\end{equation}
where
\begin{equation}
K(s,t) = \int_0^\infty dy\, \frac{1}{\xi_y} \left( \frac{\xi_y-1}{\xi_y+1} \right)^t y^{s-1} = \frac{\Gamma(2s+1)\Gamma(t-s)}{\Gamma(s+t+1)},
\end{equation}
and the contour $\gamma$ lies in the fundamental strip $-1/2 < \re s < t$.

The angular kernel in Eq.~\eqref{eq:A_integral} has a Mellin transform that can be written in closed form:
\begin{align}
M(s) &= \int_0^\infty dy\, \frac{y^{s+1}}{\sqrt{(x+y+1)^2 - 4xy}} \nonumber\\
&= (1+x)^{s+1} B(s+2,-1-s)\, {}_2F_1\!\left( s+2,-s-1,1;\frac{x}{1+x} \right) .
\end{align}
The two pieces of the numerator have disjoint convergence strips, so the Mellin contour must be deformed separately for each term. This gives
\begin{equation}
A(x) = \frac{1}{4x} \int_0^\infty dt \left\{ K(1,t) - \frac{1}{2\pi i} \int_{\mathcal C_1} ds\, K(-s,t)M(s) + (x-1)\frac{1}{2\pi i} \int_{\mathcal C_2} ds\, K(-s,t)M(s-1) \right\},
\end{equation}
where $\mathcal C_1$ runs along $-2 < \re s < -1$ and $\mathcal C_2$ runs along $-1 < \re s < 0$~\cite{Marino2024}.

The trans-series is obtained by closing the $s$ contours and summing residues. Poles of $K(-s,t)$ produce factors $x^{-t-j} = e^{-2(t+j)/\lambda_p}$, while poles of the large-$x$ expansion of $M(s)$ produce integer powers of $1/x$. Using
\begin{equation}
\label{xtolam}
\frac{1}{x} = \frac{m^2}{p^2} = e^{-2/\lambda_p},
\end{equation}
the result is organized as
\begin{equation}
\label{eq:A_transseries}
A(x) = \int_0^\infty dt \left[ e^{-2t/\lambda_p} \left( E_0+\frac{1}{x}E_1+\frac{1}{x^2}E_2+\cdots \right) - H_0 + \frac{1}{x} H_1 - \frac{1}{x^2} H_2 - \cdots \right],
\end{equation}
where the first few functions are
\begin{align}
E_0(t) &= \frac{1}{2(1-t)}, \qquad E_1(t) = \frac{t}{\lambda_p} - \frac{1}{t} - 1 + \frac{t}{2}\left[ 1 - 2\gamma_E - \psi(t) - \psi(-t) \right] \\
H_0(t) &= 0, \qquad H_1(t) = \frac{1}{t(t^2-1)}.
\end{align}
Here we have used the conventions of Eq.~(3.31)--(3.32) of Ref.~\cite{Marino2024}, with $\mathcal E_n(t) = 2E_n(2t)$.

The leading perturbative term $E_0(t)$ has the same $t = 1$ pole as the bubble-chain Borel transform in Eq.~\eqref{eq:A_pert_final}, and its residue contributes $-e^{-2/\lambda_p}/2 = -1/(2x)$. In the full trans-series, this pole is canceled by the power-suppressed term $H_1(t)/x$, whose residue at $t = 1$ is $+1/(2x)$. This is the standard trans-series manifestation of the OPE cancellation. The geometric interpretation developed below identifies these two contributions with complementary integration cycles through a common saddle. To isolate the mechanism without the full Mellin--Barnes machinery, we next reduce Eq.~\eqref{eq:A_integral} to its leading-renormalon sector.

\section{Reduced integral for the leading renormalon}
\label{sec:theta}

The exact self-energy in Eq.~\eqref{eq:A_integral} contains more information than is needed to study the leading OPE renormalon. Therefore we introduce a reduced integral that keeps precisely the two ingredients responsible for the leading cancellation, namely the perturbative $E_0$ pole and the compensating $H_1$ power correction. The reduction isolates the leading renormalon sector in a form where the contour geometry becomes explicit.

\subsection{\texorpdfstring{$\theta$}{theta} reduction}

The simplification comes from replacing the exact angular kernel in Eq.~\eqref{Ikpm} by its leading small-$m$ form,
\begin{equation}
I_0(k,p,0) = \pi\left[ 1 + \frac{p^2-k^2}{\sqrt{(p^2-k^2)^2}} \right] = 2\pi\,\Theta(p^2-k^2) .
\end{equation}
We will call this the $\theta$ reduction. Under this replacement, Eq.~\eqref{eq:A_integral} becomes 
\begin{equation}
\label{eq:theta_approx_A0}
A_0(x) = \frac{1}{4\pi x}\int_0^x dy\,\Delta_\alpha(m^2y) = \frac{1}{2x}\int_0^x dy\, \frac{1}{\xi_y\ln\frac{\xi_y+1}{\xi_y-1}} .
\end{equation}
The upper limit $y \leq x$ and the absence of the second angular term make the Borel representation elementary. Using the Mellin representation in Eq.~\eqref{mellinstart},
\begin{equation}
A_0(x) = \frac{1}{2x}\int_0^x dy \int_0^\infty dt\, \frac{1}{\xi_y} \left( \frac{\xi_y-1}{\xi_y+1} \right)^t\,, \label{eq:red_A_thet}
\end{equation}
and the $y$ integral can be done exactly:
\begin{multline}
\int_0^x dy\, \frac{1}{\xi_y} \left( \frac{\xi_y-1}{\xi_y+1} \right)^t = \frac{2}{t^3-t} -\frac{1}{2^t}\, \frac{\left( 2+x-\sqrt{\frac{4+x}{x}} \right)^t \left( 2+t^2x+tx\sqrt{\frac{4+x}{x}} \right)}
{t(t^2-1)}
\\
= \frac{2}{t(t^2-1)} + x^{1-t}\frac{1}{1-t} + x^{-t}\frac{2(1+t)}{t} - x^{-t-1}\frac{6+7t+2t^2}{1+t} + \cdots .
\end{multline}
Using $x^{-t} = e^{-2t/\lambda_p}$, this gives
\begin{equation}
\label{eq:infinite_transseries}
A_0(x) = \int_0^\infty dt \left[ e^{-2t/\lambda_p}\underbrace{\frac{1}{2}\frac{1}{1-t}}_{E_0} + \frac{1}{x}\underbrace{\frac{1}{t(t^2-1)}}_{H_1} + e^{-2t/\lambda_p}\frac{1}{x}\underbrace{\frac{1+t}{t}}_{\text{$1/x$ sector}} + \cdots \right] .
\end{equation}
Comparing with Eq.~\eqref{eq:A_transseries}, we see that the reduced integral preserves the perturbative $E_0$ pole at $t = 1$ and the compensating $H_1$ term at order $1/x$ with their exact relative normalization. The remaining terms in the $1/x$ sector are modified by the $\theta$ reduction and should not be identified term by term with the corresponding functions in the full trans-series. Thus the $\theta$ reduction keeps the leading renormalon cancellation while discarding higher power corrections and real finite terms that will not affect the common $t = 1$ saddle.

\subsection{Apparent Landau pole in the \texorpdfstring{$y$}{y} variable}

The variable $y = k^2/m^2$ gives the usual running-coupling picture of the leading renormalon. At large $y$, the auxiliary-field propagator divided by its overall factor of $2\pi$ has the expansion
\begin{equation}
\label{DeltaExp}
\frac{\Delta_\alpha(m^2y)}{2\pi} = \frac{1}{\sqrt{1+\frac{4}{y}}\ln\frac{\sqrt{1+\frac{4}{y}}+1}{\sqrt{1+\frac{4}{y}}-1}} = \frac{1}{\ln y} - \frac{2(1+\ln y)}{y\ln^2 y} + \frac{4+7\ln y+6\ln^2 y}{y^2\ln^3 y} + \cdots .
\end{equation}
The leading term of this expansion contributes to
\begin{equation}
\label{A0lny}
A_0(x) = \frac{1}{2x}\int_0^x dy\, \frac{1}{\ln y} + \cdots .
\end{equation}
Expanding around the upper endpoint gives
\begin{equation}
\label{A0lnyseries}
\int_0^x dy\, \frac{1}{\ln y} = \int_0^x dy\, \frac{1}{\ln x-\ln(x/y)} = \sum_{n = 0}^\infty \frac{1}{\ln^{n+1}x}\int_0^x dy\, \ln^n\frac{x}{y} = x\sum_{n = 0}^\infty \left(\frac{\lambda_p}{2}\right)^{n+1} n! ,
\end{equation}
where $\lambda_p = 2/\ln x$. This reproduces the leading factorial growth in Eq.~\eqref{eq:A_pert_final}. Equivalently,
\begin{equation}
\int_0^x dy\, \frac{1}{\ln y} = \int_0^x dy \int_0^\infty dt\, y^{-t} = x\int_0^\infty dt\, \frac{e^{-2t/\lambda_p}}{1-t},
\end{equation}
which is the $E_0$ Borel singularity.

In this representation the imaginary part is associated with the singularity of $1/\ln y$ at $y = 1$, or $k^2 = m^2$, which is the Landau pole of the one-loop running coupling. More generally,
\begin{equation}
2\,\operatorname{Im}\int_0^x dy\, \frac{1}{y^r\ln^s y} = -i\oint_{|y-1| = \varepsilon} dy\, \frac{1}{y^r\ln^s y} = 2\pi\,\resy\left( \frac{1}{y^r\ln^s y} \right) = 2\pi\,\frac{(1-r)^{s-1}}{(s-1)!}.
\end{equation}
The leading term in Eq.~\eqref{DeltaExp} has residue $1$, and the subleading terms also develop singularities at $y = 1$ where the additional powers of $1/y$ provide no suppression. Their residue contributions can therefore be of the same order as that of the leading term. The first three nonzero residues are
\begin{align}
\resy\left[ \frac{1}{\ln y} \right] &= 1 \nonumber\\
\resy\left[ -\frac{2(1+\ln y)}{y\ln^2 y} \right] &= -2 \nonumber\\
\resy\left[ \frac{4+7\ln y+6\ln^2 y}{y^2\ln^3 y} \right] &= 1\,.
\label{threeterms}
\end{align}
Notably, their sum vanishes. Moreover, the residue at each subsequent power of $y^{-1}$ vanishes on its own; for example,
\begin{equation}
\operatorname*{Res}_{y = 1}\left[ -\frac{2(30\ln^3 y+37\ln^2 y+30\ln y+12)}{3y^3\ln^4 y} \right] = 0 .
\end{equation}
Thus, although the successive terms in Eq.~\eqref{DeltaExp} are suppressed by increasing powers of $1/y$ at large $y$, this suppression breaks down near the apparent Landau pole at $y = 1$. In that region, the singularities of the first three terms contribute at the same order, and their residues cancel in the combination $1 - 2 + 1 = 0$. This cancellation was observed in Ref.~\cite{Nishimura:2021lno}, although its structural origin was not apparent. In the next subsection, we show that the cancellation follows directly from a change of variables in which the exact reduced integrand decomposes into precisely three terms, whose residues at the corresponding endpoint are manifestly $1$, $-2$, and $1$.

\subsection{\texorpdfstring{$z$}{z} variable and the finite trans-series}

Now let us introduce
\begin{equation}
\label{eq:z_variable}
z = \frac{\xi_y+1}{\xi_y-1}, \quad y = \frac{(z-1)^2}{z} = z-2+\frac{1}{z}.
\end{equation}
Equivalently, at large $y$, we have $z = y+2+\cO(y^{-1})$. The endpoint $y = x$ maps to
\begin{equation}
\label{Zofx}
Z = \frac{\xi_x+1}{\xi_x-1} = \frac{\sqrt{1+\frac{4}{x}}+1}{\sqrt{1+\frac{4}{x}}-1} = x+2-\frac{1}{x}+\cdots .
\end{equation}
The Jacobian and the factor $1/\xi_y$ combine to give
\begin{equation}
\frac{dy}{\xi_y} = dz\,\left( 1-\frac{2}{z}+\frac{1}{z^2} \right),
\end{equation}
so the reduced integral in Eq.~\eqref{eq:red_A_thet}  becomes
\begin{equation}
\label{A0Ein}
2xA_0(x) = \int_1^Z \frac{dz}{\ln z}\left( 1-\frac{2}{z}+\frac{1}{z^2} \right) = -\Ein(\ln Z)-\Ein(-\ln Z)\ ,
\end{equation}
where we have multiplied both sides by $2x$ for aesthetic convenience, and where
\begin{equation}
\Ein(z) = \int_0^z dt\, \frac{1-e^{-t}}{t}\ .
\end{equation}
The function $\Ein(z)$ is entire and is related to the exponential integral by $\Ein(z) = E_1(z)+\ln z+\gamma_E$, with the usual branch convention for the logarithm. Its large-$z$ trans-series is
\begin{equation}
\label{EinTrans}
\Ein(z) = -e^{-z}\sum_{n = 0}^\infty \left(-\frac{1}{z}\right)^{n+1}n! + \ln z+\gamma_E .
\end{equation}
For a negative argument the asymptotic series is non-alternating. Resolving the branch by a lateral prescription gives
\begin{equation}
\label{EinTransNeg}
\Ein(-z\pm i0) = -e^z\sum_{n = 0}^\infty \frac{n!}{z^{n+1}} + \ln z \pm i\pi+\gamma_E\,,
\end{equation}
or equivalently
\begin{equation}
\mathcal S_\pm\left[ -e^z\sum_{n = 0}^\infty \frac{n!}{z^{n+1}} \right]
= E_1(-z\pm i0)
= \Ein(-z\pm i0)-\ln z \mp i\pi-\gamma_E .
\end{equation}

We will see that it is judicious to define
\begin{equation}
\lambda_z = \frac{2}{\ln Z}\,.
\end{equation}
Since $Z = x+2+\cO(x^{-1})$, this differs from $\lambda_p = 2/\ln x$ by non-perturbative terms:
\begin{equation}
\frac{1}{\lambda_z} = \frac{1}{\lambda_p} + e^{-2/\lambda_p} - \frac{3}{2}e^{-4/\lambda_p}+\cdots,
\end{equation}
or
\begin{equation}
\label{lambdaztop}
\lambda_z = \lambda_p - \lambda_p^2 e^{-2/\lambda_p}+\cdots .
\end{equation}
We can interpret $\lambda_z$ as the running coupling evaluated at the dressed momentum
\begin{equation}
\label{eq:pz}
p_z = \frac{p+\sqrt{p^2+4m^2}}{2}, \qquad Z = \frac{p_z^2}{m^2}.
\end{equation}
Using Eqs.~\eqref{A0Ein}--\eqref{EinTransNeg}, the trans-series in $\lambda_z$ is
\begin{equation}
\label{A0ser}
2xA_0(x) = \frac{\lambda_z}{2} e^{2/\lambda_z}\sum_{n = 0}^\infty n!\left(\frac{\lambda_z}{2}\right)^n + 2\ln\frac{\lambda_z}{2} - 2\gamma_E \mp i\pi - \frac{\lambda_z}{2} e^{-2/\lambda_z}\sum_{n = 0}^\infty (-1)^n n!\left(\frac{\lambda_z}{2}\right)^n .
\end{equation}
This trans-series terminates after the second non-perturbative sector. If the same expression is re-expanded in $\lambda_p$, the relation in Eq.~\eqref{lambdaztop} generates infinitely many sectors. Since $\lambda_z$ and $\lambda_p$ differ only by non-perturbative terms, the leading perturbative series is unchanged:
\begin{equation}
A_0(x) = \frac{\lambda_p}{4x}e^{2/\lambda_p}\sum_{n = 0}^\infty n!\left(\frac{\lambda_p}{2}\right)^n = \frac{\lambda_p}{4}\sum_{n = 0}^\infty n!\left(\frac{\lambda_p}{2}\right)^n,
\end{equation}
in agreement with Eq.~\eqref{eq:A_pert_final}. Related coupling redefinitions, which leave the perturbative asymptotic series unchanged while removing the Landau pole and reorganizing the nonperturbative power corrections, were previously studied in the large-$N$ $O(N)$ nonlinear sigma model~\cite{BenekeBraunKivel1998}.

The Borel representation in the $z$ variable is especially transparent. Since $z > 1$ on the whole integration domain,
\begin{equation}
\label{eq:Mellin_z}
\frac{1}{\ln z} = \int_0^\infty dt\, z^{-t}.
\end{equation}
Therefore
\begin{align}
2xA_0(x) &= \int_1^Z dz \int_0^\infty dt\, z^{-t}\left( 1-\frac{2}{z}+\frac{1}{z^2} \right) \nonumber\\
&= \int_0^\infty dt \left[ Z e^{-2t/\lambda_z}\frac{1}{1-t} + \frac{2}{t(t^2-1)} + e^{-2t/\lambda_z}\frac{2}{t} - \frac{1}{Z}e^{-2t/\lambda_z}\frac{1}{1+t} \right].
\label{A0Mellin}
\end{align}
The first term is the non-alternating perturbative sector, while the last term gives the alternating sector. The middle two terms are regular at $t = 0$ only after being combined. With the $t = 1$ pole laterally prescribed,
\begin{equation}
\int_0^\infty dt \left[ \frac{2}{t(t+1)(t-1\pm i0)} + \frac{2e^{-2t/\lambda_z}}{t} \right] = 2\ln\frac{\lambda_z}{2} - 2\gamma_E \mp i\pi .
\end{equation}
The imaginary part in this middle contribution cancels the lateral ambiguity of the non-alternating perturbative sector. Equivalently, in the full integrand of Eq.~\eqref{A0Mellin}, the $t = 1$ pole cancels between the first term and the $2/[t(t^2-1)]$ term.

The residue cancellation seen in the $y$ variable is now transparent. Near $z = 1$, the function $1/\ln z$ has residue $1$, so the three terms in the integrand have residues
\begin{equation}
\reszz\left[\frac{1}{\ln z}\right] = 1, \quad \reszz\left[-\frac{2}{z\ln z}\right] = -2, \quad \reszz\left[\frac{1}{z^2\ln z}\right] = 1\,,
\end{equation}
which sum to zero. This explains the three nonzero residues in Eq.~\eqref{threeterms} as the large-$y$ image of the exact cancellation $1 - 2 + 1 = 0$ in the $z$ variable.

The same change of variables also explains why the $H_1$ term was hidden in the large-$y$ expansion. The map sends $y \in [0,x]$ to $z \in [1,Z]$, and the Mellin representation converges uniformly on $z \geq 1$, so the lower endpoint $z = 1$ directly produces
\begin{equation}
\label{H1appears}
-\frac{1}{1-t} - \frac{2}{t} + \frac{1}{1+t} = \frac{2}{t(t^2-1)} = 2H_1(t),
\end{equation}
where the factor of $2$ reflects the normalization of $2xA_0$. In the original $y$ variable, the Mellin representation of the large-$y$ expansion is not uniformly convergent on $0 < y < 1$. Analytically continuing from the convergent region discards the lower-endpoint contribution, so the $H_1$ sector is absent from the na\"{i}ve large-$y$ expansion even though the exact propagator is regular at the endpoint, with $\Delta_\alpha(0)/(2\pi) = 1/2$. 

The above clarifies that the location of the Landau pole and the renormalon singularity are distinct phenomena~\cite{Beneke:1998ui,BenekeBraunKivel1998,Grunberg:1995vx,Dokshitzer:1995af}. In the leading $y$-space approximation, $1/\ln y$ has a pole at $y = 1$, or $k^2 = m^2$. In the $z$ variable, the same physical point maps to
\begin{equation}
z = \frac{\sqrt{5}+1}{\sqrt{5}-1}
\end{equation}
which is an ordinary interior point where $\ln z \neq 0$. The endpoint $z = 1$ instead corresponds to $y = 0$. The full reduced integrand is nevertheless smooth there because the Jacobian factor $(1-1/z)^2$ cancels the pole of $1/\ln z$. Thus the location of the Landau-pole-like singularity depends on how the UV expansion is completed by power corrections. The invariant object (with respect to asymptotic series in the perturbative expansion) is the $t = 1$ Borel singularity; below we will show how it corresponds to a saddle of a reduced two-dimensional integral.

The $z$ representation now gives a two-dimensional integral with a simple exponent. Setting $\rho = \ln z$, Eq.~\eqref{A0Ein} becomes
\begin{equation}
\label{eq:Eintrho}
2xA_0(x) = \int_0^\infty dt \int_0^{\ln Z} d\rho\, e^{(1-t)\rho}\left( 1 - 2e^{-\rho} + e^{-2\rho} \right).
\end{equation}
The three terms can be combined by shifting the $t$ integration variable, and the result is
\begin{equation}
\label{eq:strip}
2xA_0(x) = \left[ \int_0^1 dt - \int_1^2 dt \right]\int_0^{\ln Z} d\rho\, e^{(1-t)\rho}.
\end{equation}
The Picard--Lefschetz analysis below studies the relative cycles associated with this integral.  We observe that the above is an integral over two rectangles in the $(t,\rho)$ plane with common exponent $e^{- S(t,\rho)}$ where $S(t,\rho) = (t-1)\rho$, and that the point $(t,\rho) = (1,0)$ is the saddle of this reduced action.


\section{Picard--Lefschetz geometry}
\label{sec:thimble}

We now analyze the relative cycles of the reduced integral in Eq.~\eqref{eq:strip}. The geometry is slightly different from standard Picard--Lefschetz decompositions around isolated saddles, since in this case the perturbative series is generated by the boundary corner $(t,\rho) = (0,\ln Z)$ while its Stokes jump is controlled by the critical point at $(t,\rho) = (1,0)$. This critical point is the renormalon saddle. The goal of this section is to identify the perturbative boundary thimble and the renormalon half-thimble that carries its lateral ambiguity.

\begin{figure}[t!]
\centering
\begin{tikzpicture}[>=stealth, scale=1.2]
  \fill[blue!12] (0,0) rectangle (2.4,3.0);
  \draw[thick, blue!60!black] (0,0) rectangle (2.4,3.0);
  \fill[red!8] (2.4,0) rectangle (4.8,3.0);
  \draw[thick, red!50!black] (2.4,0) rectangle (4.8,3.0);
  \draw[->] (-0.5,0) -- (5.6,0) node[right] {$t$};
  \draw[->] (0,-0.4) -- (0,3.7) node[above] {$\rho$};
  \draw (2.4,0.08) -- (2.4,-0.08) node[below] {$1$};
  \draw (4.8,0.08) -- (4.8,-0.08) node[below] {$2$};
  \draw (0.08,3.0) -- (-0.08,3.0) node[left] {$\ln Z$};
  \fill[red!70!black] (2.4,0) circle (3pt);
  \node[below right, red!70!black, font=\small] at (2.55, 0.08) {renormalon};
  \node[below right, red!70!black, font=\small] at (2.55,-0.22) {saddle};
  \fill[blue!70!black] (0,3.0) circle (3pt);
  \node[above right, blue!70!black, font=\small] at (0.05,3.05) {dominant corner};
  \node[blue!60!black, font=\small] at (1.2,1.5) {$S<0$};
  \node[blue!60!black, font=\small] at (1.2,1.0) {(non-alternating)};
  \node[red!50!black, font=\small] at (3.6,1.5) {$S>0$};
  \node[red!50!black, font=\small] at (3.6,1.0) {(alternating)};
\end{tikzpicture}
\caption{The reduced integration domain in the $(t,\rho)$ plane. The finite strip $0 \leq \rho \leq \ln Z$ splits into two rectangles. The left rectangle, $0 < t < 1$, contains the dominant corner that generates the non-alternating perturbative series. The right rectangle, $1 < t < 2$, gives the first alternating sector. The shared boundary point $(t,\rho) = (1,0)$ is the renormalon saddle.}
\label{fig:two_rectangles}
\end{figure}
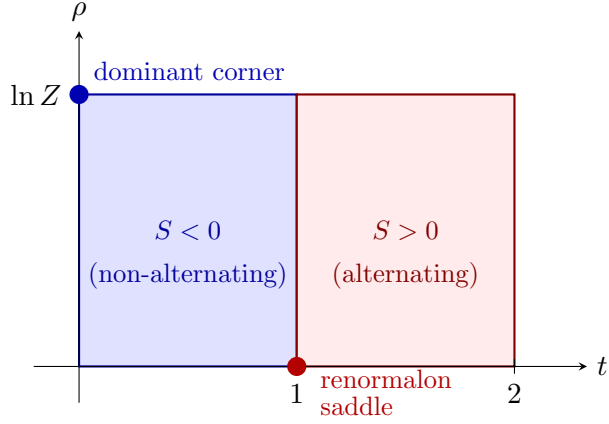

The geometry of Eq.~\eqref{eq:strip} is shown in Figure~\ref{fig:two_rectangles}. The rectangle $0 < t < 1$ has reduced action $S < 0$ and gives the dominant non-alternating asymptotic series contained in $-\Ein(-2/\lambda_z)$. The rectangle $1 < t < 2$ has $S > 0$ and gives the subdominant alternating sector contained in $-\Ein(2/\lambda_z)$.

\subsection{Perturbative boundary thimble}

We first analyze the dominant rectangle, $0 < t < 1$. Since $\ln Z = 2/\lambda_z$, it is convenient to normalize its vertical extent by rescaling $\rho \rightarrow 2\rho/\lambda_z$. The rescaled coordinate then ranges over $0 < \rho < 1$, and the exponent takes the form $e^{-2S/\lambda_z}$ with $S(t,\rho) = (t-1)\rho$. The dominant contribution comes from the corner $(t,\rho) = (0,1)$, where $S = -1$. Introducing coordinates $t = u$ and $\rho = 1-v$ gives the action $S(u,v) = -1+u+v-uv$, and so the contribution of the dominant rectangle is
\begin{equation}
I_A = \frac{2}{\lambda_z}\int_0^1 du \int_0^1 dv\, e^{-\frac{2}{\lambda_z}S(u,v)} = -\Ein(-2/\lambda_z).
\end{equation}
The perturbative expansion is obtained by replacing the finite neighborhood of the corner by the quadrant $u, v \geq 0$. Expanding the interaction term $uv$ and extending both integrations to infinity gives
\begin{equation}
I_A^\text{pert}
= \frac{2}{\lambda_z}e^{2/\lambda_z}
\int_0^\infty du \int_0^\infty dv\, e^{-\frac{2}{\lambda_z}(u+v)}
\sum_{n = 0}^\infty \frac{1}{n!}\left(\frac{2uv}{\lambda_z}\right)^n
= e^{2/\lambda_z}\sum_{n = 0}^\infty n!\left(\frac{\lambda_z}{2}\right)^{n+1}\,,
\end{equation}
which is exactly the non-alternating perturbative series in Eq.~\eqref{A0ser}.

Since the dominant contribution comes from a corner of the rectangle rather than from an interior critical point, the relevant object is a relative steepest-descent class associated with the asymptotic corner. We will refer to this class as the perturbative boundary thimble. The explicit contours used below are convenient convergent representatives of this relative homology class; we do not require these representatives themselves to be gradient-flow surfaces of $S$.

A useful representative is obtained by taking $u=e^{i\theta_1}r$ and $v=e^{i\theta_2}s$, with $r,s\geq0$, and choosing the phases so that the real part of $u+v-uv$ is positive at infinity. For example, $\theta_1=\theta_2=\pi/3$ gives $u+v-uv=e^{i\pi/3}(r+s)-e^{2i\pi/3}rs$, so $\operatorname{Re}(u+v-uv)=(r+s+rs)/2>0$. The integral is therefore absolutely convergent on this representative.

We denote the resulting contours by $\Gamma_\pm$, where the sign records whether the induced contour in the $u$ plane passes above or below $u = 1$. Their integrals give the two lateral Borel sums of the perturbative series:
\begin{equation}
\label{eq:lateral_borel}
\mathcal{S}_\pm[I_A^\text{pert}] = \frac{2}{\lambda_z}e^{2/\lambda_z}\int_{\Gamma_\pm} du\,dv\, e^{-\frac{2}{\lambda_z}(u+v-uv)} = -E_1\!\left(-\frac{2}{\lambda_z}\pm i0\right).
\end{equation}
To see the Borel representation directly, we first perform the $v$ integral. Since $u+v-uv = u+(1-u)v$, integration along the chosen $v$-contour gives $\frac{2}{\lambda_z}\int dv\,e^{-\frac{2}{\lambda_z}(1-u)v} = \frac{1}{1-u}$. The boundary-thimble integral therefore reduces to
\begin{equation}
e^{2/\lambda_z}\int_0^{\infty e^{\pm i0}}du\,\frac{e^{-2u/\lambda_z}}{1-u}\,,
\end{equation}
which is precisely the lateral Borel integral of $I_A^\text{pert}$. The pole at $u = 1$ is the leading renormalon singularity, and the two contours passing above or below it give the two lateral sums.

For comparison, the subdominant rectangle gives the alternating, Borel-summable sector of the trans-series for $2xA_0(x)$. Expanding about the corner $(t,\rho) = (2,1)$ by writing $t = 2+u$ and $\rho = 1+v$, one finds $S(u,v) = 1+u+v+uv$. The corresponding boundary-thimble integral is
\begin{equation}
I_C = \frac{2}{\lambda_z}e^{-2/\lambda_z}\int_0^\infty du \int_0^\infty dv\, e^{-\frac{2}{\lambda_z}(u+v+uv)}
= -\operatorname{Ei}\!\left(-\frac{2}{\lambda_z}\right),
\end{equation}
whose expansion is alternating. This sector is Borel summable and thus is not responsible for the leading ambiguity.

\subsection{Renormalon half-thimble and finite cap}

The Stokes phenomenon of the non-alternating series is controlled by the ordinary critical point of the reduced action at $(t,\rho) = (1,0)$, with Hessian $\partial_i\partial_j S =
\begin{pmatrix}
0 & 1 \\
1 & 0
\end{pmatrix}$ having eigenvalues $\pm 1$. Unlike the dominant corner, this point is a genuine Morse saddle.

A local thimble through the saddle is parametrized by a complex coordinate $w$ as $t = 1 + iw$ and $\rho = -i\overline w$, and so along this contour the action $S = |w|^2$ is Gaussian.  Writing $w = a+ib$, the pullback of the holomorphic two-form is $dt\wedge d\rho = -2i\,da\wedge db$; reversing the orientation changes the sign. Thus the full renormalon thimble $\mathcal J_\pm$ gives
\begin{equation}
I_R^\pm = \frac{2}{\lambda_z}\int_{\mathcal J_\pm} dt\,d\rho\, e^{-\frac{2}{\lambda_z}S(t,\rho)} = \pm 2\pi i\,,
\end{equation}
where the sign labels the two possible orientations of the thimble.

For the original integration chain, the saddle lies on its boundary at $\rho=0$. The local Morse contribution therefore enters with half the weight of the corresponding full Gaussian thimble. Equivalently, after choosing the lateral deformation of the boundary chain, only one of the two local halves of $\mathcal J_\pm$ contributes. The selected half is fixed by the orientation of the deformed relative cycle, rather than by the real projection of $\mathcal J_\pm$ into the original rectangle. Its contribution is
\begin{align}
I_{R,\text{half}}^\pm = \pm i\pi\,.
\end{align}
This is precisely the imaginary part of the lateral Borel resummation of the perturbative boundary thimble in Eq.~\eqref{eq:lateral_borel}. Thus the perturbative ambiguity is the half-thimble contribution of the renormalon saddle.

The half-thimble accounts for the local Stokes jump, but it does not by itself reconstruct the original rectangular integral. The reason is that the perturbative boundary thimble knows only about the asymptotic expansion at the dominant corner, whereas the original rectangle has finite boundaries. Recovering the finite rectangular integral therefore requires additional side and cap contributions.

\begin{figure}[t!]
\centering
\includegraphics[width=0.45\textwidth]{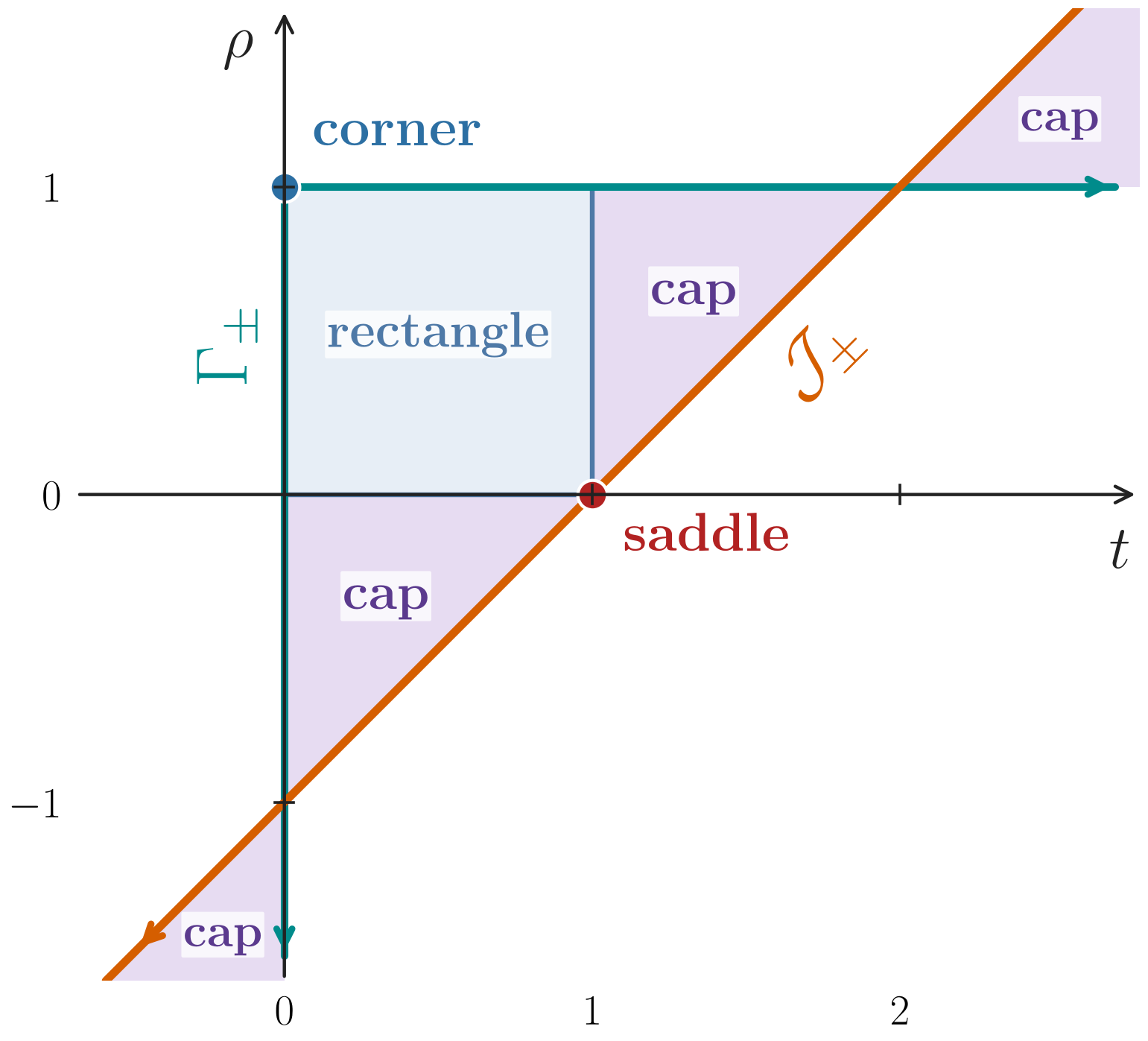}
\caption{Projection of the contour decomposition onto the real $(t,\rho)$ plane. The shaded rectangle is the original finite integration domain for the dominant sector. The contour $\Gamma_\pm$ is a convergent representative of the perturbative relative steepest-descent class and gives the lateral Borel sum, while the half-thimble $\mathcal J_\pm$ passes through the renormalon saddle at $(t,\rho) = (1,0)$.  The cap regions are the finite surfaces swept out when the side contours are deformed into the half-thimble. The half-thimble carries the local Stokes jump, whereas the cap contributes real finite terms that depend on the global completion of the contour.}
\label{fig:contour_decomposition}
\end{figure}

For the dominant rectangle, $I_A = -\Ein(-2/\lambda_z)$ and the perturbative boundary thimble gives
\begin{equation}
\mathcal S_\pm[I_A^\text{pert}] = -E_1\!\left(-\frac{2}{\lambda_z}\pm i0\right) = \operatorname{Ei}\!\left(\frac{2}{\lambda_z}\right)\pm i\pi .
\end{equation}
Accordingly, the difference between the boundary thimble and the original rectangle is
\begin{equation}
\mathcal S_\pm[I_A^\text{pert}] - I_A = \gamma_E+\ln\frac{2}{\lambda_z}\pm i\pi .
\end{equation}
The half-renormalon thimble accounts for the imaginary part of this difference. We use the term ``cap'' for a finite two-chain completing the deformation between the side contours of the boundary-thimble representative and the local half-thimble. Its detailed parametrization is not fixed by the local saddle geometry; its integral is determined by the difference between the finite rectangular integral and the lateral boundary-thimble integral and gives the remaining real contribution $\gamma_E+\ln(2/\lambda_z)$. Figure~\ref{fig:contour_decomposition} illustrates this decomposition as a projection onto the real $(t,\rho)$ plane. The rectangle represents the original finite integration domain, the boundary thimble represents the perturbative Borel sum, and the half-thimble passes through the renormalon saddle. The cap fills the finite region between the side contours and the half-thimble and therefore supplies the real finite term that is not fixed by the local saddle geometry.

\begin{figure}[t!]
\centering
\includegraphics[width=0.7\textwidth]{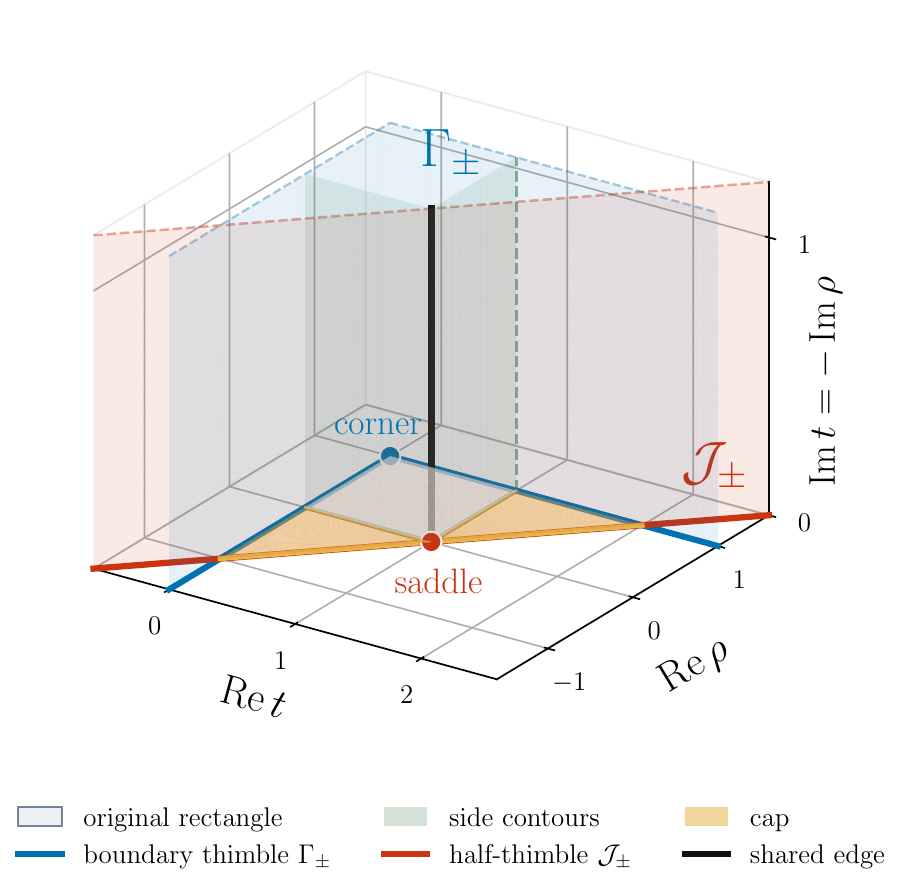}
\caption{Projection of the relevant contour surfaces in the complexified $(t,\rho)$ space onto $(\operatorname{Re}t,\,\operatorname{Re}\rho,\,\operatorname{Im}t=-\operatorname{Im}\rho)$. The rectangle lies in the real plane. The boundary thimble $\Gamma_\pm$ and the side contours extend into the imaginary direction, while the half-thimble $\mathcal J_\pm$ passes through the renormalon saddle. The black line is the shared local edge, and the gold cap fills the finite region between the side contours and the half-thimble.}
\label{fig:3d_contours}
\end{figure}

Equivalently, the side contours and the half-thimble have the same local Stokes discontinuity at the saddle, but they are not homologous by themselves because their remaining boundaries differ. The side contours extend along the Stokes walls $\rho = 0$ and $t = 1$, whereas the half-thimble runs diagonally into the region where the integrand decays. Their oriented difference therefore has a residual boundary, which is filled by a finite cap; the integral over this cap gives the real finite term above. Thus the half-thimble determines the local ambiguity, while the cap contains the additional real contribution associated with the completion of the perturbative cycle away from the saddle. The corresponding geometry in the complexified integration space is shown in Figure~\ref{fig:3d_contours}: the side contours and the half-thimble share the local edge through the saddle, while the cap joins their non-shared boundaries.

The preceding decomposition separates the local contribution of the renormalon saddle from the global completion of the contour. The Stokes discontinuity, and hence the imaginary ambiguity, is fixed by the local geometry near $(t,\rho) = (1,0)$, whereas the real finite terms also depend on the cap that completes the contour away from the saddle. We now turn to the OPE interpretation, in which the perturbative Wilson coefficient and the renormalized dimension-two condensate define complementary cycles through this renormalon saddle.

\section{OPE and the condensate cycle}
\label{sec:OPE}

The Picard--Lefschetz analysis above represents the perturbative contribution to the reduced self-energy by a boundary thimble and identifies its Stokes discontinuity with the half-thimble through the renormalon saddle. Connecting this geometry to the OPE requires identifying the operator matrix element whose ambiguity cancels that of the perturbative Wilson coefficient. That matrix element is the condensate of the dimension-two scalar four-fermion operator $V$. The discussion begins with the operator basis relevant to the vector self-energy form factor $A(p^2)$ and the derivation of the Wilson coefficient of $V$. It then turns to the renormalization of $\langle V\rangle$, including the ultraviolet subtractions required to define its scale integral after the cutoff is removed. Expressing the renormalized condensate in the same $(t,\rho)$ variables as the perturbative contribution then makes it possible to compare the two integration cycles directly and to show that their sum reconstructs the original real integration cycle.

\subsection{Operator basis and Wilson coefficients}

The operator product expansion organizes the short-distance behavior of the fermion two-point function as
\begin{equation}
\label{eq:OPE_operator}
\psibar(0)\psi(x) = C_0(x)\,\opone + \sum_n C_n(x)\,\cO_n(0) + \cdots ,
\end{equation}
where spin and flavor indices are suppressed. Eq.~\eqref{eq:OPE_operator} is an asymptotic expansion as $x \to 0$, rather than a convergent equality between nonlocal operators. The state-independent Wilson coefficients $C_n(x)$ are fixed by perturbative matching and multiply renormalized local operators, which may mix with other operators carrying the same quantum numbers. The long-distance and state-dependent information resides in the matrix elements of these renormalized operators. In the Gross--Neveu model, the ambiguity associated with the leading self-energy renormalon appears in the Wilson coefficient of the identity, while the compensating ambiguity resides in the renormalized matrix element of a dimension-two operator.

For the leading $t = 1$ cancellation, the relevant part of the OPE reduces to the identity coefficient and the dimension-two scalar operator $V$. The coefficient $C_0$ is the perturbative propagator and carries the infrared-renormalon ambiguity, while the Wilson coefficient of $V$ gives the leading noncontact contribution proportional to $1/p^2$. The matching calculation below derives this structure from the operator basis and its renormalization, rather than assuming the identification of $V$ from the required power counting alone.

There is a useful way to see the same dimension-two sector in the auxiliary-field basis. Using the constraint equation $\sigma = -g\psibar\psi$, the operator $V$ is equivalent to $\sigma^2/g$. At leading order in large $N$, the massive-vacuum fermion propagator in the large momentum region has the expansion
\begin{equation}
\frac{-i\slashed p+m}{p^2+m^2} = -\frac{i\slashed p}{p^2}+\frac{m}{p^2}+\frac{i\slashed p}{p^4}m^2-\frac{m^3}{p^4}+\cdots .
\end{equation}
Since $\langle\sigma\rangle = m$ and $\langle\sigma^2\rangle = m^2 + \cdots$, the large-momentum expansion selects the same dimension-two scalar sector represented by $V$.\footnote{The same expansion also gives the spinor-valued Wilson coefficient $\widetilde C_V(p)_{\alpha\beta}
= ig\left[\frac{(\slashed p)_{\beta\alpha}}{(p^2)^2}\right]_{\bar\mu^2}$, which will be derived below.} The interaction normalization chosen for $V$ removes an otherwise extraneous factor of $g$ from the condensate contribution and gives the simplest form for the Wilson coefficient after projection onto $A(p^2)$.

The OPE is naturally formulated in position space. Fourier transforming the Wilson coefficients,
\begin{equation}
\widetilde C_n(p) = \int d^2x\,e^{ip\cdot x}\,C_n(x),
\end{equation}
gives the corresponding momentum-space expansion, provided that local terms are retained. In particular, a constant Wilson coefficient in position space becomes a contact term at $p = 0$ and may be missed by a direct large-$p$ expansion.

We now enumerate the operator basis needed through dimension two. Loop integrals are regulated in $d = 2 - 2\varepsilon$ dimensions, while the spinor algebra is kept in the physical two-dimensional Clifford algebra. In two dimensions, this algebra is spanned by
\begin{equation}\label{eq:clifford_basis}
\Gamma_A \in \{\opone,\gamma^\mu,\gamma_5\},
\end{equation}
where $\gamma_5 \equiv -i\gamma^0\gamma^1$, $\gamma_5^2 = 1$, and $\{\gamma_5,\gamma^\mu\} = 0$. Completeness implies the Fierz identity
\begin{equation}\label{eq:fierz}
\delta_{\beta\gamma}\delta_{\delta\alpha} = \frac{1}{2}\left[ \delta_{\beta\alpha}\delta_{\delta\gamma} + (\gamma^\mu)_{\beta\alpha}(\gamma_\mu)_{\delta\gamma} + (\gamma_5)_{\beta\alpha}(\gamma_5)_{\delta\gamma} \right],
\end{equation}
or equivalently
\begin{equation}\label{eq:fierz_psi}
\psibar_\alpha\psi_\beta = \frac{1}{2}\delta_{\beta\alpha}\,\psibar\psi + \frac{1}{2}(\gamma^\mu)_{\beta\alpha}\,\psibar\gamma_\mu\psi + \frac{1}{2}(\gamma_5)_{\beta\alpha}\,\psibar\gamma_5\psi .
\end{equation}
Since $\psi$ has mass dimension $1/2$, a convenient basis through dimension two is
\begin{equation}
\cO_0 = \opone,\quad \cO_1^A = \psibar\Gamma_A\psi,\quad \cO_{2,\nu}^{A} = \psibar\Gamma_A\partial_\nu\psi,\quad \cO_2^{AB} = (\psibar\Gamma_A\psi)(\psibar\Gamma_B\psi),
\end{equation}
with $A,B \in \{\opone,\gamma^\mu,\gamma_5\}$.

In the free theory, normal ordering suffices to define coincident composite operators. Using
$\psibar_{\alpha}(0)\psi_{\beta}(x) = \,:\!\psibar_{\alpha}(0)\psi_{\beta}(x)\!: + D_{\beta\alpha}(x,0)$
and Taylor expanding the normal-ordered product around $x = 0$ gives
\begin{equation}\label{OPEexpand}
\psibar_\alpha(0)\psi_\beta(x) = D_{\beta\alpha}(x,0) + \frac{1}{2}\delta_{\beta\alpha}\,\psibar\psi(0) + \frac{1}{2}(\gamma^\mu)_{\beta\alpha}\,\psibar\gamma_\mu\psi(0) + \frac{1}{2}(\gamma_5)_{\beta\alpha}\,\psibar\gamma_5\psi(0) + x^\mu\psibar_\alpha\partial_\mu\psi_\beta(0) + \cdots .
\end{equation}
Thus $C_0(x) = D(x,0)$ and $C_1^A = \Gamma^A/2$ at tree level. Here and throughout the main-text discussion we suppress the external flavor indices and use the corresponding flavor-summed singlet normalization. (Appendix~\ref{app:OPE_details} instead keeps fixed external flavor indices explicitly; in that convention the same singlet coefficient carries the factor $\delta^{ij}/N$.) In the interacting theory, normal ordering is no longer enough, because loop corrections generate additional UV divergences. The Wilson coefficients must therefore be determined by matching with a regulator and renormalized together with the operators.

The identity coefficient for $\psibar_\alpha(0)\psi_\beta(x)$ is extracted by taking the expectation value of Eq.~\eqref{eq:OPE_operator} in the perturbative vacuum:
\begin{equation}\label{eq:C0_matching}
C_0(x)_{\alpha\beta} = \langle 0|\psibar_\alpha(0)\psi_\beta(x)|0\rangle = D_{\beta\alpha}(x,0).
\end{equation}
Including the bubble-chain resummation, the corresponding momentum-space coefficient is
\begin{equation}
\widetilde C_0(p)_{\alpha\beta} = \frac{-i(\slashed p)_{\beta\alpha}}{p^2+i\slashed p\,\Sigma(p)} \,,
\end{equation}
where at leading order in $1/N$, $\Sigma(p) = -i\slashed p\,A(p^2)/N$ with $A^\text{pert}(p^2)$ given in Eq.~\eqref{eq:A_pert_final}. Thus $C_0$ contains the full perturbative bubble-chain expansion and is the source of the leading renormalon divergence. Equivalently, the leading identity coefficient can be read off from the large-momentum expansion of the massive-vacuum propagator above; this is a useful check of the state independence of Wilson coefficients.

The coefficients of the dimension-one bilinears are determined by matching matrix elements with two additional soft external fermions, or, in diagrammatic terms, by computing the corresponding amputated four-point function. The matching condition is
\begin{align}\label{eq:C1_matching_main}
&\langle 0|\psibar_\alpha(0)\psi_\beta(x)\widetilde{\psibar}_\gamma(p_a)\widetilde\psi_\delta(p_b)|0\rangle \nonumber\\
&\qquad = C_0(x)_{\alpha\beta}\,\langle 0|\widetilde{\psibar}_\gamma(p_a)\widetilde\psi_\delta(p_b)|0\rangle + \sum_A C_1^A(x)_{\alpha\beta}\,\langle 0|\cO_1^A\,\widetilde{\psibar}_\gamma(p_a)\widetilde\psi_\delta(p_b)|0\rangle + \cdots .
\end{align}
At order $g$, the amputated four-point function in the full theory is generated by a single $\sigma$ exchange connecting the hard bilocal insertion to the soft external fermions, while the OPE reproduces the same contribution through a local operator insertion:
\begin{equation}\label{eq:C1_diagram}
\vcenter{\hbox{\CintLocalFull}} \qquad \text{and} \qquad \vcenter{\hbox{\CintLocalOPE}} .
\end{equation}
In dimensional regularization the OPE-side graph is scaleless. The detailed dimension-one coefficients and their renormalization are not needed for the leading $A$-form-factor cancellation below, because dimension-one operators do not contribute to the requisite $\slashed x$ structure with a Lorentz-invariant vacuum expectation value.

The dimension-two four-fermion Wilson coefficients require a six-point matching calculation with four additional soft legs. The first connected full-theory contributions appear at order $g^2$ and contain two $\sigma$ exchanges; schematically,
\begin{equation}
C_{\psibar\psi\psibar\psi} \sim \vcenter{\hbox{\CtwoLOfull}}\,.
\end{equation}
In the dimensional-regularization matching prescription used here, the lower-dimensional OPE-side contributions at zero soft momentum are scaleless and vanish. Appendix~\ref{app:OPE_details} gives the scalar-channel matching for the spinor-valued Wilson coefficient. Projecting this coefficient onto the vector self-energy form factor $A(p^2)$ gives Eq.~\eqref{eq:CVp_A_correct}, which enters the OPE cancellation analyzed below.

\subsection{Operator renormalization and reduced form-factor OPE}
\label{sec:mixing_main}

The bare operators and Wilson coefficients must be renormalized. The elementary fields and coupling obey
\begin{equation}
\psi_0 = Z_\psi^{1/2}\psi,\quad g_0 = \mu^{2\varepsilon}Z_g\,g,
\end{equation}
with $Z_\psi = 1 + \cO(g^2)$ and $Z_g = 1-Ng/(2\pi\varepsilon)+\cO(g^2)$. At dimension two, renormalization mixes operators with the same quantum numbers, and Appendix~\ref{app:OPE_details} gives the mixing needed to render the Wilson coefficient projected onto $A(p^2)$ finite. For the leading renormalon cancellation, the relevant consequences are that operator renormalization removes the poles in the dimension-two coefficients and that, at nonzero external momentum and at the leading order in $1/N$ considered here, $V$ is the only scalar-singlet dimension-two operator contributing a noncontact term to the $A$ form factor; the traced derivative operator contributes only through contact terms at this order.

Projecting the momentum-space fermion two-point function onto its $\slashed p$ Dirac structure isolates the OPE for the vector self-energy form factor $A(p^2)$. The momentum-space Green's function in the massive vacuum is
\begin{equation}
S(p) = \int d^dx\,e^{ip\cdot x}\,\langle m|\psibar(0)\psi(x)|m\rangle = \frac{1}{i\slashed p+m-\Sigma(p)}\,,
\end{equation}
and projecting with $\slashed p$ and using $\Sigma = -i\slashed p\,A/N+mB$ gives
\begin{equation}\label{trace1}
\tr[\slashed p\,S(p)] = -\frac{2i\,p^2(1+A/N)}{p^2(1+A/N)^2+m^2(1-B)^2} = -\frac{2i}{1+A/N} + \cO\!\left(\frac{m^2}{p^2}\right).
\end{equation}
The sign and denominator follow from $S^{-1} = i\slashed p(1+A/N)+m(1-B)$. The same projection can be implemented in position space by defining
\begin{equation}\label{eq:A_x_def}
\langle m|\psibar(0)\psi(x)|m\rangle = \mathcal A(x^2)\,\slashed x + \mathcal B(x^2)\,\opone,
\end{equation}
with
\begin{equation}
\mathcal A(x^2) = \frac{1}{2x^2}\tr\!\left[ \slashed x\,\langle m|\psibar(0)\psi(x)|m\rangle \right].
\end{equation}
Fourier transforming and tracing gives
\begin{equation}\label{eq:trpS_position}
\tr[\slashed p\,S(p)] = 2\!\int d^2x\,e^{ip\cdot x}\,(p\cdot x)\,\mathcal A(x^2)\,,
\end{equation}
and combining with Eq.~\eqref{trace1} we find
\begin{equation}\label{eq:A_from_calA}
A(p^2) = N\left[ 1 - i\!\int d^2x\,e^{ip\cdot x}\,(p\cdot x)\,\mathcal A(x^2) \right]
\end{equation}
at leading order in $1/N$, after dropping the purely kinematic $\cO(m^2/p^2)$ terms in the propagator projection. Below we nevertheless retain the $\cO(m^2/p^2)$ OPE contribution associated with the leading renormalon. These statements are consistent: the classical contribution $m^2/g$ to $\langle V\rangle$, multiplied by its Wilson coefficient $Ng/p^2$, reproduces the corresponding kinematic mass correction, whereas the fluctuation contribution $\Phi^{(2)}$ is the piece that carries the leading renormalon ambiguity studied below.

The form factor $\mathcal A(x^2)$ receives contributions only from operators whose Wilson coefficients contain a $\slashed x$ Dirac structure and whose vacuum expectation values are nonzero. The identity operator contributes through its perturbative coefficient. Although the dimension-one scalar and pseudoscalar bilinears can have nonvanishing scalar expectation values, their coefficients are proportional to $\opone$ and $\gamma_5$, respectively, and therefore enter $\mathcal B(x^2)$ rather than $\mathcal A(x^2)$. The vector bilinear has the required Dirac structure, but rotational invariance forces its vacuum expectation value to vanish. Consequently, no dimension-one operator contributes to $A(p^2)$.

At dimension two, the relevant scalar operators may be chosen as
\begin{equation}\label{KVdefs}
K \equiv i\psibar\slashed{\partial}\psi = i\,\delta_{\mu\nu}\cO_2^{\gamma^\nu,\mu},\qquad V = g\,\cO_2^{\opone,\opone}.
\end{equation}
Indeed, in two dimensions,
\begin{equation}
\partial_{p_\nu}\left[\frac{\slashed p\gamma^\nu\slashed p}{(p^2)^2}\right] = 0
\end{equation}
for $p \neq 0$, so the traced vector-derivative coefficient has no nonlocal contribution at finite momentum.

The reduced position-space OPE is
\begin{equation}\label{eq:A_OPE_reduced}
\mathcal A(x^2)=\mathcal A_{\rm pert}(x^2)+C_K(x^2)\,\langle m|K|m\rangle+C_V(x^2)\,\langle m|V|m\rangle+\cdots .
\end{equation}
The coefficient of $K$ contributes only through contact terms after Fourier transformation and will not be needed below. The noncontact coefficient relevant to the leading renormalon cancellation is
\begin{equation}\label{eq:C2A_reduced}
C_V(x^2)=-\frac{g}{8\pi}\ln(x^2\mu^2)+\cO(g/N,g^2)\,,
\end{equation}
where the sign and normalization of $C_V$ are fixed by the projection onto $A$. Away from contact terms,
\begin{equation}
\int d^2x\,e^{ip\cdot x}\ln(\mu^2x^2) = -4\pi\left[\frac{1}{p^2}\right]_{\bar\mu^2},
\end{equation}
and the extra factor of $(p\cdot x)$ in Eq.~\eqref{eq:A_from_calA} gives the projected coefficient below.

It is useful to distinguish the spinor-matrix Wilson coefficient in the two-point function from the scalar coefficient after projection onto $A$. The matrix coefficient of $V$ is
\begin{equation}
\widetilde C_V(p)_{\alpha\beta} = ig\left[\frac{(\slashed p)_{\beta\alpha}}{(p^2)^2}\right]_{\bar\mu^2} + \text{contact terms} + \cO(g/N,g^2).
\end{equation}
Projecting this coefficient with Eq.~\eqref{eq:A_from_calA} gives the coefficient that appears in the $A$-form-factor OPE:
\begin{equation}\label{eq:CVp_A_correct}
\widetilde C_V^{\,A}(p^2) = Ng\left[\frac{1}{p^2}\right]_{\bar\mu^2} + \cO(g,g^2N).
\end{equation}
The superscript $A$ distinguishes this scalar coefficient from the spinor-matrix coefficient $\widetilde C_V(p)_{\alpha\beta}$.

The momentum-space OPE projected onto the form factor $A(p^2)$ takes the form
\begin{equation}\label{eq:Ap2OPE}
A(p^2) = A_{\rm pert}(p^2) + \widetilde C_K^{\,A}(p^2)\,\langle m|K|m\rangle + \widetilde C_V^{\,A}(p^2)\,\langle m|V|m\rangle + \cdots .
\end{equation}
Since $\widetilde C_K^{\,A}(p^2)$ contributes only through a contact term, it vanishes at nonzero external momentum. The leading power correction relevant for the renormalon cancellation is therefore
\begin{equation}
A(p^2) = A_{\rm pert}(p^2) + Ng\left[\frac{1}{p^2}\right]_{\bar\mu^2}\langle m|V|m\rangle + \cdots,\qquad p \neq 0 .
\end{equation}
This identifies the renormalized matrix element of $V$ as the dimension-two contribution that cancels the $t=1$ ambiguity at the leading order in $1/N$ and in the scalar-singlet channel considered here, with the normalization required for the cancellation fixed by the projected Wilson coefficient. Other four-fermion Dirac and flavor structures arise only through subleading large-$N$ contractions and do not participate in the leading cancellation analyzed below.

\subsection{Hard-cutoff comparison and additive renormalization}
\label{sec:hard_cutoff_ope}

The hard-cutoff calculation makes the ultraviolet subtractions in the condensate contour explicit and ties them to ordinary composite-operator renormalization, including the additive mixing of $V$ with the identity. The cutoff serves only as a UV regulator, not as a Wilsonian factorization scale. When it is applied consistently to the connected matching calculation and to the operator matrix elements, it changes local terms and the relation between bare and renormalized quantities, but leaves unchanged the finite-momentum Wilson coefficient that enters the leading renormalon cancellation.

For $V$, regulating the matching calculation with a hard cutoff gives
\begin{equation}
C_{V,\bare}^{(\Lambda)}(x^2) = -\frac{g_0}{8\pi}\left[\ln\!\left(\frac{\Lambda^2x^2}{4}\right) + 2\gamma_E\right] + \cO(g_0/N,g_0^2)\,.
\end{equation}
Here the massless Fourier kernel is understood with the same infrared subtraction implicit in the matching prescription. The unsubtracted integral $\int^\Lambda d^2k\,e^{-ik\cdot x}/k^2$ is infrared divergent in two dimensions; only its nonlocal $x$ dependence is relevant for the Wilson coefficient, while the infrared-sensitive local pieces cancel against the corresponding scaleless OPE-side contributions.

Renormalization at the subtraction scale $\mu$ gives the hard-cutoff coefficient
\begin{equation}
C_V^{\rm hc}(x^2;\mu) = -\frac{g(\mu)}{8\pi}\left[\ln(\mu^2x^2) + 2\gamma_E - 2\ln2\right] + \cO(g/N,g^2)\,,
\end{equation}
where the superscript ${\rm hc}$ denotes the hard-cutoff scheme. The nonlocal logarithmic term therefore agrees with the $\overline{\rm MS}$ result, while the two schemes differ only by a finite local contribution:
\begin{equation}
C_V^{\rm hc}(x^2;\mu) = C_V^{(\overline{\rm MS})}(x^2;\mu) - \frac{g}{8\pi}(2\gamma_E - 2\ln2) + \cO(g/N,g^2)\,.
\end{equation}
Changing the cutoff profile modifies this finite constant but leaves the coefficient of $\ln(\mu^2x^2)$ unchanged.

In momentum space, the corresponding $A$-projected coefficient is
\begin{equation}
\widetilde C_V^{\,A,{\rm hc}}(p^2;\mu) = Ng\left[\frac{1}{p^2}\right]_{\mu^2}^{\rm hc} + \cO(g,g^2N),
\end{equation}
with
\begin{equation}
\left[\frac{1}{p^2}\right]_{\mu^2}^{\rm hc} = \left[\frac{1}{p^2}\right]_{\bar\mu^2}+\text{local contact term}.
\end{equation}
For $p \neq 0$, the hard-cutoff distribution reduces to the ordinary function $1/p^2$ and therefore agrees with its $\overline{\rm MS}$ counterpart. The two schemes differ only by a contact term supported at $p = 0$.

The logarithmic mixing between $K$ and $V$ that appears in the connected matching calculation is convention dependent at the level of individual matrix entries, since those entries change when one passes between the relations expressing bare operators in terms of renormalized operators and their inverse. We therefore characterize this mixing by its effect: it removes the $\ln(\Lambda^2/\mu^2)$ dependence from $C_{V,\bare}^{(\Lambda)}$ and yields the finite coefficient displayed above. The only additional hard-cutoff renormalization datum needed below is the leading additive mixing of $V$ with the identity.

The additive mixing of $V$ with the identity is especially transparent in the auxiliary-field formulation. At tree level, $D_\sigma^{(0)}(k^2) = g$, so the perturbative vacuum graph with a single insertion of $V$ behaves as
\begin{equation}
\langle0|V|0\rangle_{\rm pert} \sim \frac{1}{g}\int^\Lambda\frac{d^2k}{(2\pi)^2}\,D_\sigma(k^2) \sim \frac{\Lambda^2}{4\pi}.
\end{equation}
The quadratic divergence cannot be removed by multiplicative renormalization of $V$ or by mixing with $K$; it requires an additive subtraction proportional to the identity. The renormalized operator therefore takes the form
\begin{equation}
V_R(\mu) = Z_{VV}^{\rm hc}(\Lambda,\mu)V_{\bare}+Z_{VK}^{\rm hc}(\Lambda,\mu)K_{\bare}+Z_{V\opone}^{\rm hc}(\Lambda,\mu)\opone,
\end{equation}
with
\begin{equation}
Z_{V\opone}^{\rm hc}(\Lambda,\mu) \sim -\frac{\Lambda^2}{4\pi}+\cO(g\Lambda^2\ln\Lambda).
\label{eq:mix_ct_1V}
\end{equation}
Because this divergence belongs to the vacuum matrix element of the local operator, it does not appear in the connected matching graphs used to determine the nonlocal Wilson coefficient. The subtraction in Eq.~\eqref{eq:mix_ct_1V} is instead the hard-cutoff realization of the real ultraviolet subtraction that will later remove the divergent part of the condensate scale integral.

The hard cutoff $\Lambda$ introduced above is only a UV regulator and is removed after the composite operator has been renormalized. It should therefore not be identified with a Wilsonian factorization scale $\mu_F$. In a Wilsonian OPE, $\mu_F$ is kept finite and separates the momentum region assigned to the Wilson coefficient from the region assigned to the matrix element. The low-momentum contribution that produces the infrared renormalon in the continuum Wilson coefficient is then excluded from that coefficient and reappears through the $\mu_F$ dependence of the corresponding matrix element~\cite{Shifman:1998rb,Beneke:1998ui,Shifman:2013uka,Hoang:2009yr}. Here we instead use the continuum-renormalized OPE, in which the UV regulator is removed and the renormalon ambiguity remains visible in the separate Wilson-coefficient and condensate contributions.

\subsection{Four-fermion condensate as a scale integral}

At the leading order in $1/N$ considered here, the OPE identifies the dimension-two condensate $\langle V\rangle$ as the noncontact scalar-singlet contribution that cancels the leading renormalon ambiguity in $A(p^2)$. To connect that operator statement with the contour geometry developed above, we now rewrite the condensate as an integral over the same scale variable $z$ that appears in the reduced self-energy. The resulting comparison will show that the Wilson-coefficient contribution and the condensate are built from the same kernel but differ in their integration ranges: the reduced self-energy terminates at the external scale $Z$, whereas the local condensate extends to arbitrarily large $z$ and therefore requires ultraviolet renormalization. This common kernel is what later allows the two contributions to be represented as integration cycles of the same reduced action.

The reduced self-energy in Section~\ref{sec:theta} is governed by the kernel
\begin{equation}
\frac{dz}{\ln z}\left(1-\frac{2}{z}+\frac{1}{z^2}\right),
\end{equation}
integrated over the finite interval $1 \leq z \leq Z$. The four-fermion condensate is controlled by the same kernel, but its scale integral extends to $z = \infty$. Defining
\begin{equation}
\Phi(m) \equiv \langle m|V|m\rangle,
\end{equation}
and expanding the auxiliary-field representation of $V$ about the massive saddle as $\sigma = m+\frac{\alpha}{\sqrt N}$ gives
\begin{equation}
\Phi(m) = \frac{1}{g}\langle m|\sigma^2|m\rangle = \frac{m^2}{g} + \frac{1}{\pi\lambda}\langle m|\alpha^2|m\rangle + \cdots ,
\end{equation}
where the term linear in $\alpha$ vanishes because there are no tadpoles about the true vacuum. The leading fluctuation contribution is
\begin{equation}
\Phi^{(2)}(m) = \frac{1}{\pi\lambda}\langle m|\alpha^2|m\rangle = \vcenter{\hbox{\sigmasquaredvev}} = \frac{1}{\pi\lambda}\int\!\frac{d^dk}{(2\pi)^d}\,\Delta_\alpha(k^2) + \cdots \,,
\end{equation}
and writing the integral in terms of $y = k^2/m^2$ gives
\begin{equation}\label{Phi2y}
\Phi^{(2)}(m) = \kappa\int_0^\infty dy\,\frac{1}{\xi_y\ln\frac{\xi_y+1}{\xi_y-1}}
\end{equation}
for $\kappa = \frac{m^2}{2\pi\lambda}$. Changing to the $z$ variable gives
\begin{equation}\label{Phi2z}
\Phi^{(2)}(m) = \kappa\int_1^\infty\frac{dz}{\ln z}\left( 1-\frac{2}{z}+\frac{1}{z^2} \right) = \kappa\int_1^\infty dz\int_0^\infty dt\,z^{-t}\left( 1-\frac{2}{z}+\frac{1}{z^2} \right).
\end{equation}
Thus the condensate is governed by the same $z$-space kernel as the reduced self-energy, but its scale integral is not cut off by the external hard momentum. This makes the OPE split concrete: the original reduced self-energy integrates over $1 \leq z \leq Z$, while the local condensate integrates over $1 \leq z < \infty$ and must be renormalized.

To compare the condensate with the contour representation of the reduced self-energy, we rescale the $A$-form-factor OPE by $2x = 2p^2/m^2$. Since the Wilson coefficient of $V$ is $\widetilde C_V^{\,A}(p^2) = Ng/p^2$, its coefficient in this normalization is
\begin{equation}
\widehat C_V = \frac{2p^2}{m^2}\,\frac{Ng}{p^2} = \frac{2Ng}{m^2} = \frac{1}{\kappa},
\end{equation}
where $\kappa = m^2/(2\pi\lambda)$ and $Ng = \pi\lambda$. The factor $\widehat C_V = 1/\kappa$ cancels the prefactor in the fluctuation contribution $\Phi^{(2)}(m)$, so that
\begin{equation}
\widehat C_V\,\Phi^{(2)}(m)
= \int_1^\infty \frac{dz}{\ln z}\left(1-\frac{2}{z}+\frac{1}{z^2}\right).
\end{equation}
Strictly speaking, $\widetilde C_V^{\,A}$ was obtained from the OPE of the full form factor $A(p^2)$, whereas the contour construction is formulated for the reduced observable $A_0(x)$. Its use here is justified by the defining property of the $\theta$ reduction: it preserves the perturbative $E_0$ pole and the dimension-two $H_1$ term with their exact relative normalization. We are not asserting that $A_0$ inherits the complete OPE of $A$ beyond this reduced leading-renormalon sector.

The condensate therefore enters with the same kernel and unit normalization as the reduced self-energy. Schematically, the corresponding OPE takes the form
\begin{equation}\label{eq:ope_2xA0}
2xA_0(x) = \widehat C_0(x) + \widehat C_V\,\Phi^{(2)}(m) + \cdots .
\end{equation}
Here $\widehat C_0$ is the asymptotic contribution generated by the perturbative boundary thimble, while the condensate scale integral supplies the second contour needed to reconstruct the finite strip. The equality $\widehat C_V\kappa = 1$ ensures that the perturbative and condensate contributions are integrals of the same holomorphic kernel with the relative coefficient required for their complex tails to cancel. The leading OPE ambiguity can consequently be interpreted as a relation between integration cycles, rather than only as a numerical cancellation between independently computed residues.

The large-$z$ behavior of Eq.~\eqref{Phi2z} makes the condensate ultraviolet divergent. Writing $\rho = \ln z$ suggests combining the three terms by shifting the $t$ integration variable, as in Eq.~\eqref{eq:strip}, which would give
\begin{equation}\label{Phi2formrhot}
\Phi^{(2)}(m) = \kappa\left[ \int_0^1 dt - \int_1^2 dt \right]\int_0^\infty d\rho\,e^{(1-t)\rho}.
\end{equation}
This rearrangement is not yet legitimate, however, because the $\rho$ integral converges only for $\operatorname{Re}t > 1$, where
\begin{equation}
\int_0^\infty d\rho\,e^{(1-t)\rho} = \frac{1}{t-1}.
\end{equation}
Formally continuing the result outside its domain of convergence gives
\begin{equation}
\left[ \int_0^1 dt - \int_1^2 dt \right]\frac{1}{t-1} = \int_0^\infty dt\left[ \frac{1}{t-1} - \frac{2}{t} + \frac{1}{t+1} \right] = \int_0^\infty dt\,\frac{2}{t(t^2-1)}.
\end{equation}
The resulting kernel is $\frac{2}{t(t^2-1)} = 2H_1(t)$, where $H_1(t)$ is the dimension-two term in the large-$x$ trans-series of Eq.~\eqref{eq:A_transseries}; the factor of $2$ reflects the normalization of $2xA_0(x)$ used in the reduced contour analysis. Besides the expected pole at $t = 1$, it contains an endpoint singularity at $t = 0$ that encodes the ultraviolet divergence of the unrenormalized scale integral. A regulator must therefore be introduced before the terms are rearranged; only after the ultraviolet divergences have been subtracted can the cutoff be removed and a well-defined Borel representation obtained.

To regulate the ultraviolet region before rearranging the scale integral, we impose a dimensionless hard cutoff $z \leq \Lambda$.\footnote{The hard cutoff is chosen for convenience. A smooth regulator, such as $e^{-z/\Lambda}$, would give the same renormalized condensate up to finite scheme-dependent terms in the Borel representation and is closer in spirit to lattice-motivated prescriptions such as gradient flow~\cite{Beneke:2025hlg,Zhang:2025mer}.} The regulated fluctuation contribution is then
\begin{equation}
\Phi_\Lambda^{(2)}(m) = \kappa\int_0^\infty dt\int_1^\Lambda dz\,z^{-t}\left(1-\frac{2}{z}+\frac{1}{z^2}\right) = \kappa\int_0^\infty dt\,H_{1,\Lambda}(t),
\end{equation}
where
\begin{equation}
H_{1,\Lambda}(t) = \frac{\Lambda^{1-t}}{1-t} + \frac{2\Lambda^{-t}}{t} - \frac{\Lambda^{-1-t}}{1+t} + \frac{1}{t-1} - \frac{2}{t} + \frac{1}{1+t}.
\end{equation}
For every finite $\Lambda$, the apparent singularities at $t = 0,1,-1$ cancel within $H_{1,\Lambda}(t)$, so the regulated condensate is real and requires no lateral prescription. Its large-$\Lambda$ behavior is
\begin{align}\label{eq:vev_cutoff_asymptotic}
\Phi_\Lambda^{(2)}(m)
&= -\kappa\,\Ein(-\ln\Lambda) - \kappa\,\Ein(\ln\Lambda) \nonumber\\
&= \kappa\left[ \frac{\Lambda}{\ln\Lambda} \left( 1+\frac{1}{\ln\Lambda} +\cO((\ln\Lambda)^{-2})
\right) -2\ln\ln\Lambda +\cO(\Lambda^0)\right].
\end{align}
The leading term, $\kappa\Lambda/\ln\Lambda$, is the large-$N$ bubble-resummed power divergence canceled by the additive identity counterterm $Z_{V\opone}^{\rm hc}$ in Eq.~\eqref{eq:mix_ct_1V}. The logarithmic divergence $-2\kappa\ln\ln\Lambda$ is removed by the logarithmic composite-operator subtraction that defines the dimension-two operator at a subtraction scale $\mu$. In the reduced scale-integral description used here, we characterize this subtraction by its effect on the condensate kernel rather than by displaying the complete dimension-two operator-mixing matrix. At the level of the cutoff Borel kernel, these two subtractions act as
\begin{equation}\label{eq:ct_log}
H_{1,\Lambda}(t) \longrightarrow H_{1,\Lambda}(t) -\frac{\Lambda^{1-t}}{1-t} +\frac{2(\mu^{-t}-\Lambda^{-t})}{t}.
\end{equation}
Writing the renormalization directly as a subtraction from $H_{1,\Lambda}(t)$ fixes the sign convention unambiguously. The two terms implement, respectively, the additive identity subtraction of the power divergence and the logarithmic composite-operator subtraction. Both remove real ultraviolet divergences from the finite-cutoff condensate. The lateral ambiguity appears only after these subtractions have been made and the limit $\Lambda \to \infty$ exposes the remaining pole at $t = 1$.

After the power and logarithmic ultraviolet divergences have been subtracted, the remaining cutoff dependence vanishes as $\Lambda \to \infty$. The term $\Lambda^{1-t}/(1-t)$ is the large-$N$ Borel-space representation of the additive mixing of $V$ with the identity; because the finite part of this subtraction is scheme dependent, its identification with $Z_{V\opone}^{\rm hc}$ is understood up to finite local terms. Removing the cutoff then gives the renormalized kernel
\begin{equation}\label{eq:H1_ren_explicit}
H_1^{\rm ren}(t;\mu) = \frac{1}{t-1} - \frac{2(1-\mu^{-t})}{t} + \frac{1}{1+t} = \frac{2}{t(t^2-1)} + \frac{2\mu^{-t}}{t}.
\end{equation}
The subtraction has removed the apparent singularity at $t = 0$, since
\begin{equation}
-\frac{2(1-\mu^{-t})}{t} = -2\ln\mu + \cO(t),
\end{equation}
but the pole at $t=1$ remains. To resolve the associated Stokes discontinuity, we define the two lateral continuations of the renormalized condensate by contours passing above or below this pole:
\begin{equation}\label{eq:vev_ren_lateral_final}
\Phi_\pm^{(2),{\rm ren}}(m) = \kappa\int_0^\infty dt\left[\frac{1}{t-1\pm i0} - 2\frac{1-\mu^{-t}}{t} + \frac{1}{1+t}\right].
\end{equation}
The two choices give
\begin{equation}\label{eq:vev_imag_final}
\operatorname{Im}\Phi_\pm^{(2),{\rm ren}}(m) = \mp\kappa\pi = \mp\frac{m^2}{2\lambda}.
\end{equation}
One may instead define the real principal-value prescription
\begin{equation}
\Phi_{\rm PV}^{(2),{\rm ren}}(m)=\frac{1}{2}\left[\Phi_+^{(2),{\rm ren}}(m)+\Phi_-^{(2),{\rm ren}}(m)\right].
\end{equation}
The reason for retaining the two lateral continuations separately is that they carry the Stokes discontinuity and admit the relative-cycle interpretation developed below; their average gives the principal-value prescription.

Combining the perturbative ambiguity in Eq.~\eqref{E:todisplay1} with the Wilson coefficient of $V$ in the OPE for $A(p^2)$, given in Eq.~\eqref{eq:CVp_A_correct}, the dimension-two condensate contribution gives
\begin{equation}
\operatorname{Im}\!\left[\mathcal S_\pm(A_{\rm pert})\right] + \widetilde C_V^{\,A}(p^2)\,\operatorname{Im}\!\left[\Phi_\pm^{(2),{\rm ren}}(m)\right] = \pm\frac{\pi m^2}{2p^2} \mp\frac{Ng}{p^2}\frac{m^2}{2\lambda} = 0,
\end{equation}
where $Ng = \pi\lambda$. The ambiguity of the perturbative identity coefficient is therefore canceled exactly by the ambiguity of the renormalized matrix element of $V$, with the relative normalization fixed by the Wilson coefficient. This establishes the leading OPE cancellation directly at the level of the renormalized coefficient and matrix element. The remaining task is to show that the equality of their imaginary parts reflects a common contour geometry, rather than only a numerical agreement between their residues.

\subsection{Renormalized condensate contour and homological cancellation}

The cancellation established above fixes the relative imaginary parts of the perturbative Wilson coefficient and the renormalized condensate, but it does not yet exhibit the contour relation responsible for that equality. The purpose of this subsection is to derive the condensate cycle directly from the cutoff scale integral, keeping track of how the ultraviolet subtractions alter its boundary at large $\rho$. Once the cutoff has been removed, the remaining contour can be compared in the same $(t,\rho)$ space with the perturbative boundary thimble. Their complex tails then cancel, and their sum reconstructs the finite real strip that defines the reduced self-energy.

Before the cutoff is removed, the fluctuation contribution takes the form
\begin{equation}\label{eq:vev_3exp_reg}
\Phi_\Lambda^{(2)}(m) = \kappa\int_0^\infty dt\left[\int_0^{\ln\Lambda}d\rho\,e^{(1-t)\rho} - 2\int_0^{\ln\Lambda}d\rho\,e^{-t\rho} + \int_0^{\ln\Lambda}d\rho\,e^{-(1+t)\rho}\right].
\end{equation}
Because all three $\rho$ integrals run over the same finite real interval, the cutoff-regulated condensate is real. The third term remains convergent when $\Lambda \to \infty$ for every $t > 0$. The second term contains the logarithmic ultraviolet divergence removed by the renormalization of $V$: the subtraction in Eq.~\eqref{eq:ct_log} removes the interval $\ln\mu \leq \rho \leq \ln\Lambda$, leaving the finite integral over $0 \leq \rho \leq \ln\mu$.

\begin{figure}[t]
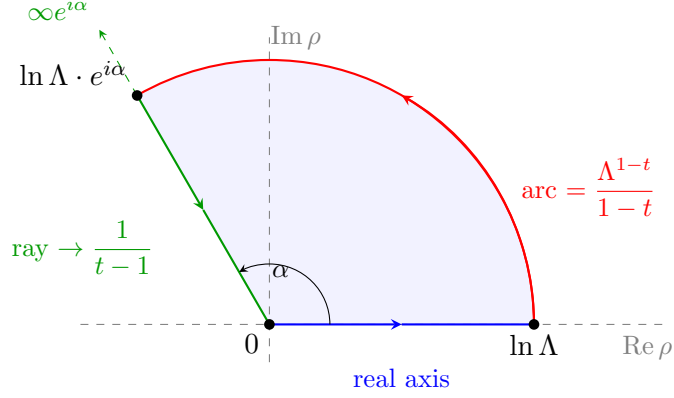

\centering
\rhotriangle
\caption{The triangular contour in the complex $\rho$ plane for $0 < t < 1$. The real-axis integral (blue) equals the ray integral (green) plus the arc (red). As $\Lambda\to\infty$ the ray converges to $1/(t-1)$; the arc becomes the power divergence $\Lambda^{1-t}/(1-t)$, which is removed by the identity-operator counterterm.}
\label{fig:rho_triangle}
\end{figure}

The remaining ambiguity arises from the first integral. For $t > 1$, the limit $\Lambda \to \infty$ converges along the positive real $\rho$ axis, whereas for $0 < t < 1$ the integrand grows exponentially. Once the power divergence has been removed, the renormalized integral is defined by continuing the $\rho$ contour into a direction along which $e^{(1-t)\rho}$ decays. The triangular contour in Figure~\ref{fig:rho_triangle} makes this continuation explicit: the original real-axis segment is equal to the rotated contour plus the connecting arc, and the arc contributes exactly the power-divergent term canceled by the additive identity counterterm.

A contour representation that keeps the lateral continuation explicit is
\begin{equation}\label{eq:vev_ren_trho}
\Phi_\pm^{(2),{\rm ren}}(m) = \lim_{\epsilon\to0^+}\kappa\int_0^{\infty e^{\pm i\epsilon}} dt\left[\int_{\mathcal C_{\rho,\pm}(\epsilon)}d\rho\,e^{(1-t)\rho} - 2\int_0^{\ln\mu}d\rho\,e^{-t\rho} + \int_0^\infty d\rho\,e^{-(1+t)\rho}\right],
\end{equation}
where $\mathcal C_{\rho,\pm}(\epsilon)$ begins at $\rho = 0$ and extends to complex infinity along a direction for which $e^{(1-t)\rho}$ decays. For the comparison with the perturbative boundary thimble, we choose $\mathcal C_{\rho,\pm}(\epsilon)$ to correspond to the path $\rho = -i e^{\mp i\epsilon/2}s$ for $0 \leq s < \infty$. The correlated deformations of the $t$ and $\rho$ contours ensure convergence at complex infinity and specify whether the $t$ contour passes above or below the pole at $t = 1$. Performing the $\rho$ integrations and taking $\epsilon\to0^+$ reproduces Eq.~\eqref{eq:vev_ren_lateral_final}. The second and third terms are unambiguous, whereas the first passes through the renormalon saddle and carries the condensate ambiguity that cancels the ambiguity of the perturbative boundary thimble.

We can now compare the perturbative Wilson coefficient and the condensate as contour integrals in the same complexified $(t,\rho)$ space. The complete reduced observable contains the finite strip
\begin{equation}\label{eq:A0_3exp_strip}
2xA_0(x) = \int_0^\infty dt\left[ \int_0^{\ln Z}d\rho\,e^{(1-t)\rho} - 2\int_0^{\ln Z}d\rho\,e^{-t\rho} + \int_0^{\ln Z}d\rho\,e^{-(1+t)\rho} \right].
\end{equation}
Only the first exponential has the $t = 1$ renormalon pole; the other two are regular at $t = 1$ and affect only real finite terms.  Accordingly, the relative-cycle identity below concerns this renormalon-carrying first term. The remaining two terms are part of the complete reduced observable but are not included in the contour identity for the leading $t=1$ ambiguity.

\begin{figure}[t!]
\centering
\includegraphics[width=0.44\textwidth]{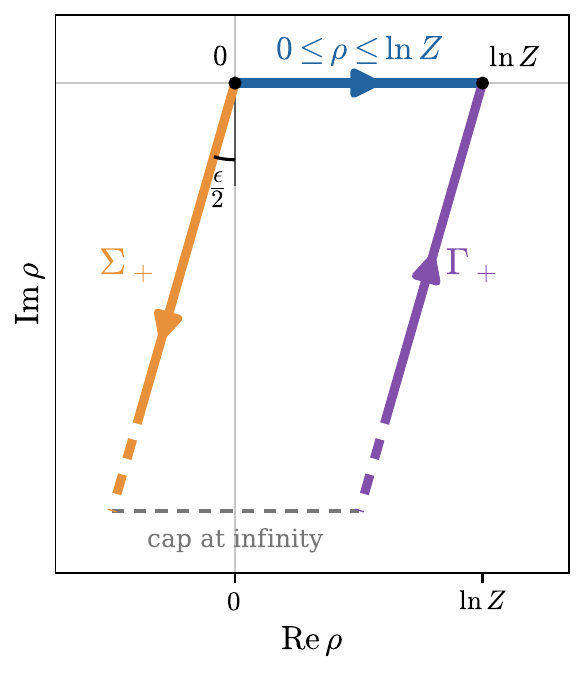}
\hspace{0.025\textwidth}
\raisebox{.15\height}{\includegraphics[width=0.44\textwidth]{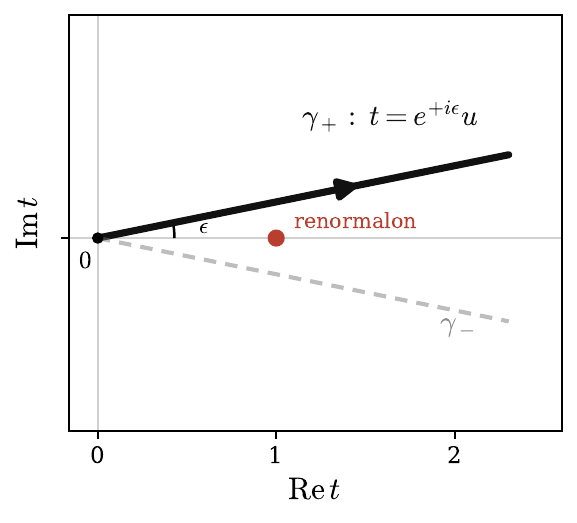}}
\caption{Geometric origin of the renormalon cancellation, shown in the $\rho$ plane and in the $t$ plane. Left: the real strip is part of the finite rectangle $\rho \in [0,\ln Z]$, $t \in \mathbb R^+$ representing $2xA_0$. The condensate cycle $\Sigma_+$ is anchored at $(0,0)$, while the perturbative thimble $\Gamma_+$ is anchored at $(0,\ln Z)$; the dashed arcs at infinity give no contribution. Right: the common lateral $t$ contour avoids the renormalon pole at $t = 1$. The condensate cycle and perturbative thimble combine to give the strip in the left panel, after which the $t$ contour can be returned to the real axis.}
\label{fig:homology}
\end{figure}

Observe that the term proportional to $e^{(1-t)\rho}$ carries the pole at $t = 1$. In the $2xA_0$ normalization, the perturbative contribution from this term is represented by a contour anchored at the upper boundary $\rho = \ln Z$:
\begin{align}\label{eq:C0_lateral_specific}
\widehat C_{0,\pm} &= -\int_0^{\infty e^{\pm i\epsilon}} \!\! dt\int_{\ln Z}^{\ln Z-i e^{\mp i\epsilon/2}\infty}d\rho\,e^{(1-t)\rho} \nonumber\\
&= \int_{\Gamma_\pm}dt\wedge d\rho\,e^{(1-t)\rho}.
\end{align}
The minus sign is the orientation inherited from the upper boundary of the finite strip. The representative $\Gamma_+$ is shown in Figure~\ref{fig:homology}.

The corresponding part of the renormalized condensate is anchored at the lower boundary $\rho = 0$:
\begin{equation}\label{eq:contour_ren_Vop}
\widehat C_V\,\Phi_\pm^{(2),{\rm ren}}(m) \supset \int_0^{\infty e^{\pm i\epsilon}}\!\!dt \int_0^{-i e^{\mp i\epsilon/2}\infty}\!\!d\rho\, e^{(1-t)\rho} = \int_{\Sigma_\pm}dt\wedge d\rho\,e^{(1-t)\rho},
\end{equation}
which defines the condensate cycle $\Sigma_\pm$, also displayed in Figure~\ref{fig:homology}.

Because $\widehat C_V\kappa = 1$, the perturbative boundary-thimble contribution and the corresponding condensate contribution enter the OPE with the same normalization. Their sum is
\begin{align}\label{eq:ope_homology}
\widehat C_{0,\pm} + \widehat C_V\,\Phi_\pm^{(2),{\rm ren}} &\supset \int_0^{\infty e^{\pm i\epsilon}}dt\left(-\int_{\ln Z}^{\ln Z-i e^{\mp i\epsilon/2}\infty} + \int_0^{-i e^{\mp i\epsilon/2}\infty}\right)d\rho\,e^{(1-t)\rho} \nonumber\\
&= \int_0^{\infty e^{\pm i\epsilon}}dt\int_0^{\ln Z}d\rho\,e^{(1-t)\rho}.
\end{align}
The two contours have the same asymptotic direction at complex infinity, so the arc joining their endpoints gives no contribution. Their oriented difference therefore reduces to the finite segment $0 \leq \rho \leq \ln Z$. After the $\rho$ integration, the factor $(Z^{1-t}-1)/(1-t)$ is regular at $t = 1$, allowing the $t$ contour to be returned to the positive real axis. The perturbative and condensate cycles thus reconstruct the finite real strip for the renormalon-carrying term, making the cancellation of their imaginary parts a consequence of the contour identity rather than a separate matching of residues.

Figure~\ref{fig:homology} depicts the contour identity in Eq.~\eqref{eq:ope_homology}. In the $\rho$ plane, the condensate cycle begins at $\rho = 0$ and the perturbative thimble begins at $\rho = \ln Z$, while both extend to complex infinity in the same direction; their oriented difference is therefore the finite segment $0 \leq \rho \leq \ln Z$. In the $t$ plane, the two cycles share the same choice of contour above or below the renormalon pole at $t = 1$.

Eq.~\eqref{eq:ope_homology} gives a Picard--Lefschetz realization of the renormalon cancellation~\cite{SVZ1979b,SVZ1979,Mueller:1984vh,Maiani:1991az,Beneke:1998ui,Luke:1994xd}. The perturbative Wilson coefficient and the renormalized condensate are represented by two relative cycles of the same holomorphic two-form, anchored at the upper and lower boundaries of the finite strip and passing through the renormalon saddle at $(t,\rho) = (1,0)$. Their common local geometry fixes the magnitude of the Stokes discontinuity, while the opposite orientations of the two boundaries fix its relative sign. When the cycles are added, their complex tails cancel and the finite real strip is recovered. The usual cancellation between the perturbative ambiguity and the condensate ambiguity reflects the fact that the OPE separates a single real integration cycle into two contour contributions whose individual lateral ambiguities disappear when the physical observable is reconstructed.

\section{Path-integral interpretation of the renormalon saddle}
\label{sec:PI}

The preceding sections introduced $t$ and $\rho$ as coordinates of the reduced contour integral governing the leading OPE cancellation. We now give these variables a direct interpretation in the large-$N$ auxiliary-field theory. The logarithmic coordinate $\rho$ is naturally encoded in the momentum dependence of the quadratic fluctuations about the massive saddle, and $t$ can be interpreted as a suitably weighted radial coordinate of the auxiliary-field Fourier mode carrying the exchanged momentum. This reproduces the same reduced $(t,\rho)$ representation introduced above and allows us to determine what the renormalon critical point represents in the original auxiliary-field variables.

At order $1/N$, the self-energy contains a single auxiliary-field propagator, so the interpretation below is modewise. The point $(t,\rho) = (1,0)$ will consequently be interpreted as a boundary critical point of this reduced mode-space representation, rather than as an additional solution of the full equation $\delta S_{\rm eff}/\delta\sigma=0$.

To connect the reduced $(t,\rho)$ representation to the underlying field theory, we begin directly from the large-$N$ auxiliary-field path integral. After integrating out the fermions, the fermion two-point function is an expectation value in the $\sigma$ functional integral:
\begin{equation}
S(x,0)\equiv\langle\psi(x)\psibar(0)\rangle=\frac{\displaystyle\int\mathcal D\sigma\,(\slashed{\partial}+\sigma)^{-1}(x,0)e^{-S_{\rm eff}[\sigma]}}{\displaystyle\int\mathcal D\sigma\,e^{-S_{\rm eff}[\sigma]}},
\qquad
S_{\rm eff}[\sigma]=\int d^2x\,\frac{\sigma^2}{2g}-N\Tr\ln(\slashed{\partial}+\sigma).
\label{eq:PI_full_functional}
\end{equation}
Expanding about the massive saddle,
\begin{equation}
\sigma=m+\frac{\alpha}{\sqrt N},
\end{equation}
gives
\begin{equation}
\left(\slashed{\partial}+m+\frac{\alpha}{\sqrt N}\right)^{-1}=S_m-\frac{1}{\sqrt N}S_m\alpha S_m+\frac{1}{N}S_m\alpha S_m\alpha S_m+\cdots,
\qquad
S_m(q)=\frac{1}{i\slashed q+m}.
\end{equation}
The linear term has vanishing expectation value by the gap equation. At order $1/N$, the $\alpha$ measure may be replaced by its Gaussian approximation about the massive saddle, and the self-energy therefore takes the form
\begin{equation}
\Sigma^{(1)}(p)=\frac{1}{N}\int\frac{d^2k}{(2\pi)^2}\,S_m(p+k)\,\langle\alpha(k)\alpha(-k)\rangle_G,
\label{eq:PI_selfenergy_functional}
\end{equation}
where
\begin{equation}
\langle\mathcal O[\alpha]\rangle_G\equiv\frac{\displaystyle\int\mathcal D\alpha\,\mathcal O[\alpha]e^{-S_m^{(2)}[\alpha]}}{\displaystyle\int\mathcal D\alpha\,e^{-S_m^{(2)}[\alpha]}},
\qquad
S_m^{(2)}[\alpha]=\frac{1}{2}\int\frac{d^2q}{(2\pi)^2}\,\alpha(q)\Gamma_m^{(\alpha)}(q^2)\alpha(-q).
\end{equation}
Thus the dressed propagator appearing in Eq.~\eqref{eq:Sigma_diagram} is directly the Gaussian path-integral contraction
\begin{equation}
\Delta_\alpha(k^2)=\langle\alpha(k)\alpha(-k)\rangle_G.
\end{equation}

\subsection{Inverse propagators and scale coordinate from the fermion bubble}

The two inverse propagators relevant below were derived in Section~\ref{sec:effective_action}. Equivalently, they are the quadratic kernels of the auxiliary-field effective action about the corresponding backgrounds. About the chirally symmetric configuration $\sigma = 0$, the renormalized inverse propagator at nonzero Euclidean momentum is
\begin{equation}\label{eq:PI_perturbative_kernel}
\Gamma_0^{(\sigma)}(k^2) = \frac{1}{g(\mu)} + \frac{N}{2\pi}\ln\frac{k^2}{\mu^2} = \frac{N}{2\pi}\ln\frac{k^2}{m^2}.
\end{equation}
Its inverse is the running auxiliary-field propagator that generates the perturbative bubble-chain expansion.

The path integral interpretation instead uses the full inverse propagator about the massive saddle. Writing $\sigma = m + \alpha/\sqrt N$ as before, the inverse propagator of $\alpha$, at leading order in $1/N$ with its full dependence on $k^2/m^2$ retained, is
\begin{equation}
\Gamma_m^{(\alpha)}(k^2) = \frac{1}{2\pi}\,\xi_k\ln\frac{\xi_k + 1}{\xi_k - 1}, \qquad \xi_k = \sqrt{1 + \frac{4m^2}{k^2}}.
\end{equation}

It is useful to isolate the logarithmic coordinate that already appeared in the reduced contour problem. Define \begin{equation}\label{eq:PI_rho_definition} \rho(k)\equiv\ln\frac{\xi_k+1}{\xi_k-1}=2\operatorname{arsinh}\frac{k}{2m}. \end{equation} The massive-vacuum inverse propagator then takes the form \begin{equation}\label{eq:PI_massive_hessian} \Gamma_m^{(\alpha)}(k^2)=\frac{1}{2\pi}\,\xi_k\rho(k). \end{equation} Since $z=e^{\rho(k)}$, this is precisely the scale coordinate $\rho=\ln z$ introduced in Section~\ref{sec:theta}.

The same variable gives
\begin{equation}\label{eq:PI_rho_uniformization}
\frac{k^2}{m^2} = 4\sinh^2\frac{\rho(k)}{2} = e^{\rho(k)} - 2 + e^{-\rho(k)}, \qquad \xi_k = \coth\frac{\rho(k)}{2}.
\end{equation}
When $k^2/m^2 = x$, it reaches the upper endpoint $\rho = \ln Z$.

To compare the massive-vacuum inverse propagator with its perturbative counterpart, write the unrescaled fluctuation as $\delta\sigma = \alpha/\sqrt N$. Its inverse propagator is
\begin{equation}\label{eq:PI_massive_sigma_kernel}
\Gamma_m^{(\delta\sigma)}(k^2) = N\Gamma_m^{(\alpha)}(k^2) = \frac{N}{2\pi}\,\xi_k\rho(k).
\end{equation}
At large momentum,
\begin{equation}\label{eq:PI_kernel_UV_match}
\xi_k\rho(k) = \ln\frac{k^2}{m^2} + \frac{2m^2}{k^2}\left(1 + \ln\frac{k^2}{m^2}\right) + \cO\!\left(\frac{m^4}{k^4}\ln\frac{k^2}{m^2}\right).
\end{equation}
The first term reproduces the perturbative inverse propagator in Eq.~\eqref{eq:PI_perturbative_kernel}, and the remaining terms begin the sequence of power corrections omitted by that ultraviolet approximation. Their resummation into the full massive-vacuum inverse propagator removes its apparent zero at $k^2 = m^2$. In the infrared,
\begin{equation}
\rho(k) = \frac{k}{m} + \cO\!\left(\frac{k^3}{m^3}\right), \qquad \xi_k = \frac{2m}{k} + \cO\!\left(\frac{k}{m}\right),
\end{equation}
so $\xi_k\rho(k) \to 2$ and
\begin{equation}
\Gamma_m^{(\alpha)}(0) = \frac{1}{\pi}.
\end{equation}

The second coordinate comes from radializing the Gaussian Fourier mode whose contraction appears in the self-energy. To make this explicit, place the theory temporarily in a finite volume $\mathcal V$ and write
\begin{equation}
\alpha(x) = \frac{1}{\sqrt{\mathcal V}}\sum_q\alpha_qe^{iq\cdot x}, \qquad \alpha_{-q} = \alpha_q^*.
\end{equation}
At fixed exchanged momentum $k$, the order-$1/N$ self-energy involves the Gaussian contraction $\langle\alpha_k \alpha_{-k}\rangle$. Since the quadratic action is diagonal in momentum space, the integrations over all modes $q\neq\pm k$ factor identically between the numerator and denominator of this expectation value and cancel. Thus, for each $k$, the functional integral relevant to this contraction reduces to the ordinary Gaussian integral over the single mode pair $\{k,-k\}$. For a nonzero momentum pair $\{k,-k\}$, the reality condition on the original field contour gives $\alpha_{-k} = \alpha_k^*$.\footnote{The radial derivation applies directly for $k\neq0$, where a real field supplies a complex Fourier-mode pair. The endpoint $k=0$ is reached by continuity in the subsequent momentum integral.} The contribution of this pair to the quadratic action is
\begin{equation}
S_{m,k}^{(2)} = \Gamma_m^{(\alpha)}(k^2)|\alpha_k|^2 = \frac{\xi_k\rho(k)}{2\pi}|\alpha_k|^2\,.
\end{equation}
The normalized Gaussian contraction is
\begin{equation}
\Delta_\alpha(k^2)=\frac{\displaystyle\int_{\mathbb C}\frac{d^2\alpha_k}{\pi}\,|\alpha_k|^2e^{-\Gamma_k|\alpha_k|^2}}{\displaystyle\int_{\mathbb C}\frac{d^2\alpha_k}{\pi}\,e^{-\Gamma_k|\alpha_k|^2}}=\frac{1}{\Gamma_k}, \quad \Gamma_k=\Gamma_m^{(\alpha)}(k^2).
\label{eq:PI_single_mode_contraction}
\end{equation}
For a complex Gaussian mode one also has
\begin{equation}
\frac{1}{\Gamma_k}=\int_{\mathbb C}\frac{d^2\alpha_k}{\pi}\,e^{-\Gamma_k|\alpha_k|^2}=\int_0^\infty ds\,e^{-\Gamma_k s}.
\label{eq:PI_single_mode_partition}
\end{equation}
Thus, at this order, the propagator can itself be represented as the partition integral of the single active Fourier mode. The last equality follows by writing $s = |\alpha_k|^2$ in polar coordinates.

Substituting this one-mode representation into the $\theta$-reduced self-energy makes the reduction of the functional integral explicit. From Eq.~\eqref{eq:theta_approx_A0},
\begin{equation}
2xA_0(x) = \frac{2}{m^2}\int_{|k|<p}\frac{d^2k}{(2\pi)^2}\,\Delta_\alpha(k^2),
\end{equation}
and hence
\begin{equation}
2xA_0(x) = \frac{2}{m^2}\int_{|k|<p}\frac{d^2k}{(2\pi)^2}\int_{\mathbb C}\frac{d^2\alpha_k}{\pi}\,e^{-S_{m,k}^{(2)}},
\qquad
S_{m,k}^{(2)} = \Gamma_m^{(\alpha)}(k^2)|\alpha_k|^2.
\label{eq:PI_one_mode_A0}
\end{equation}
Thus the reduced observable is an integral over the momentum and amplitude of the single exchanged auxiliary-field mode, obtained from the full Gaussian functional integral after all spectator modes have canceled.

Keeping fixed the scale coordinate $\rho(k) = \ln z(k)$ introduced above, define the rescaled mode and its radial coordinate by
\begin{equation}
b_{\pm k} = \sqrt{\xi_k}\,\alpha_{\pm k},\qquad t = \frac{b_kb_{-k}}{2\pi} = \frac{\xi_k\alpha_k\alpha_{-k}}{2\pi}. 
\end{equation}
On the original real field contour, $b_{-k}=b_k^*$ and hence $t=|b_k|^2/(2\pi)\geq0$, while the quadratic action becomes 
\begin{equation}\label{eq:PI_mode_action} 
S_{m,k}^{(2)}=t\rho(k).
\end{equation}
The same change of variables can now be applied directly to the measure in Eq.~\eqref{eq:PI_one_mode_A0}. After integrating over the angular direction of $k$,
\begin{equation}
\frac{2}{m^2}\frac{d^2k}{(2\pi)^2}=\frac{dy}{2\pi}=\frac{2\sinh\rho}{2\pi}\,d\rho,
\end{equation}
while after integrating over the phase of the complex mode,
\begin{equation}
\frac{d^2\alpha_k}{\pi}=d|\alpha_k|^2=\frac{2\pi}{\xi_k}\,dt.
\end{equation}
Their product is
\begin{equation}
\frac{2\sinh\rho}{\xi_k}\,d\rho\,dt=2\sinh\rho\,\tanh\frac{\rho}{2}\,d\rho\,dt=\left(e^\rho-2+e^{-\rho}\right)d\rho\,dt.
\label{eq:PI_full_jacobian}
\end{equation}

There is a reparametrization freedom in factorizing the quadratic kernel into a momentum coordinate and a radial coordinate. Here $\rho(k)$ is not chosen by this factorization: it is the spectral coordinate $\rho=\ln z$ already introduced in Section~\ref{sec:theta}. Once this $\rho(k)$ is held fixed, the normalization of $t$ above is fixed by requiring $S_{m,k}^{(2)}=t\rho(k)$. With this choice, 
\begin{equation}\label{eq:PI_exact_mellin} 
\frac{\Delta_\alpha(k^2)}{2\pi}=\frac{1}{\xi_k}\int_0^\infty dt\,e^{-t\rho(k)} = \frac{1}{\xi_k\rho(k)}. 
\end{equation}
Thus $t$ is precisely the Mellin parameter introduced earlier; when the hard-momentum endpoint is isolated, it becomes the Borel coordinate of the perturbative sector.

After complexification, $\alpha_k$ and $\alpha_{-k}$ become independent and $t=\xi_k\alpha_k\alpha_{-k}/(2\pi)$ extends holomorphically away from the original real field contour. The complex contours considered in Sections~\ref{sec:thimble} and~\ref{sec:OPE} are therefore cycles of this one-mode integral, which is obtained exactly from the large-$N$ Gaussian functional integral at order $1/N$ after the spectator modes are canceled. We do not claim that these cycles lift to Lefschetz thimbles of the full interacting auxiliary-field functional integral; in particular, a critical point of the reduced action need not satisfy the full saddle equation $\delta S_{\rm eff}/\delta\sigma=0$.

\subsection{Reduced action and the Borel coordinate}

Substituting Eq.~\eqref{eq:PI_exact_mellin} into the reduced self-energy reproduces the $(t,\rho)$ representation derived in Section~\ref{sec:theta},
\begin{equation}
\label{eq:PI_exact_trho}
2xA_0(x) = \int_0^\infty dt\int_0^{\ln Z}d\rho\,e^{-t\rho}\left(e^\rho-2+e^{-\rho}\right).
\end{equation}
The new point here is its direct path-integral origin: $e^{-t\rho}$ is the Gaussian weight of the active auxiliary-field mode, while $e^\rho-2+e^{-\rho}$ is the combined Jacobian from the radial momentum measure and the change from the ordinary mode amplitude to the rescaled radial coordinate $t$. The renormalon-sensitive first term is consequently governed by the reduced action
\begin{equation}
S(t,\rho)=(t-1)\rho.
\end{equation}
As shown in Sections~\ref{sec:theta}--\ref{sec:thimble}, isolating the hard-scale endpoint identifies $t$ with the Borel coordinate and gives the critical point $(t,\rho) = (1,0)$. We now use the mode-space interpretation above to determine what this critical point represents in the original auxiliary-field variables.

The reduced action has the critical-point equations
\begin{equation}
\partial_t S = \rho = 0, \qquad \partial_\rho S = t - 1 = 0,
\end{equation}
so its critical point is $(t,\rho) = (1,0)$. This point should be distinguished from the hard corner $(t,\rho) = (0,\ln{Z})$ that generates the perturbative expansion. At the hard corner, the exchanged momentum is of order the external momentum and the massive-vacuum Hessian reduces to its ultraviolet form. The boundary saddle at $(1,0)$ instead controls the Stokes discontinuity of the perturbative expansion generated at the hard corner.

The relation between the reduced boundary $\rho = 0$ and the original auxiliary-field mode follows from
\begin{equation}
k = m\rho + \cO(\rho^3), \qquad \xi_k = \frac{2}{\rho} + \cO(\rho).
\end{equation}
At fixed $t$, the definition of the radial coordinate gives
\begin{equation}\label{eq:PI_boundary_mode_scaling}
\alpha_k \alpha_{-k} = \frac{2\pi t}{\xi_k} = \pi t\rho + \cO(\rho^3).
\end{equation}
Equivalently, since $k=m\rho+\cO(\rho^3)$, at the critical value $t=1$ one has
\begin{equation}
|\alpha_k|^2=\frac{\pi k}{m}+\cO(k^3/m^3).
\end{equation}
Thus the reduced saddle is the approach to the corner $(k,\alpha_k)=(0,0)$ of the one-mode integral along a definite scaling trajectory. Both the exchanged momentum and the amplitude of the original auxiliary-field mode vanish as the reduced saddle is approached, namely $k \to 0$ and $\alpha_k \alpha_{-k} \to 0$, while the weighted radius remains fixed, $\frac{\xi_k \alpha_k \alpha_{-k}}{2\pi} = t$. Equivalently, the rescaled mode $b_k = \sqrt{\xi_k}\alpha_k$ retains the finite radius $b_k b_{-k} = 2\pi t$, although the transformation from $\alpha_k$ to $b_k$ becomes singular as $k \to 0$. The point $(t,\rho) = (1,0)$ therefore represents a boundary limit of the original auxiliary-field mode space in which momentum and ordinary mode amplitude vanish together. Its finite radial coordinate refers to the rescaled variable, not to a finite-amplitude configuration of the original $\alpha$ field.

The exact reduced integrand is nevertheless regular at this boundary:
\begin{equation}
\frac{dy}{\xi_y\rho(y)} = \frac{y(\rho)}{\rho}\,d\rho = \left(\rho + \cO(\rho^3)\right)d\rho.
\end{equation}
The critical point governs a Stokes discontinuity only after the exact Jacobian $e^\rho - 2 + e^{-\rho}$ has been separated into the perturbative and power-suppressed contributions in Eq.~\eqref{eq:PI_exact_trho}. Before this decomposition, the complete real integral is nonsingular at $\rho = 0$.

It is useful to compare $t$ with the ordinary, unweighted radius of the same mode pair,
\begin{equation}
t_0 = \frac{\alpha_k \alpha_{-k}}{2\pi}.
\end{equation}
The two are related by
\begin{equation}
t = \xi_k t_0.
\end{equation}
At large momentum, $\xi_k \to 1$ and
\begin{equation}
\rho(k) = \ln\frac{k^2}{m^2} + \cO\!\left(\frac{m^2}{k^2}\right),
\end{equation}
so $t$ approaches the ordinary mode radius and $\rho(k)$ approaches the logarithmic momentum variable of the perturbative running-coupling description. The infrared behavior is different: keeping $t$ fixed while $k \to 0$ requires $t_0 \to 0$. The renormalon saddle therefore has finite $t$ even though the ordinary mode radius $t_0$ vanishes.

The same distinction separates the renormalon saddle from the apparent Landau pole. The perturbative inverse propagator
\begin{equation}
\Gamma_0^{(\sigma)}(k^2) = \frac{N}{2\pi}\ln\frac{k^2}{m^2}
\end{equation}
vanishes at $k^2 = m^2$, whereas the massive-vacuum inverse propagator
\begin{equation}
\Gamma_m^{(\delta\sigma)}(k^2) = \frac{N}{2\pi}\xi_k\rho(k)
\end{equation}
is positive for real Euclidean momentum and remains finite as $k \to 0$. The reduced contour geometry is therefore organized around the regular endpoint $\rho = 0$, corresponding to $k = 0$, rather than around the interior zero produced by the ultraviolet approximation.

The critical value $t = 1$ also carries the scaling required by the compensating OPE operator. Restoring the overall factor $1/(2x)$ in $A_0$, the upper-endpoint contribution is proportional to
\begin{equation}
\frac{Z}{2x}e^{-2t/\lambda_z},
\end{equation}
where $Z/x = 1 + \cO(x^{-1})$. The nonperturbative weight is therefore
\begin{equation}
e^{-2t/\lambda_z} = Z^{-t}
\end{equation}
up to a prefactor regular in the large-$x$ expansion. At the renormalon saddle $t = 1$,
\begin{equation}
Z^{-1} = \frac{m^2}{p_z^2} = \frac{m^2}{p^2} + \cO\!\left(\frac{m^4}{p^4}\right),
\end{equation}
which is precisely the power suppression associated with the dimension-two operator $V$.

The perturbative background, the massive saddle, and the renormalon saddle therefore play distinct roles. Expansion about $\sigma = 0$ defines the perturbative Wilson coefficient and generates its asymptotic series. The spectral function in the quadratic fluctuations about $\sigma = m$ identifies the previously introduced momentum coordinate $\rho$; once this coordinate is fixed, the quadratic form gives the corresponding rescaled radial variable $t$. The critical point $(t,\rho) = (1,0)$ controls the Stokes jump of the perturbative boundary cycle, but is a boundary critical point of the reduced mode integral rather than an additional saddle of the full auxiliary-field effective action.

\section{Comparison with the \texorpdfstring{$O(N)$}{O(N)} sigma model}
\label{sec:ON}

The leading renormalon cancellation in the two-dimensional large-$N$ $O(N)$ nonlinear sigma model is governed locally by the same reduced action $(t - 1)\rho$ that controls the Gross--Neveu cancellation described above, although the global kernels and contour geometries differ. The $O(N)$ model provided one of the earliest exactly solvable settings in which renormalon cancellation could be analyzed within the OPE~\cite{David:1982qv,David1984,David1986,Novikov1985,BenekeBraunKivel1998,Beneke:1998ui}. We focus on the feature directly parallel to Gross--Neveu: the leading (positive-axis) renormalon arises in an auxiliary-field exchange contribution, and its ambiguity is canceled by the condensate of a dimension-two operator. Appendix~\ref{app:ON_details} contains additional details.

The model can be formulated in terms of an $N$-component field $n^a$ and a Lagrange multiplier $\alpha$ enforcing the nonlinear constraint:
\begin{equation}
\label{eq:ON_action_alpha}
S = \frac{1}{2}\int d^2x\,\left[(\partial_\mu n^a)^2 + \frac{\alpha}{\sqrt N}\left(n^a n^a - \frac{N}{g}\right)\right].
\end{equation}
At large $N$, integrating out the $n^a$ fields yields a saddle with
\begin{equation}
\label{eq:ON_stationary}
\langle m|\alpha|m\rangle = \sqrt N\,m^2, \qquad m^2 = \mu^2e^{-4\pi/g(\mu)}.
\end{equation}
Equivalently, defining $\lambda(\mu) = g(\mu)/(4\pi)$ gives
\begin{equation}
\lambda(\mu) = \frac{1}{\ln(\mu^2/m^2)}.
\end{equation}

The auxiliary-field contour differs from its Gross--Neveu counterpart. In Gross--Neveu, the auxiliary field is introduced by a Hubbard--Stratonovich transformation, whereas $\alpha$ acts as a Lagrange multiplier imposing the nonlinear constraint. Its microscopic contour runs along an imaginary direction and is deformed through the real large-$N$ saddle, with quadratic fluctuations integrated along the corresponding steepest-descent contour. This contour in the complexified $\alpha$-field space is distinct from the reduced renormalon cycles discussed in Sections~\ref{sec:thimble} and~\ref{sec:OPE}, which instead live in the complexified $(t,\rho)$ space.

Expanding about the saddle as $\alpha = \sqrt N\,m^2 + \widehat\alpha$, the propagator of the fluctuation $\widehat\alpha$ takes the closed form~\cite{Novikov1985,BenekeBraunKivel1998}
\begin{equation}
\label{eq:ON_alpha_prop}
\overline{\Delta}_\alpha(p^2) = 4\pi\,\frac{\sqrt{p^2(p^2 + 4m^2)}}{\ln\!\dfrac{\sqrt{p^2 + 4m^2} + \sqrt{p^2}}{\sqrt{p^2 + 4m^2} - \sqrt{p^2}}}.
\end{equation}
Introducing the same spectral variable used in the Gross--Neveu analysis,
\begin{equation}
\xi = \sqrt{1 + \frac{4m^2}{p^2}},
\end{equation}
the above becomes
\begin{equation}
\overline{\Delta}_\alpha(p^2) = 4\pi p^2\xi\left[\ln\frac{\xi + 1}{\xi - 1}\right]^{-1}.
\end{equation}
The $O(N)$ propagator therefore contains the same logarithmic spectral function as its Gross--Neveu counterpart, multiplied by a different momentum-dependent prefactor, reflecting the different spin content of the fields that have been integrated out.

The quantity analogous to the Gross--Neveu self-energy is the order-$1/N$ correction generated by a single $\widehat\alpha$ exchange,
\begin{equation}
\label{eq:ON_Sigma_one_alpha}
\overline\Sigma(p^2) = \frac{1}{N}\int\frac{d^2k}{(2\pi)^2}\, \frac{\overline{\Delta}_\alpha(k^2)}{(p+k)^2 + m^2}.
\end{equation}
Its large-momentum expansion was derived in Ref.~\cite{BenekeBraunKivel1998}. Defining $\lambda(p) = \frac{1}{\ln(p^2/m^2)}$, the leading perturbative sector can be written as
\begin{equation}
\label{eq:ON_self_energy_borel}
\overline\Sigma^{(0)}(p^2) = -\frac{p^2}{N}\int_0^\infty dt\, e^{-t/\lambda(p)}\left[\psi(2-t) + \psi(t-1) + 2\gamma_E - \frac{1}{\lambda(p)}\right],
\end{equation}
where $\psi(z) = \Gamma'(z)/\Gamma(z)$. The integrand has positive-axis poles at $t = 1,2,\ldots$. Near the leading pole,
\begin{equation}
\psi(t-1) = -\frac{1}{t-1} - \gamma_E + \cdots,
\end{equation}
so the singular part of Eq.~\eqref{eq:ON_self_energy_borel} is
\begin{equation}
\overline\Sigma^{(0)}(p^2)\big|_{t \simeq 1} = \frac{p^2}{N}\int_0^\infty dt\, \frac{e^{-t/\lambda(p)}}{t-1}.
\end{equation}
Let $\mathcal C_{\rm up}$ and $\mathcal C_{\rm down}$ denote contours passing above and below the pole, respectively. Since
\begin{equation}
e^{-1/\lambda(p)} = \frac{m^2}{p^2},
\end{equation}
the corresponding imaginary parts are
\begin{equation}
\operatorname{Im}\overline\Sigma^{(0)}_{\rm up/down}(p^2) = \mp\frac{\pi m^2}{N},
\end{equation}
with the upper and lower signs corresponding to $\mathcal C_{\rm up}$ and $\mathcal C_{\rm down}$. The leading ambiguity is therefore of order $m^2/N$ in the self-energy and, after expanding the full propagator, of order $m^2/(Np^4)$ in the propagator.

The dimension-two operator $\alpha$ supplies the OPE contribution associated with the leading pole. Expanding the fundamental propagator (i.e.~the propagator for $n^a$) in a slowly varying $\alpha$ background gives
\begin{equation}
\frac{1}{p^2 + \alpha/\sqrt N} = \frac{1}{p^2} - \frac{\alpha}{\sqrt N\,p^4} + \cdots,
\end{equation}
so its Wilson coefficient is
\begin{equation}
\label{eq:ON_Calpha}
C_\alpha(p^2) = -\frac{1}{\sqrt N\,p^4}.
\end{equation}
With the contour convention introduced above, the renormalized condensate has the lateral imaginary part
\begin{equation}
\label{eq:Im_alpha_vev_ON}
\operatorname{Im}\langle m|\alpha|m\rangle_{\rm up/down} = \pm\frac{\pi m^2}{\sqrt N}.
\end{equation}
The corresponding condensate is
\begin{equation}
\langle m|\alpha|m\rangle_{\rm up/down} = \sqrt N\,m^2\left(1 \pm \frac{i\pi}{N} + \cO(N^{-2})\right),
\end{equation}
in agreement with the standard large-$N$ result in dimensional regularization after translating between conventions for the two lateral contours~\cite{David1984,David1986,BenekeBraunKivel1998}. Multiplying by the Wilson coefficient gives
\begin{equation}
C_\alpha(p^2)\operatorname{Im}\langle m|\alpha|m\rangle_{\rm up/down} = \mp\frac{\pi m^2}{Np^4}.
\end{equation}
The ambiguity of the identity coefficient follows from expanding the full propagator:
\begin{equation}
\frac{1}{p^2 + m^2 + \overline\Sigma(p^2)} = \frac{1}{p^2 + m^2} - \frac{\overline\Sigma(p^2)}{(p^2 + m^2)^2} + \cdots.
\end{equation}
Using the imaginary part of the self-energy obtained above yields
\begin{equation}
\operatorname{Im}C_{\mathbf 1,{\rm up/down}}(p^2) = \pm\frac{\pi m^2}{Np^4} + \cO\!\left(\frac{m^4}{Np^6}\right).
\end{equation}
The identity coefficient and the $\alpha$ condensate therefore carry equal and opposite lateral ambiguities.

Appendix~\ref{app:ON_details} shows that the part of the $\alpha$ condensate carrying the leading positive-axis singularity can be written using the same logarithmic spectral coordinate
\begin{equation}
\rho_k = \ln\frac{\xi_k + 1}{\xi_k - 1}
\end{equation}
and a Borel coordinate normalized so that the singularity lies at $t = 1$. In a neighborhood of this singularity, the perturbative contribution and the condensate can both be represented by the reduced action $(t - 1)\rho$, with opposite lateral orientations. Their global contour geometries remain different from the Gross--Neveu construction: the $O(N)$ angular integral does not reduce to the two-rectangle strip of Section~\ref{sec:theta}, and the perturbative identity coefficient contains an infinite sequence of positive-axis poles. The comparison therefore identifies the common local saddle responsible for the leading cancellation without equating the complete reduced kernels or their full integration cycles.

\section{Discussion}
\label{sec:discussion}

We have thus shown that the leading IR renormalon in the fermion self-energy of the Gross--Neveu theory can be identified with the boundary critical point $(t,\rho) = (1,0)$ of the reduced action $S(t,\rho) = (t-1)\rho$, distinct from the hard-momentum corner $(0,\ln Z)$ that generates the perturbative expansion. This renormalon saddle is a critical point of the reduced contour problem rather than an additional saddle of the original auxiliary-field effective action.

The contour identity contains more information than the equality of the perturbative and condensate residues. The two relative cycles share the same lateral $t$ contour and extend toward complex infinity in the same asymptotic direction, but they are anchored at opposite ends of the finite $\rho$ interval and carry opposite orientations along their complex tails. When the two OPE contributions are added, these tails cancel and the real strip $0 \leq \rho \leq \ln{Z}$ is recovered. Performing the $\rho$ integral expresses the same cancellation algebraically through the regular combination $(Z^{1-t} - 1)/(1-t)$. The vanishing of the total Stokes discontinuity is therefore a consequence of recombining the two prescription-dependent relative cycles into the original real contour.  The local half-thimble fixes the Stokes discontinuity, and real finite terms depend on the global completion of the contour away from the renormalon saddle.

The same construction also shows that the renormalon saddle is more robust than the apparent Landau pole: coupling redefinitions can remove the latter and reorganize the non-perturbative power corrections without changing the $t = 1$ Borel singularity or its Stokes discontinuity.

Several extensions of the present construction would be interesting. By restricting ourselves to the leading renormalon using the $\theta$ approximation, we have not addressed the geometry of the subleading renormalons and their corresponding OPE contributions. Extending the construction to the exact self-energy would require determining how the additional Mellin--Barnes poles and higher-dimensional operator sectors are represented by relative cycles, and whether their Stokes discontinuities are controlled by additional boundary critical points of a more general reduced contour problem.

Finite-$N$ corrections pose a complementary difficulty. They introduce interactions among the auxiliary-field modes, so the modewise Gaussian radialization that produces the coordinate $t$ may no longer apply directly. Determining whether a corresponding reduced saddle persists once these mode interactions are included would distinguish features tied specifically to the leading large-$N$ Gaussian theory from those that survive in the full auxiliary-field path integral.

A useful test of the generality of this local mechanism is provided by the large-$N$ $O(N)$ nonlinear sigma model. We find that in the vicinity of the leading IR renormalon in the scalar propagator of the theory, the perturbative contribution and the dimension-two condensate are again governed locally by the reduced action $(t - 1)\rho$, with opposite lateral orientations. The global contour problems are nevertheless different: the perturbative scalar propagator does not reduce to the rectangular-strip representation of the Gross--Neveu reduced self-energy, its perturbative Wilson coefficient contains an infinite sequence of IR renormalons, and the microscopic Lagrange-multiplier contour defines a separate steepest-descent problem. The comparison therefore points to a common local structure governing the cancellation of the leading Borel singularity by a dimension-two condensate, while leaving the global organization of the relevant cycles dependent on the theory.

Gauge theories such as QCD provide a natural setting in which to ask whether this contour picture extends more generally. The OPE again organizes renormalon ambiguities in terms of gauge-invariant local operators, suggesting that an analogous geometric cancellation may exist. Establishing such a description, however, is likely to be more difficult because comparably explicit non-perturbative expressions are rarely available. We emphasize that our construction is purely for renormalons in observables that admit an OPE; a significant generalization will be needed for addressing renormalons in Lagrangian parameters such as heavy quark pole mass, which we leave to future work.

Taken together, these results identify the leading IR renormalon in the fermion self-energy of Gross--Neveu theory with a boundary critical point of the reduced mode-space action. The perturbative Wilson coefficient and the renormalized four-fermion condensate appearing in its OPE define complementary relative cycles associated with this critical point, and their separate imaginary parts are the Stokes discontinuities of those cycles. Recombining the two contributions reconstructs the finite real strip of the reduced leading renormalon integral, causing the corresponding Stokes discontinuities to cancel.

\section*{Acknowledgements}

We acknowledge the use of Claude Opus, Claude Fable, GPT 5.5, and GPT 6 to assist with analytical derivations and to suggest edits to the manuscript. All AI-assisted derivations were independently checked by the authors, and all results and arguments were written up and verified by the authors. JC is supported by a Fellowship from the Alfred P.~Sloan Foundation. This work was supported in part by the U.S. Department of Energy, Office of Science, under grant DE-SC0013607. This work was performed in part at the Erwin-Schrödinger International Institute for Mathematics and Physics at the University of Vienna during the program ``New Paradigms for Harnessing Quantum Field Theory at Colliders'' (2026); AB thanks the centre for their hospitality during the completion of the work.


\appendix

\section{Details of the OPE calculation}
\label{app:OPE_details}

Here we compute the short-distance expansion of the fermion two-point function $\psibar_\alpha(0)\psi_\beta(x)$. We work throughout with the ordering $\psibar_\alpha(0)\psi_\beta(x)$; with this convention the free contraction is denoted $D_{\beta\alpha}(x,0)$. The general form of the OPE is
\begin{equation}\label{eq:ope_app_general}
\psibar_\alpha(0)\psi_\beta(x) = C_{\opone}(x)_{\alpha\beta}\opone + \sum_A C_A(x)_{\alpha\beta}\,\cO_A(0) + \cdots\,,
\end{equation}
where the sum runs over all local $U(N)$-singlet operators, organized in order of increasing mass dimension. To ease notation we have suppressed flavor indices in Eq.~\eqref{eq:ope_app_general}, where on the LHS we consider the two point $\psibar_\alpha^i(0)\psi_\beta^j(x)$ with $i, j$ as the flavor indices that are not summed over. On the RHS of the same equation we will have overall factors of $\delta^{ij}$ entering as part of the Wilson coefficients. We only consider flavor singlets as these operators are the only ones with non-vanishing vacuum expectation values. The momentum-space Wilson coefficients are defined by the Fourier transform
\begin{equation}
\label{E:Fourierconv}
\widetilde C_n(p) = \int d^dx\, e^{ip\cdot x} C_n(x).
\end{equation}
Note that the OPE is defined in position space and the operators in the OPE are evaluated at $x=0$ rather than at $p=0$.

Since $\psibar_\alpha\psi_\beta$ carries spinor indices, the Wilson coefficients are $2\times2$ matrices in spinor space, and the OPE must be decomposed in a complete basis of such matrices. In two dimensions, the Clifford algebra is spanned by three independent structures
\begin{equation}
\Gamma_A \in \{\opone,\gamma^\mu,\gamma_5\},
\end{equation}
where $\gamma_5 \equiv -i\gamma^0\gamma^1$ satisfies $\gamma_5^2 = 1$ and $\{\gamma_5,\gamma^\mu\}=0$. A matrix representation is given by the Pauli matrices $\gamma^0=\sigma_1$, $\gamma^1=\sigma_2$, $\gamma_5=\sigma_3$. Any $2\times2$ matrix can be expanded in this basis. The completeness of the basis implies the Fierz identity
\begin{equation}\label{eq:app_fierz}
\delta_{\beta\gamma}\delta_{\delta\alpha} = \frac{1}{2}\left[\delta_{\beta\alpha}\,\delta_{\delta\gamma} + (\gamma^\mu)_{\beta\alpha}\,(\gamma_\mu)_{\delta\gamma} + (\gamma_5)_{\beta\alpha}\,(\gamma_5)_{\delta\gamma}\right],
\end{equation}
which decomposes a product of Kronecker deltas in spinor space into the three Dirac channels. Equivalently,
\begin{equation}
\psibar_\alpha\psi_\beta = \frac{1}{2}\delta_{\beta\alpha}\,\psibar\psi + \frac{1}{2}(\gamma^\mu)_{\beta\alpha}\,\psibar\gamma_\mu\psi + \frac{1}{2}(\gamma_5)_{\beta\alpha}\,\psibar\gamma_5\psi .
\end{equation}
To decompose the flavor structure it will also be helpful to use the Fierz identity
\begin{equation}\label{eq:app_fierz_UN}
\delta^{ib}\delta^{ja} = \frac{1}{N}\delta^{ij}\delta^{ab} + 2\sum_A (T^A)^i{}_j (T^A)^a{}_b ,
\end{equation}
where the sum is taken over the $N^2-1$ generators in the fundamental representation, normalized as $\tr(T^A T^B)=\delta^{AB}/2$. This allows us to isolate the trace and traceless components in the product of flavor indices.

Since $\psi$ has mass dimension $1/2$, we can enumerate the operators in increasing numbers of fields and derivatives. Up to dimension two, a basis of $U(N)$-singlet operators is
\begin{equation}
\cO_0 = \opone,\quad \cO_1^A = \psibar\Gamma_A\psi,\quad \cO_{2,\nu}^{A} = \psibar\Gamma_A\partial_\nu\psi,\quad \cO_2^{AB} = (\psibar\Gamma_A\psi)(\psibar\Gamma_B\psi),
\end{equation}
where $A,B$ run over $\opone,\gamma^\mu,\gamma_5$.

We work in $d = 2-2\varepsilon$ dimensions throughout, continuing loop momenta while keeping the physical two-dimensional spinor algebra. We use the auxiliary-field Lagrangian in Eq.~\eqref{eq:GN_sigma},
\begin{equation}
\cL_\sigma = \psibar\slashed{\partial}\psi + \frac{1}{2g}\sigma^2 + \sigma\psibar\psi.
\end{equation}
We recall that the equations of motion are
\begin{equation}
\slashed{\partial}\psi + \sigma\psi = 0,\quad \sigma = -g\psibar\psi.
\end{equation}
In particular, the $\sigma$ equation of motion can be used to express any operator involving $\sigma$ as an operator built from $\psibar\psi$, so that we do not need to enumerate operators with $\sigma$ separately.

The OPE is an operator identity, valid around any background and inside any correlation function. We consider two backgrounds: the perturbative vacuum $|0\rangle$ and the non-perturbative vacuum $|m\rangle$ where the fermion acquires a dynamical mass $m$. The Wilson coefficients $C_n(x)$ are c-number functions that encode short-distance physics. They depend only on the separation $x$ and the coupling, and carry no information about the vacuum. The OPE separates short-distance physics, encoded in Wilson coefficients, from long-distance physics, encoded in condensates such as $\langle m|\cO_n|m\rangle$.

Composite operators with multiple fields at the same spacetime point require a prescription to be well-defined. In the free theory, normal ordering suffices. There we can use
\begin{align}
\psibar_\alpha(0)\psi_\beta(x)=\,:\psibar_\alpha(0)\psi_\beta(x)\!:+ D_{\beta\alpha}(x,0)
\end{align}
and then Taylor expand the normal-ordered product around $x=0$. Using the Fierz identity in Eq.~\eqref{eq:app_fierz}, retaining only $U(N)$-singlet composite operators via Eq.~\eqref{eq:app_fierz_UN}, and making all indices explicit for clarity, we get
\begin{align}
\psibar_\alpha^i(0)\psi_\beta^j(x) &= D_{\beta\alpha}^{ij}(x,0)+\frac{1}{2N}\delta^{ij}\delta_{\beta\alpha}\,\psibar\psi(0)+\frac{1}{2N}\delta^{ij}(\gamma^\mu)_{\beta\alpha}\,\psibar\gamma_\mu\psi(0) \nonumber\\
&\quad +\frac{1}{2N}\delta^{ij}(\gamma_5)_{\beta\alpha}\,\psibar\gamma_5\psi(0)+\frac{1}{N}\delta^{ij}x^\mu\,\psibar_\alpha\partial_\mu\psi_\beta(0)+\cdots . \label{eq:tay_ope}
\end{align}
Thus $C_0(x)=D(x,0)$, $C_1^A=\delta^{ij}\Gamma^A/(2N)$, and so on in the free theory. The propagator is only nonzero when the flavors match, so $D_{\beta\alpha}^{ij}(x,0)=\delta^{ij}D_{\beta\alpha}(x,0)$. In the interacting theory, normal ordering is insufficient and the operator definition is tied to the renormalization scheme.

Only operators whose Wilson coefficients contain a $\slashed x$ structure can contribute to the form factor $\mathcal A(x^2)$ defined by the $\slashed x$ projection of the two-point function. Dimension-one scalar and pseudoscalar bilinears have coefficients proportional to $\opone$ and $\gamma_5$, so they do not contribute to $\mathcal A(x^2)$. The vector bilinear can carry a vector Dirac structure, but its vacuum expectation value vanishes by Lorentz invariance. Thus no dimension-one condensate contributes to $A(p^2)$. At dimension two, scalar operators can appear with matrix coefficients proportional to $\slashed x\ln(x^2\mu^2)$; after extracting the scalar form factor $\mathcal A(x^2)$ this gives logarithmic scalar Wilson coefficients.

For later use it is helpful to single out the scalar operators embedded in the full OPE. These are the operators that can take vacuum expectation values without violating Lorentz invariance. We define
\begin{equation}
\cO_1 \equiv \psibar\psi,\quad K \equiv i\psibar\slashed{\partial}\psi,\quad \cO_{4f}\equiv(\psibar\psi)^2,\quad V\equiv g\,\cO_{4f}\,.
\end{equation}
When we later project onto the scalar channel relevant for vacuum matrix elements, it is convenient to write
\begin{equation}
\psibar(0)\psi(x) = C_0(x)\opone + C_1(x)\,\cO_1(0) + C_2^K(x)\,K(0) + C_2^V(x)\,V(0) + \cdots .
\end{equation}
Here the Wilson coefficients are scalar form-factor coefficients: the Dirac structure and the implicit flavor indices have already been projected. This scalar-channel normalization differs from the full open-index Clifford-basis coefficients above. In the six-point matching below it is convenient to keep the indices open temporarily; the coefficient of the unweighted four-fermion operator $\cO_{4f}$ is then converted to the coefficient of $V=g\cO_{4f}$ before being inserted into the scalar-channel OPE. The parity-odd operator $\psibar\gamma_5\psi$ does not contribute to the scalar-channel OPE, while the vector operator $\psibar\gamma^\mu\psi$ and the derivative operators are already included in the full OPE above. The free-theory matching gives $C_0(x)=D_F(x,0)$, $C_1^{(0)}=1$, $C_2^{\mu(0)}=x^\mu$, and $C_2^{K(0)}=0$.

\subsection{Wilson coefficients}

Since the OPE is an operator identity, it holds inside any correlation function. The Wilson coefficients are determined by matching: we compute correlation functions of both sides with progressively more external fields, and identify coefficients order by order. The Wilson coefficients are perturbative; the condensate expectation values are determined independently.

\subsubsection{Two-point functions}

In dimensional regularization, the perturbative vacuum expectation value of any renormalized local operator of positive dimension vanishes order by order, because the corresponding vacuum graphs are scaleless. Therefore $\langle0|\cO_n|0\rangle=0$ for all nontrivial operators in the OPE. Taking the perturbative vacuum expectation value of Eq.~\eqref{eq:ope_app_general} isolates the Wilson coefficient of the identity operator:
\begin{equation}\label{eq:app_C0_matching}
C_0(x)_{\alpha\beta}^{ij} = \langle0|\psibar_\alpha^i(0)\psi_\beta^j(x)|0\rangle = D_{\beta\alpha}^{ij}(x,0).
\end{equation}
At tree level,
\begin{equation}
D_{\beta\alpha}^{ij}(x,0) = -\delta^{ij}\frac{\Gamma(d/2)}{2\pi^{d/2}}\frac{(\slashed x)_{\beta\alpha}}{(x^2)^{d/2}}.
\end{equation}
With the Fourier convention in~\eqref{E:Fourierconv}, we find
\begin{equation}
\widetilde D_{\beta\alpha}^{ij}(p) = -i\,\delta^{ij}\frac{(\slashed p)_{\beta\alpha}}{p^2},
\end{equation}
in agreement with the momentum-space identity coefficient below. In the perturbative massless expansion, where $\Sigma(p)=-i\slashed p\,A(p^2)/N$, the bubble-chain resummed identity coefficient may be written as
\begin{equation}
\widetilde C_0(p)_{\alpha\beta}^{ij} = \delta^{ij}\frac{-i(\slashed p)_{\beta\alpha}}{p^2+i\slashed p\,\Sigma(p)}.
\end{equation}
Since $C_0$ equals the full perturbative propagator, it encodes the bubble-chain resummation and is the source of the leading renormalon divergence.

\subsubsection{Four-point functions}

At dimension one all the operators have two fermion fields, $\cO_1^A=\psibar\Gamma_A\psi$. To extract their Wilson coefficients, we consider the mixed position/momentum space amputated four-point function
\begin{multline}
\label{eq:C1_matching}
\langle0|\psibar_\alpha^i(0)\psi_\beta^j(x)\;\psibartilde{k}{\gamma}(p_a)\widetilde{\psi}_\delta^l(p_b)|0\rangle_{\rm amp}
= C_0(x)_{\alpha\beta}^{ij}\,\langle0|\psibartilde{k}{\gamma}(p_a)\widetilde{\psi}_\delta^l(p_b)|0\rangle_{\rm amp} \\
+\sum_A C_1^A(x)_{\alpha\beta}^{ij}\,\langle0|\cO_1^A(0)\;\psibartilde{k}{\gamma}(p_a)\widetilde{\psi}_\delta^l(p_b)|0\rangle_{\rm amp}+\cdots .
\end{multline}
Here the fields $\psibartilde{k}{\gamma}(p_a)$ and $\widetilde\psi_\delta^l(p_b)$ are in momentum space while the fields we are bringing together are in position space, and all external legs are amputated. The OPE is an expansion in the soft quantities $x\cdot p_a$ and $x\cdot p_b$. The ellipsis includes higher-dimension operators with two fermion fields such as $\cO_{2,\mu}^A$.

The first term on the right-hand side is the contribution of the identity operator. The second term contains the dimension-one bilinears. Operators with four fermion fields, such as $\cO_{4f}$, cannot contribute to this matrix element and must be extracted separately from a six-point function.

\subsubsection*{Order $g^0$}

In the free theory, the amputated four-point function with open spinor and flavor indices is
\begin{align}
\label{eq:freeleft_indexed}
\langle0|\psibar_\alpha^i(0)\psi_\beta^j(x)\psibartilde{k}{\gamma}(p_a)\widetilde{\psi}_\delta^l(p_b)|0\rangle_{\rm amp} &= \vcenter{\hbox{\freeleftB}}+\vcenter{\hbox{\freeleftA}} \nonumber\\
&= D_{\beta\alpha}^{ij}(x,0)(2\pi)^d\delta^{(d)}(p_a+p_b)\delta_{\gamma\delta}\delta^{kl}+e^{ip_a\cdot x}\delta_{\beta\gamma}\delta_{\delta\alpha}\delta^{il}\delta^{jk}.
\end{align}
On the OPE side, the amputated matrix elements of the local operators are
\begin{align}
\label{eq:app_mat_el_op_tree}
\langle0|\opone\;\psibartilde{k}{\gamma}(p_a)\widetilde{\psi}_\delta^l(p_b)|0\rangle_{\rm amp}^{(0)} &= (2\pi)^d\delta^{(d)}(p_a+p_b)\delta_{\gamma\delta}\delta^{kl},\\
\langle 0|[\psibar\Gamma_A\psi](0)\;\psibartilde{k}{\gamma}(p_a)\widetilde{\psi}_\delta^l(p_b)|0\rangle_{\rm amp}^{(0)}
  &=\; (\Gamma_A)_{\delta\gamma} \delta^{k l}\,\\
\langle 0|[\psibar\Gamma_A\partial_\mu\psi](0)\;\psibartilde{k}{\gamma}(p_a)\widetilde{\psi}_\delta^l(p_b)|0\rangle_{\rm amp}^{(0)}
  &=\; i(p_{a
  })_{\mu}\,(\Gamma_A)_{\delta\gamma} \delta^{k l}\,.
\end{align}
The first diagram in Eq.~\eqref{eq:freeleft_indexed} is reproduced by $C_0\cdot\opone$. For the second diagram, we use the spin and flavor Fierz identities and expand $e^{ip_a\cdot x}=1+ip_{a\mu}x^\mu+\cO(x^2)$:
\begin{equation}
\label{eq:free_fierz_expand}
e^{ip_a\cdot x}\delta_{\beta\gamma}\delta_{\delta\alpha}\delta^{il}\delta^{jk} = \frac{1}{2N}\delta^{ij}\delta^{kl}\sum_{A\in\{\opone,\gamma^\mu,\gamma_5\}}(\Gamma_A)_{\beta\alpha}(\Gamma_A)_{\delta\gamma}\left(1+ip_{a\mu}x^\mu+\cdots\right).
\end{equation}
Here we have discarded higher orders in $x^2$ and traceless flavor contributions. Thus
\begin{equation}
\label{eq:free_wilson}
C_0(x)_{\alpha\beta}^{ij}=D_{\beta\alpha}^{ij}(x,0),\quad C_1^A(x)_{\alpha\beta}^{ij}=\frac{1}{2N}\delta^{ij}(\Gamma_A)_{\beta\alpha},\quad C_2^{A,\mu}(x)_{\alpha\beta}^{ij}=\frac{1}{2N}\delta^{ij}x^\mu(\Gamma_A)_{\beta\alpha}.
\end{equation}
These results agree with the Taylor expansion in Eq.~\eqref{eq:tay_ope} and serve as a cross-check on the diagrammatic approach.

\subsubsection*{Order $g$ dimension one}

At order $g\sim\lambda/N$, the first correction to the amputated four-point function comes from a single $\sigma$ exchange. The free $\sigma$ propagator is a contact interaction,
\begin{equation}
\label{eq:app_prop_sigma}
D^\sigma(y,z)=\langle0|\sigma(y)\sigma(z)|0\rangle=g\,\delta^{(2)}(y-z),
\end{equation}
so at this order the two Yukawa insertions collapse to a single integration point. The amputated matrix element with open spinor and flavor indices is
\begin{align}
\label{eq:M1_open}
\mathcal M^{(1)} &= \vcenter{\hbox{\CintLocalFull + \CintLocalFullB}} \nonumber\\
&= g\mu^{2\varepsilon}\int d^dy\,e^{ip_{ab}\cdot y}D_{\beta\rho}^{jn}(x-y)D_{\sigma\alpha}^{im}(y)\left(\delta_{\rho\sigma}\delta_{\gamma\delta}\delta^{mn}\delta^{kl}-\delta_{\rho\delta}\delta_{\gamma\sigma}\delta^{ml}\delta^{kn}\right) \nonumber\\
&= -g\mu^{2\varepsilon}\int\frac{d^dk}{(2\pi)^d}\frac{e^{-i(k-p_{ab})\cdot x}(\slashed k-\slashed p_{ab})_{\beta\rho}\slashed k_{\sigma\alpha}}{(k-p_{ab})^2k^2}\delta^{jn}\delta^{im}\left(\delta_{\rho\sigma}\delta_{\gamma\delta}\delta^{mn}\delta^{kl}-\delta_{\rho\delta}\delta_{\gamma\sigma}\delta^{ml}\delta^{kn}\right),
\end{align}
where $p_{ab}=p_a+p_b$. The dimension-one Wilson coefficients are extracted at zeroth order in $p_{ab}$. For the two terms in parentheses this gives, respectively,
\begin{align}
-g\mu^{2\varepsilon}\int\frac{d^dk}{(2\pi)^d}\frac{e^{-ik\cdot x}\slashed k_{\beta\rho}\slashed k_{\sigma\alpha}}{(k^2)^2}\delta_{\rho\sigma}\delta_{\gamma\delta}\delta^{jn}\delta^{im}\delta^{mn}\delta^{kl} &= -g\mu^{2\varepsilon}\int\frac{d^dk}{(2\pi)^d}\frac{e^{-ikx}}{k^2}\delta_{\beta\alpha}\delta_{\gamma\delta}\delta^{ij}\delta^{kl},\\
g\mu^{2\varepsilon}\int\frac{d^dk}{(2\pi)^d}\frac{e^{-ik\cdot x}\slashed k_{\beta\rho}\slashed k_{\sigma\alpha}}{(k^2)^2}\delta_{\rho\delta}\delta_{\gamma\sigma}\delta^{jn}\delta^{im}\delta^{ml}\delta^{kn} &= g\mu^{2\varepsilon}\int\frac{d^dk}{(2\pi)^d}\frac{e^{-ikx}}{k^2}\frac{\slashed k_{\beta\delta}\slashed k_{\gamma\alpha}}{k^2}\delta^{il}\delta^{jk}.
\end{align}
The first term gives $-gG(x)\delta_{\beta\alpha}\delta_{\gamma\delta}\delta^{ij}\delta^{kl}$, where
\begin{equation}
\label{eq:G_kernel_first}
G(x)=\mu^{2\varepsilon}\int\frac{d^dk}{(2\pi)^d}\frac{e^{-ikx}}{k^2}=\frac{\mu^{2\varepsilon}\Gamma(-\varepsilon)}{(4\pi)^{1-\varepsilon}}\left(\frac{x^2}{4}\right)^\varepsilon=-\frac{1}{4\pi}\left(\frac{1}{\varepsilon}+\ln\frac{\bar\mu^2x^2}{4}+2\gamma_E\right)+\cO(\varepsilon),
\end{equation}
with $\bar\mu^2=4\pi e^{-\gamma_E}\mu^2$. For the second term, we use the Fierz identities to write it as
\begin{equation}
\frac{g}{2N}\mu^{2\varepsilon}\delta^{ij}\delta^{kl}\sum_A(\Gamma_A)_{\gamma\delta}\int\frac{d^dk}{(2\pi)^d}\frac{e^{-ik\cdot x}}{(k^2)^2}(\slashed k\Gamma_A\slashed k)_{\beta\alpha}+\cdots ,
\end{equation}
where the ellipsis denotes traceless flavor structures. The spinor integrals are
\begin{align}
\mu^{2\varepsilon}\int\frac{d^dk}{(2\pi)^d}\frac{e^{-ik\cdot x}}{(k^2)^2}(\slashed k\opone\slashed k)_{\beta\alpha} &= G(x)\delta_{\beta\alpha},\\
\mu^{2\varepsilon}\int\frac{d^dk}{(2\pi)^d}\frac{e^{-ik\cdot x}}{(k^2)^2}(\slashed k\gamma_5\slashed k)_{\beta\alpha} &= -G(x)(\gamma_5)_{\beta\alpha},\\
\mu^{2\varepsilon}\int\frac{d^dk}{(2\pi)^d}\frac{e^{-ik\cdot x}}{(k^2)^2}(\slashed k\gamma^\nu\slashed k)_{\beta\alpha} &= 2\varepsilon\,G(x)\frac{x^\nu(\slashed x)_{\beta\alpha}}{x^2}.
\label{eq:app_eqint3}
\end{align}
Therefore
\begin{equation}
\label{eq:M1_decomp}
\mathcal M^{(1)}\big|_{p_{ab}=0} = -gG(x)\left[\left(1-\frac{1}{2N}\right)\delta_{\beta\alpha}\delta_{\gamma\delta}+\frac{1}{2N}(\gamma_5)_{\beta\alpha}(\gamma_5)_{\gamma\delta}-\frac{\varepsilon}{N}\frac{(\slashed x)_{\beta\alpha}(\slashed x)_{\gamma\delta}}{x^2}\right]\delta^{ij}\delta^{kl}.
\end{equation}

On the OPE side, we consider graphs where we insert the various fermion bilinear operators at $x=0$:
\begin{equation}
\label{eq:Clambda_O1_matrix}
\mathcal M^{(1)}_\cO = \langle0|\cO(0)\,\psibartilde{k}{\gamma}(p_a)\widetilde\psi_\delta^l(p_b)|0\rangle = \vcenter{\hbox{\CintLocalOPE}}+\vcenter{\hbox{\CintLocalOPEB}} .
\end{equation}
The identity operator gives only the disconnected graphs already computed at order $g^0$. For $\cO_1^A=\psibar\Gamma_A\psi$, the first diagram, with a closed fermion loop, gives
\begin{equation}
\label{eq:app_m1_O1a}
\mathcal M^{(1)}_{1,\cO_1^A} = -Ng\mu^{2\varepsilon}\int\frac{d^dk}{(2\pi)^d}\frac{\tr[\slashed k\Gamma_A\slashed k]}{(k^2)^2}\delta_{\gamma\delta}\delta^{kl} = \begin{cases} -2Ng\,G(0)\delta_{\delta\gamma}\delta^{kl}, & A=\opone,\\ 0, & A=\gamma_5,\\ 0, & A=\gamma^\mu . \end{cases}
\end{equation}
For the second diagram, where the operator fields contract with the external legs, one finds
\begin{equation}
\label{eq:app_m2_O1a}
\mathcal M^{(1)}_{2,\cO_1^A} = g\mu^{2\varepsilon}\int\frac{d^dk}{(2\pi)^d}\frac{(\slashed k\Gamma_A\slashed k)_{\delta\gamma}}{(k^2)^2}\delta^{kl} = \begin{cases} g\,G(0)\delta_{\delta\gamma}\delta^{kl}, & A=\opone,\\ -g\,G(0)(\gamma_5)_{\delta\gamma}\delta^{kl}, & A=\gamma_5,\\ g\,\dfrac{2\varepsilon}{2-2\varepsilon}G(0)(\gamma^\mu)_{\delta\gamma}\delta^{kl}, & A=\gamma^\mu . \end{cases}
\end{equation}
Putting these together gives
\begin{equation}
\mathcal M^{(1)}_{\cO_1^A} = \begin{cases} -g\,G(0)(2N-1)\delta_{\delta\gamma}\delta^{kl}, & A=\opone,\\ -g\,G(0)(\gamma_5)_{\delta\gamma}\delta^{kl}, & A=\gamma_5,\\ g\,\dfrac{2\varepsilon}{2-2\varepsilon}G(0)(\gamma^\mu)_{\delta\gamma}\delta^{kl}, & A=\gamma^\mu . \end{cases}
\end{equation}
At $x=0$, the corresponding momentum integrals are scaleless and vanish in dimensional regularization. Thus the bare Wilson coefficients at order $g$ are obtained from the full-theory graphs in Eq.~\eqref{eq:M1_decomp}. Including the free-theory result gives
\begin{align}
\label{eq:C1_g}
C_1^{\opone}(x)_{\alpha\beta} &= \frac{1}{2N}\delta_{\beta\alpha}\left(1-gG(x)(2N-1)+\cdots\right),\nonumber\\
C_1^{\gamma_5}(x)_{\alpha\beta} &= \frac{1}{2N}(\gamma_5)_{\beta\alpha}\left(1-gG(x)+\cdots\right),\\
C_1^{\gamma_\nu}(x)_{\alpha\beta} &= \frac{1}{2N}(\gamma_\nu)_{\beta\alpha}-\frac{g}{4\pi N}\frac{x_\nu(\slashed x)_{\beta\alpha}}{x^2}+\cdots .\nonumber
\end{align}
The vector coefficient receives only a finite correction because the $2\varepsilon$ prefactor in Eq.~\eqref{eq:M1_decomp} cancels the $1/\varepsilon$ pole in $G(x)$.

In momentum space, suppressing the flavor indices,
\begin{align}
\label{eq:C1_g_p}
\widetilde C_1^{\opone}(p)_{\alpha\beta} &= \frac{1}{2N}\delta_{\beta\alpha}(2\pi)^d\delta^{(d)}(p)-\frac{g(2N-1)}{2N}\delta_{\beta\alpha}\left[\frac{1}{p^2}\right]_*^{\bar\mu}+\cdots,\nonumber\\
\widetilde C_1^{\gamma_5}(p)_{\alpha\beta} &= \frac{1}{2N}(\gamma_5)_{\beta\alpha}(2\pi)^d\delta^{(d)}(p)-\frac{g}{2N}(\gamma_5)_{\beta\alpha}\left[\frac{1}{p^2}\right]_*^{\bar\mu}+\cdots,\nonumber\\
\widetilde C_1^{\gamma_\nu}(p)_{\alpha\beta} &= \frac{1}{2N}(\gamma_\nu)_{\beta\alpha}(2\pi)^d\delta^{(d)}(p)+\frac{g}{2N}\left[\frac{\slashed p\gamma_\nu\slashed p}{(p^2)^2}\right]_{*\,\beta\alpha}^{\bar\mu}+\cdots .
\end{align}
Here $[\cdots]_*^{\bar\mu}$ denotes a star distribution: it agrees with the na\"{i}ve power law for $p\neq0$, while its contact terms at $p=0$ are fixed by $\overline{\mathrm{MS}}$ subtraction.

\subsubsection*{Order $g$ dimension two}

To match the dimension-two bilinear operators $\cO_{2,\mu}^A\equiv\psibar\Gamma_A\partial_\mu\psi$, we expand the same full-theory graph of Eq.~\eqref{eq:M1_open} to first order in the soft external momenta. In this graph the dependence is through $p_{ab}=p_a+p_b$. The right-acting derivative operator has tree-level amputated matrix element
\begin{equation}
\langle0|\cO_{2,\mu}^A(0)\,\psibartilde{k}{\gamma}(p_a)\widetilde\psi_\delta^l(p_b)|0\rangle_{\rm amp}^{(0)} = ip_{a\mu}(\Gamma_A)_{\delta\gamma}\delta^{kl}.
\end{equation}
Strictly speaking, terms proportional to $p_b$ correspond to left-acting derivative operators $(\partial_\mu\psibar)\Gamma_A\psi$.
A complete dimension-two bilinear basis should include both left- and right-acting derivatives, or an equation-of-motion/integration-by-parts convention should be stated. For the leading $A$-form-factor cancellation below, only contact terms from this sector survive at nonzero external momentum.
Expanding the $p_{ab}$-dependent integrand gives
\begin{equation}
e^{-i(k-p_{ab})\cdot x}\frac{\slashed k-\slashed p_{ab}}{(k-p_{ab})^2} = e^{-ik\cdot x}\left[\frac{\slashed k}{k^2}+p_{ab,\mu}\left(ix^\mu\frac{\slashed k}{k^2}-\frac{\gamma^\mu}{k^2}+2\frac{k^\mu\slashed k}{(k^2)^2}\right)+\cdots\right].
\end{equation}
The useful integral is
\begin{equation}
\label{eq:dim2_kderiv}
\mu^{2\varepsilon}\int\frac{d^dk}{(2\pi)^d}\frac{k^\mu e^{-ik\cdot x}}{(k^2)^2} = i\partial_x^\mu\left[-\frac{x^2}{4(1+\varepsilon)}G(x)\right] = -\frac{i x^\mu}{2}G(x).
\end{equation}
For example, in the scalar channel,
\begin{equation}
\left[\left(ix^\mu\frac{\slashed k}{k^2}-\frac{\gamma^\mu}{k^2}+2\frac{k^\mu\slashed k}{(k^2)^2}\right)\frac{\slashed k}{k^2}\right]_{\opone} = ix^\mu\frac{1}{k^2}+\frac{k^\mu}{(k^2)^2}.
\end{equation}
Using Eq.~\eqref{eq:dim2_kderiv}, the scalar linear kernel is
\begin{equation}
ix^\mu G(x)-\frac{i x^\mu}{2}G(x)=\frac{i x^\mu}{2}G(x).
\end{equation}
After stripping the tree-level factor $ip_{a\mu}$, the order-$g$ scalar derivative coefficient is therefore $x^\mu/2$ times the corresponding dimension-one order-$g$ coefficient. The same pattern holds for the pseudoscalar channel; the vector channel receives the finite tensor correction displayed below. Thus
\begin{align}
\label{eq:C2A_g}
C_2^{\opone,\mu}(x)_{\alpha\beta} &= \frac{x^\mu}{2N}\delta_{\beta\alpha}-\frac{g(2N-1)}{4N}x^\mu G(x)\delta_{\beta\alpha}+\cdots,\nonumber\\
C_2^{\gamma_5,\mu}(x)_{\alpha\beta} &= \frac{x^\mu}{2N}(\gamma_5)_{\beta\alpha}-\frac{g}{4N}x^\mu G(x)(\gamma_5)_{\beta\alpha}+\cdots,\nonumber\\
C_2^{\gamma_\nu,\mu}(x)_{\alpha\beta} &= \frac{x^\mu}{2N}(\gamma_\nu)_{\beta\alpha}-\frac{g}{8\pi N}\frac{x^\mu x_\nu}{x^2}(\slashed x)_{\beta\alpha}+\cdots .
\end{align}
In momentum space, using $\int d^dx\,e^{ipx}x^\mu f(x)=-i\partial_{p_\mu}\widetilde f(p)$, this gives
\begin{align}
\label{eq:C2A_g_p}
\widetilde C_2^{\opone,\mu}(p)_{\alpha\beta} &= -\frac{i}{2N}\delta_{\beta\alpha}(2\pi)^d\frac{\partial}{\partial p_\mu}\delta^{(d)}(p)+\frac{i g(2N-1)}{4N}\delta_{\beta\alpha}\frac{\partial}{\partial p_\mu}\left[\frac{1}{p^2}\right]_*^{\bar\mu}+\cdots,\nonumber\\
\widetilde C_2^{\gamma_5,\mu}(p)_{\alpha\beta} &= -\frac{i}{2N}(\gamma_5)_{\beta\alpha}(2\pi)^d\frac{\partial}{\partial p_\mu}\delta^{(d)}(p)+\frac{i g}{4N}(\gamma_5)_{\beta\alpha}\frac{\partial}{\partial p_\mu}\left[\frac{1}{p^2}\right]_*^{\bar\mu}+\cdots,\nonumber\\
\widetilde C_2^{\gamma_\nu,\mu}(p)_{\alpha\beta} &= -\frac{i}{2N}(\gamma_\nu)_{\beta\alpha}(2\pi)^d\frac{\partial}{\partial p_\mu}\delta^{(d)}(p)-\frac{i g}{4N}\frac{\partial}{\partial p_\mu}\left[\frac{\slashed p\gamma_\nu\slashed p}{(p^2)^2}\right]_{*\,\beta\alpha}^{\bar\mu}+\cdots .
\end{align}

\subsubsection{Six-point functions}

To match to the dimension-two four-fermion operators
\begin{equation}
\cO_2^{A,B}\equiv(\psibar\Gamma_A\psi)(\psibar\Gamma_B\psi),
\end{equation}
we must go to a six-point function with two additional soft fermion pairs:
\begin{align}
\label{eq:C2AB_matching}
\langle0|\psibar_\alpha^i(0)\psi_\beta^j(x)&\psibartilde{k}{\gamma}(p_a)\widetilde\psi_\delta^l(p_b)\psibartilde{m}{\rho}(p_c)\widetilde\psi_\sigma^n(p_d)|0\rangle_{\rm amp} \nonumber\\
&= C_0(x)_{\alpha\beta}^{ij}\langle0|\psibartilde{k}{\gamma}(p_a)\widetilde\psi_\delta^l(p_b)\psibartilde{m}{\rho}(p_c)\widetilde\psi_\sigma^n(p_d)|0\rangle_{\rm amp} \nonumber\\
&\quad +\sum_A C_1^A(x)_{\alpha\beta}^{ij}\langle0|\cO_1^A(0)\psibartilde{k}{\gamma}(p_a)\widetilde\psi_\delta^l(p_b)\psibartilde{m}{\rho}(p_c)\widetilde\psi_\sigma^n(p_d)|0\rangle_{\rm amp} \nonumber\\
&\quad +\sum_{A,\mu} C_2^{A,\mu}(x)_{\alpha\beta}^{ij}\langle0|\cO_{2,\mu}^A(0)\psibartilde{k}{\gamma}(p_a)\widetilde\psi_\delta^l(p_b)\psibartilde{m}{\rho}(p_c)\widetilde\psi_\sigma^n(p_d)|0\rangle_{\rm amp} \nonumber\\
&\quad +\sum_{A,B}C_2^{A,B}(x)_{\alpha\beta}^{ij}\langle0|\cO_2^{A,B}(0)\psibartilde{k}{\gamma}(p_a)\widetilde\psi_\delta^l(p_b)\psibartilde{m}{\rho}(p_c)\widetilde\psi_\sigma^n(p_d)|0\rangle_{\rm amp}^{(0)}+\cdots .
\end{align}
To isolate the genuine four-fermion sector, we need the term zeroth order in the soft momenta. We extract the Wilson coefficients by sending the external momenta to zero and regulating the integrals dimensionally. In this prescription, the OPE-side contributions from the bilinear operators reduce to scaleless integrals. The Wilson coefficient $C_2^{A,B}(x)$ can then be read off from the six-point functions after projecting onto the relevant flavor singlet basis.

On the OPE side, the tree-level matrix element of the four-fermion operator is
\begin{align}
\label{eq:O2AB_tree}
\langle0|\cO_2^{A,B}(0)\psibartilde{k}{\gamma}(p_a)\widetilde\psi_\delta^l(p_b)\psibartilde{m}{\rho}(p_c)\widetilde\psi_\sigma^n(p_d)|0\rangle_{\rm amp}^{(0)} &= \delta^{kl}\delta^{mn}(\Gamma_A)_{\delta\gamma}(\Gamma_B)_{\sigma\rho}+\delta^{ml}\delta^{kn}(\Gamma_A)_{\delta\rho}(\Gamma_B)_{\sigma\gamma} \nonumber\\
&\quad +\delta^{kl}\delta^{mn}(\Gamma_A)_{\sigma\rho}(\Gamma_B)_{\delta\gamma}+\delta^{ml}\delta^{kn}(\Gamma_A)_{\sigma\gamma}(\Gamma_B)_{\delta\rho}.
\end{align}
The first connected full-theory graphs appear at order $g^2$ and contain two local $\sigma$ exchanges. We only need the leading scalar singlet coefficient for the main text. The representative ordered contraction contributing to the scalar channel is
\begin{align}\label{eq:app_graphs_6pt}
\mathcal{M}^{(2)}_1 = \vcenter{\hbox{\CtwoLOfull}}\ , 
\end{align}
which yields
\begin{equation}
\label{eq:C2V_full_inta}
\mathcal M_1^{(2)} = g^2\mu^{4\varepsilon}\int d^dy_1\,d^dy_2\,e^{ip_{ab}\cdot y_1+ip_{cd}\cdot y_2}\left[D(x-y_2)D(y_2-y_1)D(y_1)\right]_{\beta\alpha}\delta_{\gamma\delta}\delta_{\rho\sigma}\delta^{ij}\delta^{kl}\delta^{mn}.
\end{equation}
The other ordered contractions reproduce the other terms in Eq.~\eqref{eq:O2AB_tree}. Subleading finite-$N$ tensor structures can be obtained by keeping the Fierz-projected crossed contractions; they are not needed for the leading scalar $A$-form-factor coefficient used in the main text.

At zero external momenta, a useful integral is
\begin{equation}
J_\mu(x) = \mu^{2\varepsilon}\int\frac{d^dk}{(2\pi)^d}\frac{k_\mu e^{-ik\cdot x}}{(k^2)^2}.
\end{equation}
Using a Schwinger parameter for $1/(k^2)^2$ gives
\begin{equation}
J_\mu(x)=i\partial_\mu\left[-\frac{x^2}{4(1+\varepsilon)}G(x)\right]=-\frac{i x_\mu}{2}G(x).
\end{equation}
Using this integral, the ordered scalar graph evaluates to
\begin{equation}
\label{eq:app_sixpta}
\mathcal M_1^{(2)} = i g^2\mu^{2\varepsilon}J_\mu(x)(\gamma^\mu)_{\beta\alpha}\delta^{ij}S_{\gamma\delta\rho\sigma}^{klmn} = \frac{g^2\mu^{2\varepsilon}}{2}G(x)(\slashed x)_{\beta\alpha}\delta^{ij}S_{\gamma\delta\rho\sigma}^{klmn},
\end{equation}
where $S_{\gamma\delta\rho\sigma}^{klmn}=\delta_{\gamma\delta}\delta_{\rho\sigma}\delta^{kl}\delta^{mn}$.

The OPE-side graphs with insertions of the lower-dimensional bilinear operators are scaleless when the external soft momenta are set to zero. Thus the leading scalar four-fermion Wilson coefficient can be read off from the scalar six-point graph. For the unweighted four-fermion operator $\cO_{4f}=(\psibar\psi)^2$, one obtains, at leading order in large $N$,
\begin{equation}
\label{eq:C4f_leading_app}
C_{\cO_{4f}}(x)_{\alpha\beta} = \frac{g^2\mu^{2\varepsilon}}{2}G(x)(\slashed x)_{\beta\alpha}+\cO(g^2/N).
\end{equation}
Using Eq.~\eqref{eq:G_kernel_first}, this is
\begin{equation}
\label{eq:C4f_leading_expanded_app}
C_{\cO_{4f}}(x)_{\alpha\beta} = -\frac{g^2}{8\pi}\left[\frac{1}{\varepsilon}+\ln(x^2\mu^2)+\text{scheme-dependent constant}\right](\slashed x)_{\beta\alpha}+\cO(g^2/N).
\end{equation}
Therefore, for the operator $V=g\cO_{4f}=g(\psibar\psi)^2$, the matrix Wilson coefficient is
\begin{equation}
\label{eq:CV_matrix_x_app}
C_V(x)_{\alpha\beta} = -\frac{g}{8\pi}\ln(x^2\mu^2)(\slashed x)_{\beta\alpha}+\text{contact/scheme-dependent terms}+\cO(g/N,g^2).
\end{equation}
Equivalently, in the scalar form factor $\mathcal A(x^2)$ defined by
\[
\langle m|\psibar(0)\psi(x)|m\rangle=\mathcal A(x^2)\slashed x+\cdots,
\]
the coefficient multiplying $\langle V\rangle$ is
\begin{equation}
\label{eq:CV_scalar_x_app}
C_V(x^2)=-\frac{g}{8\pi}\ln(x^2\mu^2)+\cdots .
\end{equation}
Fourier transforming the matrix coefficient gives, away from contact terms,
\begin{equation}
\label{eq:CV_matrix_p_app}
\widetilde C_V(p)_{\alpha\beta} = ig\left[\frac{(\slashed p)_{\beta\alpha}}{(p^2)^2}\right]_{\bar\mu^2}+\text{contact terms}.
\end{equation}
Projecting onto the $A$ form factor then gives
\begin{equation}
\label{eq:CV_A_projected_app}
\widetilde C_V^{\,A}(p^2) = -\frac{iN}{2}\tr\!\left[\slashed p\,\widetilde C_V(p)\right] = Ng\left[\frac{1}{p^2}\right]_{\bar\mu^2}+\cdots .
\end{equation}
This is the coefficient needed in the main text. The spinor-matrix coefficient $\widetilde C_V(p)_{\alpha\beta}$ and the $A$-projected scalar coefficient $\widetilde C_V^{\,A}(p^2)$ are equivalent but different objects.

\subsection{Operator renormalization}
\label{app:operator_renormalization}

We have computed the bare Wilson coefficients in the previous subsections. The elementary fields and coupling are renormalized as
\begin{equation}
\psi_0 = Z_\psi^{1/2}\psi,\quad g_0 = \mu^{2\varepsilon}Z_g\,g.
\end{equation}
At leading order in large $N$, the fermion wavefunction is not renormalized. Indeed, the only order-$g$ self-energy graph is a tadpole,
\begin{equation}
\vcenter{\hbox{\selfenergySigma}} = \vcenter{\hbox{\selfenergyTadpole}} = g\int\!\frac{d^dk}{(2\pi)^d}\,\frac{\tr[\slashed k]}{k^2} = 0,
\end{equation}
so
\begin{equation}
Z_\psi = 1+\cO(g^2),\quad Z_g = 1-\frac{gN}{2\pi\varepsilon}+\cO(g^2).
\end{equation}
Here $Z_g$ is the large-$N$ coupling renormalization already used in Eq.~\eqref{eq:Zg_gn}. Finite-$N$ corrections to $Z_g$ are not needed for the leading large-$N$ cancellation discussed in the main text.

To renormalize composite operators, we use the convention
\begin{equation}
\label{eq:app_ren_op_def}
\cO_1^A = Z_1^A\,\psibar\Gamma_A\psi = \frac{Z_1^A}{Z_\psi}\,\cO_{1,\bare}^A,\quad \cO_{1,\bare}^A = \psibar_0\Gamma_A\psi_0.
\end{equation}
The product of bare Wilson coefficients and bare operators is independent of the subtraction scale:
\begin{equation}
0 = \mu\frac{d}{d\mu}\left(C_{1,\bare}^A\cO_{1,\bare}^A\right) = \mu\frac{d}{d\mu}\left(C_{1,\bare}^A\,\frac{Z_\psi}{Z_1^A}\,\cO_1^A\right).
\end{equation}
Thus the renormalized Wilson coefficient is
\begin{equation}
C_1^A = \frac{Z_\psi}{Z_1^A}C_{1,\bare}^A.
\end{equation}
With this convention, the anomalous dimension below is the anomalous dimension of the Wilson coefficient,
\begin{equation}
\gamma_{C_1}^A = \frac{d}{d\ln\mu}\ln C_1^A = \frac{d}{d\ln\mu}\ln\frac{Z_\psi}{Z_1^A}.
\end{equation}
The operator anomalous dimension has the opposite sign in this convention.  Because $g_0 = \mu^{2\varepsilon}Z_g g$, the beta function in $d = 2 - 2\varepsilon$ begins as
\begin{equation}
\mu\frac{dg}{d\mu} = -2\varepsilon g + \cO(g^2).
\end{equation}
Thus, if $Z = 1 + ag/\varepsilon + \cdots$, then
\begin{equation}
\frac{d}{d\ln\mu}\ln Z = -2ag + \cO(g^2).
\end{equation}
Since the Wilson coefficients renormalize with the inverse operator factor, $C_R = (Z_\psi/Z)C_{\bare}$, the corresponding Wilson-coefficient anomalous dimension is $2ag + \cO(g^2)$.

\subsubsection*{Dimension-one bilinears}

The dimension-one bilinears $\cO_1^A = \psibar\Gamma_A\psi$ are renormalized by the one-loop vertex correction
\begin{equation}
\label{eq:O1_vertex_general_app}
\vcenter{\hbox{\bilinearVC}} = g\mu^{2\varepsilon}\!\int\!\frac{d^dk}{(2\pi)^d}\,\frac{(\slashed k+\slashed p\,')\Gamma_A(\slashed k+\slashed p)}{(k+p')^2(k+p)^2}.
\end{equation}
To extract the UV pole, take $p = 0$ and $p' = q$. In the scalar channel,
\begin{equation}
g\mu^{2\varepsilon}\!\int\!\frac{d^dk}{(2\pi)^d}\,\frac{(\slashed k+\slashed q)\slashed k}{(k+q)^2k^2} \;\longrightarrow\; g\mu^{2\varepsilon}\!\int\!\frac{d^dk}{(2\pi)^d}\,\frac{(k+q)\cdot k}{(k+q)^2k^2} = \frac{g}{4\pi\varepsilon}+\text{finite}.
\end{equation}
The closed-loop graph contributes only in the scalar channel and gives
\begin{equation}
-Ng\mu^{2\varepsilon}\!\int\!\frac{d^dk}{(2\pi)^d}\,\frac{\tr[(\slashed k+\slashed q)\slashed k]}{(k+q)^2k^2} = -\frac{Ng}{2\pi\varepsilon}+\text{finite}.
\end{equation}
Thus the total scalar pole is
\begin{equation}
-\frac{g(2N-1)}{4\pi\varepsilon}.
\end{equation}
For the pseudoscalar channel, the open-line graph has the opposite pole and the closed-loop graph vanishes:
\begin{equation}
g\mu^{2\varepsilon}\!\int\!\frac{d^dk}{(2\pi)^d}\,\frac{(\slashed k+\slashed q)\gamma_5\slashed k}{(k+q)^2k^2} = -\frac{g}{4\pi\varepsilon}\gamma_5+\text{finite}.
\end{equation}
The vector current has no one-loop pole, as expected from the Ward identity.

Since the renormalized operator is defined by Eq.~\eqref{eq:app_ren_op_def}, the counterterm must cancel these poles in the renormalized matrix element. Hence
\begin{equation}
\label{eq:Z1_app_corrected}
Z_1^{\opone} = 1+\frac{g(2N-1)}{4\pi\varepsilon}+\cO(g^2),\quad Z_1^{\gamma_5} = 1+\frac{g}{4\pi\varepsilon}+\cO(g^2),\quad Z_1^{\gamma^\mu} = 1+\cO(g^2).
\end{equation}
The corresponding Wilson-coefficient anomalous dimensions are
\begin{equation}
\label{eq:gammaC1_app_corrected}
\gamma_{C_1}^{\opone} = \frac{g(2N-1)}{2\pi}+\cO(g^2),\quad \gamma_{C_1}^{\gamma_5} = \frac{g}{2\pi}+\cO(g^2),\quad \gamma_{C_1}^{\gamma^\mu} = 0.
\end{equation}

\subsubsection*{Dimension-two derivative bilinears}

The derivative bilinears are extracted from the same amputated two-point function, now with an insertion of
\begin{equation}
\cO_{2,\mu}^A = \psibar\Gamma_A\partial_\mu\psi.
\end{equation}
The differentiated field carries momentum $k+p$, so the one-loop graph is
\begin{equation}
\label{eq:O2D_vertex_general_app}
\vcenter{\hbox{\bilinearVC}} = ig\mu^{2\varepsilon}\!\int\!\frac{d^dk}{(2\pi)^d}\,(k+p)_\mu\,\frac{(\slashed k+\slashed p\,')\Gamma_A(\slashed k+\slashed p)}{(k+p')^2(k+p)^2}.
\end{equation}
To isolate the right-acting derivative coefficient, we set $p' = 0$ and expand to first order in $p$. Terms proportional to $p'$ correspond to left-acting derivative operators $(\partial_\mu\psibar)\Gamma_A\psi$ and are not needed for the finite-momentum $A$-form-factor coefficient. The linearized integrand is
\begin{equation}
ig\mu^{2\varepsilon}\!\int\!\frac{d^dk}{(2\pi)^d}\frac{1}{(k^2)^2}\left[p_\mu\slashed k\Gamma_A\slashed k+k_\mu\slashed k\Gamma_A\slashed p-2\frac{k_\mu k\cdot p}{k^2}\slashed k\Gamma_A\slashed k\right].
\end{equation}

For $A=\opone$, the pole part reduces to
\begin{equation}
ig\mu^{2\varepsilon}\!\int\!\frac{d^dk}{(2\pi)^d}\frac{1}{(k^2)^2}\left[p_{\mu} k^2 -\slashed{p}\slashed{k}k_\mu \right] = ig\frac{p_\mu}{4\pi \varepsilon}  - ig\frac{\slashed{p}\gamma_\mu }{8\pi \varepsilon} +\text{finite}\ .
\end{equation}
With the convention $\epsilon_{01}=+1$, $\gamma^0=\sigma_1$, $\gamma^1=\sigma_2$, and $\gamma_5=-i\gamma^0\gamma^1$, we use
\begin{equation}
\gamma_\mu\gamma_\nu = \delta_{\mu\nu}+i\epsilon_{\mu\nu}\gamma_5.
\end{equation}
The scalar part of the open-line pole is therefore $ig\,p_\mu/(8\pi\varepsilon)$. The closed-loop scalar graph gives
\begin{equation}
-\frac{iNg}{4\pi\varepsilon}p_\mu,
\end{equation}
so the total diagonal scalar pole relative to the tree structure $ip_\mu$ is
\begin{equation}
-\frac{g(2N-1)}{8\pi\varepsilon}.
\end{equation}
Similarly, the diagonal pseudoscalar pole is
\begin{equation}
-\frac{g}{8\pi\varepsilon},
\end{equation}
while the vector derivative bilinear has no diagonal one-loop pole. Therefore
\begin{equation}
\label{eq:Z2D_app_corrected}
Z_{2,D}^{\opone} = 1+\frac{g(2N-1)}{8\pi\varepsilon}+\cO(g^2),\quad Z_{2,D}^{\gamma_5} = 1+\frac{g}{8\pi\varepsilon}+\cO(g^2),\quad Z_{2,D}^{\gamma^\nu} = 1+\cO(g^2).
\end{equation}
The corresponding Wilson-coefficient anomalous dimensions are
\begin{equation}
\label{eq:gammaC2D_app_corrected}
\gamma_{C_{2,D}}^{\opone} = \frac{g(2N-1)}{4\pi}+\cO(g^2),\quad \gamma_{C_{2,D}}^{\gamma_5} = \frac{g}{4\pi}+\cO(g^2),\quad \gamma_{C_{2,D}}^{\gamma^\nu} = 0.
\end{equation}
There are also off-diagonal entries between scalar and pseudoscalar derivative bilinears proportional to $\epsilon_{\mu\nu}$. They depend on the precise convention for $\epsilon_{\mu\nu}$ and on the choice of left- versus right-acting derivative basis. Since these entries do not enter the leading finite-momentum $A$-form-factor cancellation, we do not display them.

\subsubsection*{Dimension-two four-fermion sector}

The complete finite-$N$ four-fermion anomalous-dimension matrix is not needed for the leading large-$N$ cancellation studied in the main text. The only dimension-two operator that contributes at nonzero external momentum to the leading $A$-form-factor ambiguity is $V$. The matching calculation in the preceding subsection gives the leading scalar matrix coefficient of the unweighted operator $\cO_{4f}=(\psibar\psi)^2$:
\begin{equation}
C_{\cO_{4f}}(x)_{\alpha\beta} = -\frac{g^2}{8\pi}\left[\frac{1}{\varepsilon}+\ln(x^2\mu^2)+\text{scheme-dependent constant}\right](\slashed x)_{\beta\alpha}+\cO(g^2/N).
\end{equation}
Since $V = g\,\cO_{4f}$, the corresponding renormalized matrix Wilson coefficient of $V$ is
\begin{equation}
C_V(x)_{\alpha\beta} = -\frac{g}{8\pi}\ln(x^2\mu^2)(\slashed x)_{\beta\alpha}+\text{contact/scheme-dependent terms}+\cO(g/N,g^2).
\end{equation}
Equivalently, in the scalar form factor $\mathcal A(x^2)$ defined by
\begin{equation}
\langle m|\psibar(0)\psi(x)|m\rangle = \mathcal A(x^2)\slashed x+\cdots,
\end{equation}
the coefficient multiplying $\langle V\rangle$ is
\begin{equation}
\label{eq:CV_scalar_app_summary}
C_V(x^2) = -\frac{g}{8\pi}\ln(x^2\mu^2)+\cdots.
\end{equation}
Fourier transforming the matrix coefficient gives
\begin{equation}
\label{eq:CV_matrix_p_app_summary}
\widetilde C_V(p)_{\alpha\beta} = ig\left[\frac{(\slashed p)_{\beta\alpha}}{(p^2)^2}\right]_{\bar\mu^2}+\text{contact terms}.
\end{equation}
Projecting onto the $A$ form factor gives
\begin{equation}
\label{eq:CV_A_projected_app_summary}
\widetilde C_V^{\,A}(p^2) = -\frac{iN}{2}\tr\!\left[\slashed p\,\widetilde C_V(p)\right] = Ng\left[\frac{1}{p^2}\right]_{\bar\mu^2}+\cdots .
\end{equation}
This is the only dimension-two Wilson coefficient needed for the leading $t=1$ cancellation in the main text. The derivative-bilinear sector contributes only contact terms to $A(p^2)$ at nonzero external momentum at this order, and the full finite-$N$ dimension-two anomalous-dimension matrix will not be displayed.

\subsection{Renormalized Wilson coefficients}

The renormalized OPE is obtained by replacing bare operators with renormalized ones. To avoid a convention clash, let $\mathcal Z$ denote the inverse operator-renormalization matrix, defined by
\begin{equation}
\cO_{\bare}^{I} = \mathcal Z^{I}{}_{J}\,[\cO^J].
\end{equation}
Then the Wilson coefficients in the renormalized OPE are
\begin{equation}
\label{eq:Cren_general}
C_{R}^{J} = C_{\bare}^{I}\,\mathcal Z^{I}{}_{J}.
\end{equation}
With this convention the renormalized OPE through dimension two is
\begin{equation}
\psibar_\alpha(0)\psi_\beta(x) = C_0(x)_{\alpha\beta}\,\opone + \sum_A C_{1,R}^A(x)_{\alpha\beta}\,[\cO_1^A] + \sum_A C_{2,D,R}^{A,\mu}(x)_{\alpha\beta}\,[\cO_{2,\mu}^A] + \sum_{A,B}C_{2,4f,R}^{A,B}(x)_{\alpha\beta}\,[\cO_2^{A,B}] + \cdots .
\end{equation}
At one loop, $C_0$ is unchanged since $Z_\psi = 1+\cO(g^2)$.

For the dimension-one bilinears, the scalar and pseudoscalar poles in Eq.~\eqref{eq:C1_g} are removed by the inverse counterterms $\mathcal Z_1$. Defining the finite kernel
\begin{equation}
G_R(x)\equiv G(x)+\frac{1}{4\pi\varepsilon} = -\frac{1}{4\pi}\left(\ln\frac{\bar\mu^2x^2}{4}+2\gamma_E\right),
\end{equation}
the renormalized dimension-one coefficients are
\begin{align}
\label{eq:C1R_summary}
C_{1,R}^{\opone}(x)_{\alpha\beta} &= \frac{1}{2N}\,\delta_{\beta\alpha}\Bigl(1-g(2N-1)G_R(x)\Bigr)+\cO(g^2),\nonumber\\
C_{1,R}^{\gamma_5}(x)_{\alpha\beta} &= \frac{1}{2N}\,(\gamma_5)_{\beta\alpha}\Bigl(1-gG_R(x)\Bigr)+\cO(g^2),\nonumber\\
C_{1,R}^{\gamma_\nu}(x)_{\alpha\beta} &= \frac{1}{2N}\,(\gamma_\nu)_{\beta\alpha} - \frac{g}{4\pi N}\frac{x_\nu}{x^2}(\slashed{x})_{\beta\alpha} + \cO(g^2).
\end{align}

The derivative-bilinear sector can mix with other dimension-two operators, but the complete finite-$N$ dimension-two mixing matrix is not needed for the leading large-$N$ cancellation studied in the main text. At nonzero external momentum, the derivative-bilinear coefficients contribute only contact terms to the $A$ form factor at this order. We therefore do not display the full derivative-bilinear renormalization matrix here.

The leading nontrivial power correction to $A(p^2)$ thus comes from the scalar four-fermion operator $V$. It is useful to distinguish the unweighted operator $\cO_{4f}\equiv(\psibar\psi)^2$, so that $V=g\,\cO_{4f}$. The leading scalar six-point matching in Eq.~\eqref{eq:C4f_leading_expanded_app} gives, suppressing the overall fixed-flavor factor,
\begin{equation}
C_{\cO_{4f},\bare}(x)_{\alpha\beta} = \frac{g^2\mu^{2\varepsilon}}{2}G(x)(\slashed{x})_{\beta\alpha} + \cO(g^2/N).
\end{equation}
Using Eq.~\eqref{eq:G_kernel_first}, this becomes
\begin{equation}
C_{\cO_{4f},\bare}(x)_{\alpha\beta} = -\frac{g^2}{8\pi}\left[\frac{1}{\varepsilon}+\ln(x^2\mu^2)+\text{scheme-dependent constant}\right](\slashed{x})_{\beta\alpha} + \cO(g^2/N).
\end{equation}
After subtracting the pole, the renormalized matrix coefficient of $\cO_{4f}$ is
\begin{equation}
C_{\cO_{4f},R}(x)_{\alpha\beta} = -\frac{g^2}{8\pi}\ln(x^2\mu^2)(\slashed{x})_{\beta\alpha} + \text{scheme-dependent contact terms} + \cO(g^2/N,g^3).
\end{equation}
Since $V=g\,\cO_{4f}$, the corresponding matrix Wilson coefficient of $V$ is
\begin{equation}
\label{eq:CV_matrix_x_app_ren}
C_V(x)_{\alpha\beta} = -\frac{g}{8\pi}\ln(x^2\mu^2)(\slashed{x})_{\beta\alpha} + \text{scheme-dependent contact terms} + \cO(g/N,g^2).
\end{equation}
Thus the scalar coefficient entering the $\mathcal A(x^2)$ form factor below is finite and has the sign
\begin{equation}
\label{eq:CV_scalar_x_app_ren}
C_V(x^2) = -\frac{g}{8\pi}\ln(x^2\mu^2)+\cO(g^2).
\end{equation}

\subsection{OPE for \texorpdfstring{$A$}{A}}

We now project the OPE onto the scalar form factor $A(p^2)$. It is cleaner to first work in position space. For fixed external flavor indices, define
\begin{equation}
\langle m|\psibar_\alpha^i(0)\psi_\beta^j(x)|m\rangle = \delta^{ij}\left[\mathcal A(x^2)(\slashed{x})_{\beta\alpha}+\mathcal B(x^2)\delta_{\beta\alpha}\right].
\end{equation}
Equivalently, for a fixed flavor with no sum over $i$,
\begin{equation}
\mathcal A(x^2) = \frac{1}{2x^2}\,\tr\big[\slashed{x}\,\langle m|\psibar^i(0)\psi^i(x)|m\rangle\big].
\end{equation}
If instead one traces over flavor, the right-hand side should be divided by $N$.

Only operators whose Wilson coefficients contain a $\slashed{x}$ structure, and whose matrix elements can be nonzero in a rotationally invariant state, contribute to $\mathcal A(x^2)$. At dimension zero, the identity operator contributes through the perturbative coefficient. At dimension one, the scalar and pseudoscalar operators $\cO_1^{\opone}$ and $\cO_1^{\gamma_5}$ can have scalar expectation values, but their Wilson coefficients are proportional to $\opone$ and $\gamma_5$, so they contribute to $\mathcal B(x^2)$ rather than to $\mathcal A(x^2)$. The vector bilinear cannot acquire a Lorentz-invariant vacuum expectation value. Thus no dimension-one operator contributes to $A(p^2)$.

At dimension two, the relevant traced scalar sector may be represented by
\begin{equation}
\label{eq:KVdefs_app}
K \equiv i\psibar\slashed{\partial}\psi = i\,\delta_{\mu\nu}\cO_{2}^{\gamma^\nu,\mu},\quad V \equiv g\,\cO_2^{\opone,\opone}.
\end{equation}
The coefficient of $K$ is a contact term after Fourier transformation and is not needed for the leading finite-momentum cancellation. The coefficient of $V$ is the one computed above. The position-space scalar-form-factor OPE therefore has the form
\begin{equation}
\mathcal A(x^2) = \mathcal A_{\rm pert}(x^2) + C_K(x^2)\,\langle m|K|m\rangle + C_V(x^2)\,\langle m|V|m\rangle + \cdots,
\end{equation}
where the only non-contact coefficient needed below is
\begin{equation}
\label{eq:CV_A_position_summary_app}
C_V(x^2) = -\frac{g}{8\pi}\ln(x^2\mu^2)+\cO(g^2).
\end{equation}

Fourier transforming the matrix coefficient in Eq.~\eqref{eq:CV_matrix_x_app_ren} gives, away from contact terms,
\begin{equation}
\label{eq:CV_matrix_p_app_ren}
\widetilde C_V(p)_{\alpha\beta} = ig\left[\frac{(\slashed p)_{\beta\alpha}}{(p^2)^2}\right]_{\bar\mu^2}+\text{contact terms}.
\end{equation}
Projecting onto the $A$ form factor gives
\begin{equation}
\label{eq:CV_A_projected_app_ren}
\widetilde C_V^{\,A}(p^2) = -\frac{iN}{2}\tr\!\left[\slashed p\,\widetilde C_V(p)\right] = Ng\left[\frac{1}{p^2}\right]_{\bar\mu^2}+\cdots .
\end{equation}
Thus, for $p\neq0$, the part of the momentum-space OPE relevant for the leading renormalon cancellation is
\begin{equation}
\label{eq:A_OPE_final_app}
A(p^2) = A_{\rm pert}(p^2) + Ng\left[\frac{1}{p^2}\right]_{\bar\mu^2}\langle m|V|m\rangle + \cdots .
\end{equation}
This is the coefficient used in the main text. It is important to distinguish the spinor-matrix coefficient $\widetilde C_V(p)_{\alpha\beta}$ in Eq.~\eqref{eq:CV_matrix_p_app_ren} from the $A$-projected scalar coefficient $\widetilde C_V^{\,A}(p^2)$ in Eq.~\eqref{eq:CV_A_projected_app_ren}.

\subsection{Summary of renormalized Wilson coefficients}
\label{sec:wilson_summary}

We collect here the renormalized Wilson coefficients needed in the main text. Throughout this summary the coefficients are written for fixed external flavor indices, with the overall factor of $\delta^{ij}$ suppressed. If one instead traces over flavor, the corresponding coefficients should be multiplied by $N$. We use the finite kernel
\begin{equation}
G_R(x) \equiv G(x) + \frac{1}{4\pi\varepsilon} = -\frac{1}{4\pi}\left(\ln\frac{\bar\mu^2x^2}{4} + 2\gamma_E\right).
\end{equation}
The Fourier transform can turn smooth functions into distributions and can induce contact terms at $p = 0$. We use the convention
\begin{equation}
\int d^2x\,e^{ip\cdot x}\ln(\bar\mu^2x^2) = -4\pi\left[\frac{1}{p^2}\right]_{\bar\mu^2},
\end{equation}
which implies
\begin{equation}
\label{eq:star_dist}
\frac{d}{d\ln\bar\mu^2}\left[\frac{1}{p^2}\right]_{\bar\mu^2} = -\pi\,\delta^{(2)}(p).
\end{equation}
Indeed, differentiating the first equation gives $(2\pi)^2\delta^{(2)}(p)$ on the left-hand side. Constant shifts of the logarithm correspond to local contact terms in momentum space.

\subsubsection*{Position-space Wilson coefficients}

For fixed external flavor, the position-space coefficients through the terms needed below are
\begin{subequations}
\label{eq:Cx_summary}
\begin{align}
\text{Identity:}\quad
\cO = \opone:\quad
C_{\opone,R}(x)_{\alpha\beta}
&= D_{\beta\alpha}(x,0)
= -\frac{\Gamma(d/2)}{2\pi^{d/2}}\frac{(\slashed{x})_{\beta\alpha}}{(x^2)^{d/2}},
\label{eq:C0x}
\\[6pt]
\text{Scalar:}\quad
\cO = \psibar\psi:\quad
C_{S,R}(x)_{\alpha\beta}
&= \frac{1}{2N}\delta_{\beta\alpha}\Bigl(1 - g(2N - 1)G_R(x)\Bigr) + \cO(g^2),
\label{eq:CSx}
\\[6pt]
\text{Pseudoscalar:}\quad
\cO = \psibar\gamma_5\psi:\quad
C_{P,R}(x)_{\alpha\beta}
&= \frac{1}{2N}(\gamma_5)_{\beta\alpha}\Bigl(1 - gG_R(x)\Bigr) + \cO(g^2),
\label{eq:CPx}
\\[6pt]
\text{Vector:}\quad
\cO = \psibar\gamma^\nu\psi:\quad
C_{\gamma^\nu,R}(x)_{\alpha\beta}
&= \frac{1}{2N}(\gamma^\nu)_{\beta\alpha} - \frac{g}{4\pi N}\frac{x^\nu(\slashed{x})_{\beta\alpha}}{x^2} + \cO(g^2),
\label{eq:CVx}
\\[6pt]
\text{Scalar derivative:}\quad
\cO = \psibar\partial^\mu\psi:\quad
C_{D,S,R}^{\mu}(x)_{\alpha\beta}
&= \frac{x^\mu}{2N}\delta_{\beta\alpha} - \frac{g(2N - 1)}{4N}x^\mu G_R(x)\delta_{\beta\alpha} + \cO(g^2),
\label{eq:CDSx}
\\[6pt]
\text{Pseudoscalar derivative:}\quad
\cO = \psibar\gamma_5\partial^\mu\psi:\quad
C_{D,P,R}^{\mu}(x)_{\alpha\beta}
&= \frac{x^\mu}{2N}(\gamma_5)_{\beta\alpha} - \frac{g}{4N}x^\mu G_R(x)(\gamma_5)_{\beta\alpha} + \cO(g^2),
\label{eq:CDPx}
\\[6pt]
\text{Vector derivative:}\quad
\cO = \psibar\gamma^\nu\partial^\mu\psi:\quad
C_{D,V,R}^{\mu\nu}(x)_{\alpha\beta}
&= \frac{x^\mu}{2N}(\gamma^\nu)_{\beta\alpha} - \frac{g}{8\pi N}\frac{x^\mu x^\nu}{x^2}(\slashed{x})_{\beta\alpha} + \cO(g^2),
\label{eq:CDVx}
\\[6pt]
\text{Scalar four-fermion:}\quad
\cO = V \equiv g(\psibar\psi)^2:\quad
C_V(x)_{\alpha\beta}
&= -\frac{g}{8\pi}\ln(x^2\mu^2)(\slashed{x})_{\beta\alpha} + \cO(g/N,g^2).
\label{eq:C2Vx}
\end{align}
\end{subequations}

\noindent Equivalently, in the scalar form factor $\mathcal A(x^2)$ defined by $\langle m|\psibar(0)\psi(x)|m\rangle = \mathcal A(x^2)\slashed x + \cdots$, the coefficient multiplying $\langle V\rangle$ is
\begin{equation}
\label{eq:CV_scalar_summary}
C_V(x^2) = -\frac{g}{8\pi}\ln(x^2\mu^2) + \cO(g/N,g^2).
\end{equation}

The traced derivative operator is
\begin{equation}
K \equiv i\psibar\slashed{\partial}\psi = i\,\delta_{\mu\nu}\cO_2^{\gamma^\nu,\mu}.
\end{equation}
Its coefficient is a contact term after Fourier transformation and will not be needed for the leading finite-momentum renormalon cancellation. Indeed, away from $p = 0$ one has
\begin{align}
\partial_{p_\nu}\!\left[\frac{\slashed p\gamma^\nu\slashed p}{(p^2)^2}\right] = \frac{(2d-4)\slashed{p}}{(p^{2})^2} = 0
\end{align}
in two dimensions, so the traced vector-derivative coefficient is purely local. The leading finite-momentum power correction to $A(p^2)$ comes from $V$.

The selection rules are as follows. At dimension zero, the identity coefficient contributes to the $\slashed{x}$ structure. At dimension one, the scalar and pseudoscalar operators can have scalar expectation values, but their Wilson coefficients are proportional to $\opone$ and $\gamma_5$, so they contribute to $\mathcal B(x^2)$ rather than to $\mathcal A(x^2)$. The vector bilinear cannot acquire a Lorentz-invariant vacuum expectation value. Thus no dimension-one operator contributes to $A(p^2)$. At dimension two, $K$ is a contact term at nonzero external momentum and $V$ gives the leading nontrivial power correction.

\subsubsection*{Momentum-space Wilson coefficients}

The Fourier transform of the renormalized coefficients introduces distributions. Of course, constants in position space become $\delta^{(2)}(p)$, and factors of $x^\mu$ become $-i\partial/\partial p_\mu$ acting on the corresponding momentum-space distribution. Logarithms in position space give star distributions.

For fixed external flavor, the momentum-space coefficients are
\begin{subequations}
\label{eq:Cp_summary}
\begin{align}
\widetilde C_{\opone,R}(p)_{\alpha\beta}
&= \frac{-i(\slashed p)_{\beta\alpha}}{p^2 + i\slashed p\,\Sigma_{\rm pert}(p)},
\label{eq:C0p}
\\[6pt]
\widetilde C_{S,R}(p)_{\alpha\beta}
&= \frac{1}{2N}\delta_{\beta\alpha}(2\pi)^2\delta^{(2)}(p) - \frac{g(2N - 1)}{2N}\delta_{\beta\alpha}\left[\frac{1}{p^2}\right]_{\bar\mu^2} + \cO(g^2),
\label{eq:CSp}
\\[6pt]
\widetilde C_{P,R}(p)_{\alpha\beta}
&= \frac{1}{2N}(\gamma_5)_{\beta\alpha}(2\pi)^2\delta^{(2)}(p) - \frac{g}{2N}(\gamma_5)_{\beta\alpha}\left[\frac{1}{p^2}\right]_{\bar\mu^2} + \cO(g^2),
\label{eq:CPp}
\\[6pt]
\widetilde C_{\gamma^\nu,R}(p)_{\alpha\beta}
&= \frac{1}{2N}(\gamma^\nu)_{\beta\alpha}(2\pi)^2\delta^{(2)}(p) + \frac{g}{2N}\left[\frac{\slashed p\gamma^\nu\slashed p}{(p^2)^2}\right]_{\bar\mu^2,\beta\alpha} + \cO(g^2),
\label{eq:CVp}
\\[6pt]
\widetilde C_{D,S,R}^{\mu}(p)_{\alpha\beta}
&= -\frac{i}{2N}\delta_{\beta\alpha}(2\pi)^2\frac{\partial}{\partial p_\mu}\delta^{(2)}(p) + \frac{i g(2N - 1)}{4N}\delta_{\beta\alpha}\frac{\partial}{\partial p_\mu}\left[\frac{1}{p^2}\right]_{\bar\mu^2} + \cO(g^2),
\label{eq:CDSp}
\\[6pt]
\widetilde C_{D,P,R}^{\mu}(p)_{\alpha\beta}
&= -\frac{i}{2N}(\gamma_5)_{\beta\alpha}(2\pi)^2\frac{\partial}{\partial p_\mu}\delta^{(2)}(p) + \frac{i g}{4N}(\gamma_5)_{\beta\alpha}\frac{\partial}{\partial p_\mu}\left[\frac{1}{p^2}\right]_{\bar\mu^2} + \cO(g^2),
\label{eq:CDPp}
\\[6pt]
\widetilde C_{D,V,R}^{\mu\nu}(p)_{\alpha\beta}
&= -\frac{i}{2N}(\gamma^\nu)_{\beta\alpha}(2\pi)^2\frac{\partial}{\partial p_\mu}\delta^{(2)}(p) - \frac{i g}{4N}\frac{\partial}{\partial p_\mu}\left[\frac{\slashed p\gamma^\nu\slashed p}{(p^2)^2}\right]_{\bar\mu^2,\beta\alpha} + \cO(g^2),
\label{eq:CDVp}
\\[6pt]
\widetilde C_V(p)_{\alpha\beta}
&= ig\left[\frac{(\slashed p)_{\beta\alpha}}{(p^2)^2}\right]_{\bar\mu^2} + \text{contact terms} + \cO(g/N,g^2).
\label{eq:C2Vp}
\end{align}
\end{subequations}
In Eq.~\eqref{eq:C0p}, $\Sigma_{\rm pert}(p)$ denotes the perturbative massless self-energy entering the identity Wilson coefficient; the formula is not intended to include the massive scalar form factor $mB(p^2)$.

The spinor-matrix coefficient $\widetilde C_V(p)_{\alpha\beta}$ should not be confused with the scalar coefficient in the OPE for $A(p^2)$. Projecting with the same convention as in the main text gives
\begin{equation}
\label{eq:CV_A_projected_summary}
\widetilde C_V^{\,A}(p^2) = -\frac{iN}{2}\tr\!\left[\slashed p\,\widetilde C_V(p)\right] = Ng\left[\frac{1}{p^2}\right]_{\bar\mu^2} + \cO(g,g^2N).
\end{equation}
Thus, for $p \neq 0$, the part of the momentum-space OPE relevant for the leading renormalon cancellation is
\begin{equation}
\label{eq:A_OPE_summary}
A(p^2) = A_{\rm pert}(p^2) + Ng\left[\frac{1}{p^2}\right]_{\bar\mu^2}\langle m|V|m\rangle + \cdots.
\end{equation}

The all-orders scalar-channel expression obtained from the bubble-chain resummation uses a different scalar-channel normalization than the fixed-flavor open-index coefficients above. It should therefore not be identified directly with any single entry in Eq.~\eqref{eq:Cp_summary} without specifying the spin and flavor projection.

\subsection{Hard cutoff OPE}
\label{app:hard_cutoff_ope}

For the connected Wilson coefficients relevant at nonzero external momentum, the hard-cutoff scheme differs from dimensional regularization only by finite pieces in the counterterms and by the relation between bare and renormalized quantities. We do not need a full hard-cutoff rederivation of all Wilson coefficients. The salient points are the following.

First, connected matching graphs give the same nonlocal dependence on $x$ as in dimensional regularization. For the operator $V = g(\psibar\psi)^2$, the hard-cutoff version of the scalar $\mathcal A$-coefficient has the form
\begin{equation}
C_{V,\bare}^{(\Lambda)}(x^2) = -\frac{g_0}{8\pi}\left[\ln\!\left(\frac{\Lambda^2x^2}{4}\right) + 2\gamma_E\right] + \cO(g_0/N,g_0^2).
\end{equation}
After renormalization at the subtraction scale $\mu$,
\begin{equation}
C_V^{\rm hc}(x^2;\mu) = -\frac{g(\mu)}{8\pi}\left[\ln(\mu^2x^2) + 2\gamma_E - 2\ln2\right] + \cO(g/N,g^2).
\end{equation}
Thus
\begin{equation}
C_V^{\rm hc}(x^2;\mu) = C_V^{(\overline{\rm MS})}(x^2;\mu) - \frac{g}{8\pi}(2\gamma_E - 2\ln2) + \cO(g/N,g^2),
\end{equation}
where the difference is a finite scheme-dependent local term. For a different cutoff profile, the logarithmic coefficient is unchanged, while the constant changes.

The mixing between $K$ and $V$ is convention dependent as a matrix entry, because it depends on whether one writes bare operators in terms of renormalized operators or the inverse relation. We therefore specify it operationally: it is the subtraction that removes the $\ln(\Lambda^2/\mu^2)$ term in $C_{V,\bare}^{(\Lambda)}$ and produces the finite coefficient above. The only hard-cutoff renormalization constant whose leading coefficient is needed below is the additive identity mixing of $V$.

In momentum space, the corresponding $A$-projected coefficient is
\begin{equation}
\widetilde C_V^{\,A,{\rm hc}}(p^2;\mu) = Ng\left[\frac{1}{p^2}\right]_{\mu^2}^{\rm hc} + \cO(g,g^2N),
\end{equation}
with
\begin{equation}
\left[\frac{1}{p^2}\right]_{\mu^2}^{\rm hc} = \left[\frac{1}{p^2}\right]_{\bar\mu^2} + \text{local contact term}.
\end{equation}
For $p \neq 0$ this agrees with the na\"{i}ve power law $1/p^2$. The difference from the $\overline{\rm MS}$ star distribution is entirely a contact term at $p = 0$.

Second, the identity-operator coefficient is renormalized only through the usual field and coupling renormalization. Expressed in terms of the renormalized coupling at scale $\mu$, it has the same leading infrared renormalon as in dimensional regularization. Changing the UV regulator can shift local terms and the first few perturbative coefficients, but it does not remove the factorial large-order growth governed by the infrared region.

Finally, a hard cutoff makes visible an additive identity renormalization of $V$. In the auxiliary-field formulation, $V = g(\psibar\psi)^2 = \frac{\sigma^2}{g}$. The perturbative vacuum graph with one insertion of $V$ is
\begin{equation}
\langle0|V|0\rangle_{\rm pert} \sim \frac{1}{g}\int^\Lambda\frac{d^2k}{(2\pi)^2}\,D_\sigma(k^2).
\end{equation}
At tree level $D_\sigma^{(0)}(k^2) = g$, so
\begin{equation}
\langle0|V|0\rangle_{\rm pert}^{(0)} \sim \int^\Lambda\frac{d^2k}{(2\pi)^2} = \frac{\Lambda^2}{4\pi}.
\end{equation}
Thus the hard-cutoff renormalized operator must include an additive identity subtraction,
\begin{equation}
V_R(\mu) = Z_{VV}^{\rm hc}(\Lambda,\mu)V_{\bare} + Z_{VK}^{\rm hc}(\Lambda,\mu)K_{\bare} + Z_{V\opone}^{\rm hc}(\Lambda,\mu)\opone,
\end{equation}
with
\begin{equation}
Z_{V\opone}^{\rm hc}(\Lambda,\mu) \sim -\frac{\Lambda^2}{4\pi} + \cO(g\Lambda^2\ln\Lambda).
\end{equation}
This power-divergent subtraction is absent from connected matching graphs. It renormalizes the local operator itself and is the hard-cutoff counterpart of the real UV subtractions used in the condensate cycle.

\subsection{Comment on Mari\~{n}o--Miravitllas}

It is useful to distinguish which ingredients of the OPE cancellation are computed directly in Ref.~\cite{Marino2024}, and which are fixed by matching to the exact large-$N$ answer. In that work, the perturbative sector $\Phi_0(\lambda)$ is obtained from the bubble-chain calculation of the fermion self-energy. The OPE contributions associated with the two-quark and four-quark condensates are then computed diagrammatically, including the operator-renormalization effects needed to make the corresponding Wilson-coefficient contributions finite. This calculation reproduces the factorially divergent power series multiplying the first two power corrections in the exact trans-series.

The overall condensate constants, however, are not fixed by the Wilson-coefficient diagrams alone. They are the non-perturbative input parameters of the OPE, and in Ref.~\cite{Marino2024} they are determined by comparing the OPE trans-series with the exact large-$N$ trans-series. In their notation this matching gives
\begin{equation}
c_1 = \frac{1}{2}+\frac{3}{2}\ln 2,
\end{equation}
for the two-quark condensate sector, and the corresponding constant in the four-quark condensate sector is
\begin{equation}
d_1 = 1-\gamma_E+2\ln 2 \mp \frac{i\pi}{2}.
\end{equation}
The imaginary part of $d_1$ is therefore fixed by matching to the laterally resummed exact trans-series, with the sign correlated with the Borel prescription for the perturbative sector.

The purpose of the present paper is to give a geometric origin for this complex condensate normalization. In our formulation, the imaginary part is not introduced as an unknown matching constant. It arises from the lateral continuations of the renormalized condensate cycle itself: after the UV subtractions are performed, the two lateral contours pass on opposite sides of the same renormalon saddle that controls the perturbative Wilson coefficient. Thus the complex parts of the two lateral four-fermion condensates are the Stokes contributions required by the homological cancellation, while their average gives the principal-value prescription.

\section{Additional details for the \texorpdfstring{$O(N)$}{O(N)} sigma model}
\label{app:ON_details}

This appendix records the details behind the comparison in Section~\ref{sec:ON}. We use them only to identify the leading $t = 1$ cancellation. The full trans-series of the large-$N$ $O(N)$ sigma model is known and contains an infinite sequence of positive-axis Borel singularities~\cite{David:1982qv,David1984,David1986,Novikov1985,BenekeBraunKivel1998,Beneke:1998ui}; we do not rederive that full structure here.

\subsection{Auxiliary-field saddle and contour}

The constrained $O(N)$ model may be written as
\begin{equation}
S = \frac{1}{2}\int d^2x\,\left[(\partial_\mu n^a)^2 + \frac{\alpha}{\sqrt N}\left(n^a n^a - \frac{N}{g}\right)\right].
\end{equation}
At fixed $\alpha$, the $n^a$ fields are Gaussian. For constant $\alpha$, the effective potential per unit volume is
\begin{equation}
V_{\rm eff}(\alpha) = -\frac{\sqrt N\,\alpha}{2g(\mu)} + \frac{\sqrt N\,\alpha}{8\pi}\left[\ln\frac{\mu^2}{\alpha/\sqrt N} + 1\right],
\end{equation}
where an $\alpha$-independent vacuum-energy term has been dropped. Extremizing with respect to $\alpha$ gives the saddle
\begin{equation}
\frac{1}{g(\mu)} = \frac{1}{4\pi}\ln\frac{\mu^2}{\alpha/\sqrt N},
\end{equation}
and hence
\begin{equation}
\langle m|\alpha|m\rangle = \sqrt N\,m^2,\qquad m^2 = \mu^2 e^{-4\pi/g(\mu)}.
\end{equation}

The field $\alpha$ is a Lagrange multiplier, not a Hubbard--Stratonovich scalar. Thus the correct steepest-descent contour for the microscopic $\alpha$ integral is rotated relative to the real axis. This is visible already from the effective potential: $V_{\rm eff}''(\sqrt N m^2) = -1/(8\pi m^2)$, so the saddle is a maximum along the real $\alpha$ direction. The fluctuation propagator below is written after the conventional rotation to the steepest-descent contour. This field-space contour rotation is separate from the Picard--Lefschetz thimbles in the reduced $(t,\rho)$ plane. The latter encode the lateral prescriptions of the perturbative expansion and of the renormalized condensate, while the former is part of defining the large-$N$ saddle itself.

\subsection{Auxiliary-field exchange and the leading Borel pole}

Expanding around the saddle as $\alpha = \sqrt N m^2 + \widehat\alpha$, the $\widehat\alpha$ propagator is
\begin{equation}
\overline{\Delta}_\alpha(p^2) = 4\pi\,\frac{\sqrt{p^2(p^2+4m^2)}}{\ln\!\dfrac{\sqrt{p^2+4m^2}+\sqrt{p^2}}{\sqrt{p^2+4m^2}-\sqrt{p^2}}}.
\end{equation}
Equivalently, if
\begin{equation}
I(p^2) = \int\frac{d^2q}{(2\pi)^2}\frac{1}{(q^2+m^2)((q+p)^2+m^2)},
\end{equation}
then
\begin{equation}
I(p^2) = \frac{1}{2\pi\sqrt{p^2(p^2+4m^2)}}\ln\frac{\sqrt{p^2+4m^2}+\sqrt{p^2}}{\sqrt{p^2+4m^2}-\sqrt{p^2}},\qquad \overline{\Delta}_\alpha(p^2) = \frac{2}{I(p^2)},
\end{equation}
where the sign is the one appropriate after the $\alpha$-contour rotation.

The order-$1/N$ correction to the fundamental-field propagator is
\begin{equation}
\overline\Sigma(p^2) = \frac{1}{N}\int\frac{d^2k}{(2\pi)^2}\,\frac{1}{(p+k)^2+m^2}\,\overline{\Delta}_\alpha(k^2).
\end{equation}
Following Ref.~\cite{BenekeBraunKivel1998}, the leading perturbative sector can be represented as
\begin{equation}
\overline\Sigma^{(0)}(p^2) = -\frac{p^2}{N}\int_0^\infty dt\,e^{-t/\lambda(p)}\left[\psi(2-t)+\psi(t-1)+2\gamma_E-\frac{1}{\lambda(p)}\right],
\end{equation}
where
\begin{equation}
\lambda(p) = \frac{1}{\ln(p^2/m^2)}.
\end{equation}
Near $t = 1$, $\psi(t-1)$ has a simple pole. More explicitly,
\begin{equation}
\psi(t-1) = -\frac{1}{t-1}-\gamma_E+\cdots,\qquad \psi(2-t) = -\gamma_E+\cdots,
\end{equation}
so the leading singular part of the self-energy is
\begin{equation}
\overline\Sigma^{(0)}(p^2)\big|_{t\simeq 1} = \frac{p^2}{N}\int_0^\infty dt\,e^{-t/\lambda(p)}\frac{1}{t-1} + \text{terms regular at }t = 1.
\end{equation}
For the propagator,
\begin{equation}
G(p^2) = \frac{1}{p^2+m^2+\overline\Sigma(p^2)} = C_{\mathbf 1}(p^2)+C_\alpha(p^2)\langle\alpha\rangle+\cdots,
\end{equation}
the identity coefficient contains
\begin{equation}
C_{\mathbf 1}^{(t=1)}(p^2) = -\frac{1}{Np^2}\int_0^\infty dt\,e^{-t/\lambda(p)}\frac{1}{t-1}+\cdots .
\end{equation}
Let $\mathcal C_{\rm up}$ and $\mathcal C_{\rm down}$ denote contours passing above and below the pole, respectively. With
\begin{equation}
\frac{1}{t-1\pm i0} = \operatorname{PV}\frac{1}{t-1}\mp i\pi\delta(t-1),
\end{equation}
where the upper and lower contours correspond to $+i0$ and $-i0$, respectively, the identity coefficient has
\begin{equation}
\operatorname{Im}C_{\mathbf 1,{\rm up/down}}^{(t=1)}(p^2) = \pm\frac{\pi m^2}{Np^4}.
\end{equation}

\subsection{The \texorpdfstring{$\alpha$}{alpha} condensate}

The condensate that cancels the leading pole is $\langle\alpha\rangle$. At leading order in $1/N$,
\begin{equation}
\langle m|\alpha|m\rangle = \sqrt N\,m^2.
\end{equation}
At the next order, the constraint equation
\begin{equation}
\langle n^a(0)n^a(0)\rangle = \frac{N}{g}
\end{equation}
determines the tadpole correction. Writing $\alpha = \sqrt N m^2+\alpha^{(1)}+\cdots$ and expanding the propagator in the constraint gives
\begin{equation}
0 = -N\int\frac{d^2\ell}{(2\pi)^2}\frac{\alpha^{(1)}/\sqrt N+\overline\Sigma(\ell^2)}{(\ell^2+m^2)^2}+\cdots .
\end{equation}
Since
\begin{equation}
\int\frac{d^2\ell}{(2\pi)^2}\frac{1}{(\ell^2+m^2)^2} = \frac{1}{4\pi m^2},
\end{equation}
this implies
\begin{equation}
\alpha^{(1)} = -4\pi m^2\sqrt N\int\frac{d^2\ell}{(2\pi)^2}\frac{\overline\Sigma(\ell^2)}{(\ell^2+m^2)^2}.
\end{equation}
One can then perform the $\ell$ integral:
\begin{align}
\alpha^{(1)}
&= -\frac{4\pi m^2}{\sqrt N}\int\meas{k}{d}\meas{\ell}{d}\,\frac{\overline{\Delta}_\alpha(k^2)}{(\ell^2 + m^2)^2((\ell+k)^2 + m^2)} \nonumber\\
&= -\frac{1}{\sqrt N}\int\meas{k}{d}\,\overline{\Delta}_\alpha(k^2)\left[\frac{1}{k^2 + 4m^2}+\frac{4m^2}{\sqrt{k^2}(k^2 + 4m^2)^{3/2}}\operatorname{artanh}\sqrt{\frac{k^2}{k^2 + 4m^2}}\right] \nonumber\\
&= -\frac{m^2}{\sqrt N}\int_1^\infty dz\,\left[\left(1-\frac{1}{z}\right)^2\frac{1}{\ln z}+\frac{2(z-1)}{z(z+1)}\right],
\end{align}
where in the last line we used the same $z$ variable as in the main text. The first term in the $z$-space kernel carries the leading positive-axis singularity, while the second term is regular at $t = 1$. Using $\frac{1}{\ln z} = \int_0^\infty dt\,z^{-t}$ and writing $z = e^\rho$, the singular part of the first term is governed locally by the same reduced action $S(t,\rho) = (t - 1)\rho$ as in the Gross--Neveu analysis. The additional term $2(z-1)/[z(z+1)]$ affects the real renormalized condensate but does not modify the $t = 1$ residue.

Up to real scheme-dependent subtractions, the singular part of the renormalized condensate may therefore be written as
\begin{equation}
\langle m|\alpha|m\rangle_{\rm ren}^{(1)} = -\frac{m^2}{\sqrt N}\int_{\mathcal C_{\rm up/down}}dt\,\left[\frac{1}{t-1}+\text{terms regular at }t = 1\right].
\end{equation}
At finite cutoff the regulated condensate is real; the lateral ambiguity appears only after the ultraviolet subtractions have been made and the cutoff has been removed. The corresponding lateral imaginary parts are
\begin{equation}
\operatorname{Im}\langle m|\alpha|m\rangle_{\rm up/down} = \pm\frac{\pi m^2}{\sqrt N},
\end{equation}
or equivalently
\begin{equation}
\langle m|\alpha|m\rangle_{\rm up/down} = \sqrt N\,m^2\left(1\pm\frac{i\pi}{N}+\cO(N^{-2})\right).
\end{equation}

The Wilson coefficient of $\alpha$ follows from expanding the leading propagator in a slowly varying $\alpha$ background:
\begin{equation}
\frac{1}{p^2+\alpha/\sqrt N} = \frac{1}{p^2}-\frac{\alpha}{\sqrt N\,p^4}+\cdots,
\end{equation}
so
\begin{equation}
C_\alpha(p^2) = -\frac{1}{\sqrt N\,p^4}.
\end{equation}
The condensate contribution therefore has
\begin{equation}
C_\alpha(p^2)\operatorname{Im}\langle m|\alpha|m\rangle_{\rm up/down} = \mp\frac{\pi m^2}{Np^4},
\end{equation}
which cancels the ambiguity of the identity coefficient.

\subsection{Comment on Wilsonian treatments}

Many discussions of the $O(N)$ model use a Wilsonian factorization scale $\mu_F$ rather than the continuum-renormalized OPE used in the main text. In such a scheme, modes above $\mu_F$ are assigned to Wilson coefficients and modes below $\mu_F$ remain in the matrix elements. The separation is arbitrary, so the $\mu_F$ dependence cancels between coefficients and condensates. This reorganizes the factorially growing pieces between the coefficient and matrix element, and in such a scheme the cutoff-dependent condensate can itself have an asymptotic expansion in $g(\mu_F)$.

This Wilsonian reorganization is compatible with the continuum picture, but it is not the organization used for the homological cancellation in Sections~\ref{sec:thimble} and~\ref{sec:OPE}. There, the regulator has been removed, the Wilson coefficient and condensate are separately defined by lateral prescriptions, and their imaginary parts cancel because the corresponding relative cycles pass through the same reduced saddle.

\bibliographystyle{JHEP}
\bibliography{GNRenormalon}

\end{document}